\documentclass[american,3p]{elsarticle}
\usepackage[latin9]{inputenc}
\usepackage{geometry}
\usepackage{color}
\usepackage{babel}
\usepackage{mathrsfs}
\usepackage{amsmath}
\usepackage{amssymb}
\usepackage{graphicx}
\usepackage{esint}

\makeatletter

\newcommand{\noun}[1]{\textsc{#1}}
\providecommand{\tabularnewline}{\\}

\usepackage{pslatex}
\usepackage{enumerate}
\usepackage{fancyhdr}
\usepackage{soul}
\usepackage{cancel}
\renewcommand{\boldsymbol}[1]{\pmb{#1}}

\RequirePackage{lineno}
\usepackage{xcolor}
\usepackage{mathrsfs}
\biboptions{sort&compress}

\usepackage{babel}

\@ifundefined{showcaptionsetup}{}{%
 \PassOptionsToPackage{caption=false}{subfig}}
\usepackage{subfig}
\makeatother

\begin{document}
\begin{frontmatter}

\title{Performance portability of lattice Boltzmann methods for two-phase flows with phase change}

\author[CEA]{\noun{Werner Verdier}}

\ead{werner.verdier@cea.fr}

\address[CEA]{DES, ISAS, DM2S, STMF, LMSF, CEA, Université de Paris-Saclay, F-91191, Gif-sur-Yvette, France.}

\author[CEA2]{\noun{Pierre Kestener}}

\address[CEA2]{DRF -- Maison de la Simulation, CEA, Université de Paris-Saclay, F-91191, Gif-sur-Yvette, France.}

\ead{pierre.kestener@cea.fr}

\author[CEA]{\noun{Alain Cartalade}\corref{cor1}}

\ead{alain.cartalade@cea.fr}

\cortext[cor1]{Corresponding author. Tel.:+33 (0)1 69 08 40 67}

\begin{abstract}

Numerical codes using the lattice Boltzmann methods (LBM) for simulating
one- or two-phase flows are widely compiled and run on graphical process
units. However, those computational units necessitate to re-write
the program by using a low-level language which is suited to those
architectures (e.g. \texttt{CUDA} for GPU NVIDIA$^{\circledR}$ or
\texttt{OpenCL}). In this paper we focus our effort on the performance
portability of LBM i.e. the possibility of writing LB algorithms with
a high-level of abstraction while remaining efficient on a wide range
of architectures such as multicores x86, GPU NVIDIA$^{\circledR}$,
ARM, and so on. For such a purpose, implementation of LBM is carried
out by developing a unique code, \texttt{LBM\_saclay} written in the
C++ language, coupled with the \texttt{Kokkos} library for performance
portability in the context of High Performance Computing. In this
paper, the LBM is used to simulate a phase-field model for two-phase
flow problems with phase change. The mathematical model is composed
of the incompressible Navier-Stokes equations coupled with the conservative
Allen-Cahn model. Initially developed in the literature for immiscible
binary fluids, the model is extended here to simulate phase change
occurring at the interface between liquid and gas. For that purpose,
a heat equation is added with a source term involving the time derivative
of the phase field. In the phase-field equation a source term is added
to approximate the mass production rate at the interface. Several
validations are carried out to check step-by-step the implementation
of the full model. Finally, computational times are compared on CPU
and GPU platforms for the physical problem of film boiling.

\end{abstract}

\begin{keyword}

Lattice Boltzmann method, phase-field model, two-phase flows with
phase change, performance portability, \texttt{Kokkos} library, \texttt{LBM\_saclay},
conservative Allen-Cahn model.

\end{keyword}

\end{frontmatter}

\section{\label{sec:Introduction}Introduction}

The Lattice Boltzmann Method (LBM) \citep{BookLBM2017,BookLBM_2013}
is a very attractive method to simulate problems involving fluid flows.
Since more than ten years, numerical codes using that method are widely
compiled and run on Graphical Process Units (GPU) \citep{Li_etal_VisComp2003,Tolke_CompVisSci2008,Kuznik_etal_CAMWA2010,Zhou_etal_CMAME2012}.
The GPUs allow for a very high calculation throughput and they are particularly efficient for repetitive workloads with simple memory access patterns. These units were initially designed for image processing or graphics rendering, but LBM simulations can also benefit from their use, because the stages of streaming and collide are two simple (stencil-like) computational operations. Numerous works have demonstrated the efficiency of LBM on single GPU (e.g. \citep{Kuznik_etal_CAMWA2010}) and later on clusters of GPUs (e.g. \citep{Obrecht_CAMWA2013,Januszewski-Kostur_CPC2014}).
However, those computational
units necessitate to re-write the code by using a low-level language
which is suited to their specific architectures (e.g. \texttt{CUDA}
for GPU NVIDIA$^{\circledR}$ or \texttt{OpenCL}). In this paper we
focus our effort on the performance portability of LBM i.e. the possibility
of writing LBM algorithms with a high-level of abstraction, but by
remaining efficient on a wide range of architectures such as multicores
x86, GPU NVIDIA$^{\circledR}$, ARM, and so on.

The issue of performance
portability has already been studied and implementation of numerical
algorithms running on various architectures (GPU and so on) can be
done by directive approaches (mostly \texttt{OpenMP} or \texttt{OpenACC}).
Directive-based parallel programming solutions consist in decorating source code with comments that are interpreted by the compiler to derive the actual parallel code. They are useful when porting a legacy simulation code with a large number of lines, for which it is not reasonable to rewrite it from scratch. However, those programming models deal with computational patterns (\texttt{for} loops, reduction loops, ...) and do not provide tools for data or memory containers.
Here we present an application of a more promising approach that uses
a library-based solution which offers high-level abstract programming concepts and hardware agnostic solution for a better integration into C++ codes. Among libraries sharing the same goal of performance portability (like \texttt{RAJA} or \texttt{SYCL}),
the \texttt{Kokkos} library \citep{Carter_etal_JPDC2014} is used for
simulating two-phase flows with LBM. \texttt{Kokkos} implements
a programming model in C++ for writing performance portable applications
targeting all major High Performance Computing (HPC) platforms. Programming
tools provide abstractions for both parallel execution of code and
data management, i.e. they provide memory containers (multidimensional arrays) where the actual memory layout  will be chosen by the library during compilation. Directive-based solution does not provide such advanced features regarding memory. The \texttt{Kokkos} library can currently use \texttt{OpenMP}, \texttt{Pthreads}
and \texttt{CUDA} as backend programming models. The library has already
been applied to accelerate high-order mesh optimization in \citep{Eichstadt_etal_CPC2018}.

Because of its explicit scheme and local interactions, the LBM ideally exploits the massively parallel supercomputers based on either CPUs or GPUs or heterogeneous architectures. In this paper, we take advantage of those benefits to study two-phase flows. Several topical reviews exist in the literature for modeling two-phase
flows in LBM framework \citep{Huang-Sukop-Lu_Multiphase_Book2015,Li_etal_ProgECS2016}. The main families of methods are the color-gradient method \citep{Gunstensen_etal_PRA1991}, the pseudo-potential method \citep{Shan-Chen_PRE1993,Shan-Chen_PRE1994}, the free-energy method \cite{Swift_etal_PRE1996}, and the phase-field method \citep{Jacqmin_JCP1999}. Most of approaches consider the interface as a diffuse zone (characterized
by a thickness and a surface tension) which can be seen as a small
region of transition between bulk phases. In pseudo-potential methods
\citep{Shan-Chen_PRE1993,Shan-Chen_PRE1994} an additional force term
is added in the Navier-Stokes equations to take into account an equation
of state which is not the classical law of perfect gases \citep{Yuan-Schaefer_PoF2006}.
In that case, the density plays the role of a phase index varying
smoothly between densities of gas and liquid. Several recent applications
use that method for simulating liquid-gas phase change \citep{Li-Cheng_IJHMT2017,Li_etal_IJHMT2017}.
Another class of diffuse interface methods is the color-gradient 
model \citep{Gunstensen_etal_PRA1991} for which
two distribution functions are introduced for computation of each
phase (red and blue). In those approaches, surface tension is derived
from a recoloring step involving both distribution functions \citep{Leclaire_etal_AMM2012, Leclaire_etal_IJMF2013}.
The final approach that is commonly applied in the LBM literature is to capture the interfacial behavior through a phase-field equation.
In this paper, we follow this latter method: the phase-field theory for
two-phase flows \citep{Anderson_etal_Review_AnRevFluiMech1998}.
The phase-field method is quite similar to the free-energy lattice Boltzmann method \cite{Swift_etal_PRE1996} in the sense that both models are thermodynamically consistent and can be derived from a free-energy functional. However, in the free-energy LB approach, the density gradient appears explicitly in the free-energy functional and the phase separation is described by a non-ideal equation of state. For that purpose, the equilibrium distribution function is modified to include a non-ideal thermodynamic pressure tensor. In this paper, both fluids are considered as quasi-incompressible, i.e. we assume that the incompressibility condition holds in the bulk phases except in the interfacial zone where the mass production rate $\dot{m}'''$ acts. That mass production rate comes from the phase change between the gas and the liquid. A new function $\phi$ is introduced to track the interfacial zone where the density varies.

Two main phase-field models for interface tracking between two immiscible
fluids exist in literature: the first one is the Cahn-Hilliard (CH)
model \citep{Cahn-Hilliard_JCP1958,Jacqmin_JCP1999,Jasnow-Vinals_PhysFluids1996}
which was extensively applied in LBM literature for simulating spinodal
decomposition \citep{Kendon_etal_SpinodalDecomp_JFM2001}, buoyancy
of bubbles \citep{Zheng_etal_LargeDensityRatio_JCP2006}, drop impact
\citep{Lee-Liu_DropImpact_LBM_JCP2010}, Rayleigh-Taylor instability
\citep{Zu-He_PRE2013} and so on. The second one is a more recent
model, called the conservative Allen-Cahn (CAC) model, which was first
developed in \citep{Sun-Beckermann_JCP2007} and derived in conservative
form in \citep{Chiu-Lin_JCP2011}. The model became popular in the
LB community \citep{Geier_etal_PRE2015,Fakhari_etal_JCP2017,Mitchell_etal_IJMF2018}
and several papers compare the Cahn-Hilliard and conservative Allen-Cahn
models, e.g. \citep{Lee_Kim_MathCompSim2016} without LBM and \citep{Wang_etal_Compare-CH-CAC_PRE2016}
with LBM. In this work the CAC model is chosen for interface tracking
in order to eliminate the curvature-driven interface motion which
is implicitly contained in the CH equation (see Section \ref{sec:Phase-field_Model}).
Moreover, the CAC model involves only a second-order derivative and
does not require to compute the fourth-order derivative (Laplacian
of chemical potential) which appears in the CH equation.

In this paper, we take advantage of the simplicity of LBM to develop a
new portable code for simulating two phase flows with the coupled
Navier-Stokes/conservative Allen-Cahn (NS/CAC) model. The new code,
called \texttt{LBM\_saclay}, targets all major HPC platforms such
as multi-GPUs and multi-CPUs. In this paper, we also check the capability
of the NS/CAC model to simulate phase change problems in the vicinity
of the critical temperature. Near the critical temperature, properties
of each phase vary smoothly and the range of variation of those parameters
remains small. Several fluid flow models of phase change have already
been proposed in the literature with the Cahn-Hilliard equation \cite{Dong_etal_NHTPartA2009,Safari_etal_PRE2013}.
Following those references, the NS/CAC model is extended here by adding
a source term in both the mass balance and the CAC equations. The
source term involves the mass production rate $\dot{m}'''$ occurring
at the interface. In references \cite{Dong_etal_NHTPartA2009,Safari_etal_PRE2013},
the liquid is often considered at saturation temperature and its thermal
conductivity is neglected. Under those assumptions, $\dot{m}'''$
is calculated by a gradient operator (Fourier's law) involving only
the thermal conductivity of gas. Moreover, in order to avoid computing
the temperature equation in liquid phase (because the thermal conductivity
is neglected), a cut-off value of the phase-field is introduced beyond
which the temperature equation will not be computed \cite{Safari_etal_PRE2013}.
Here we propose an alternative way to calculate $\dot{m}'''$ that
avoids computing this gradient and avoids introducing this cut-off
value.
For that purpose, $\dot{m}'''$ will be related to the normal interface velocity and expressed as a source term close to what is done in solidification models (section \ref{sub:Alternative-forms}).
Implementation of lattice Boltzmann methods will be checked
step-by-step by considering separately solutions of the phase-field
equation, the phase-field coupled successively with a fluid flow,
and the phase-field coupled only with temperature for which the ratio
of physical properties remain low.
Finally, the aspects of two-phase flow, phase change and heat transfer are coupled to simulate the phenomena of film boiling \cite{Review_FilmBoilingIJHMT2017}.

This paper is organized as follows. Section \ref{sec:Phase-field_Model}
presents the continuous mathematical model based on the conservative
Allen-Cahn equation which is extended to handle phase change. The
model derivation will be reminded, as well as definition of the chemical
potential and interpolation methods for kinematics viscosities and
densities. Section \ref{sec:Lattice-Boltzmann-methods} presents the
Lattice Boltzmann schemes based on the Bhatnagar-Gross-Krook (BGK)
collision operator for each equation. 
That collision operator is chosen because of its simplicity of implementation. Several improvements exist such as the two-relaxation-times (TRT) and the multiple-relaxation-times (MRT). Their benefits will be quickly discussed in that section.
Computation of gradient and
Laplacian operators that are involved in equations of phase-field
and fluid flow will also be specified. Details on numerical implementation
with the \texttt{Kokkos} library and various optimizations of LBM
kernel will be discussed in Section \ref{sub:KernelOptimization}.
In Section \ref{sec:Validations}, several basic code verifications are presented
to check the implementation of each equation step-by-step.
In Section \ref{sec:FilmBoiling}, two purely qualitative simulations will be presented on the two-dimensional test case of film boiling. The first one will illustrate the capability of the model to simulate the detachment of bubbles on nodes and antinodes. The second one will illustrate the influence of the Jacob number on their detachment and shape. Here, we give a comparison of the code performance running on two architectures (CPU Intel and GPU NVIDIA$^{\circledR}$). Finally, Section \ref{sec:Conclusion} and three appendices will conclude this paper.

\section{\label{sec:Phase-field_Model}Two-phase flow with mass transfer}

\subsection{Phase change model}
A single component fluid is considered, which can be either in a liquid ($l$) or gas ($g$) phase.
The system is then composed with two incompressible
fluids with constant densities $\rho_{l}$ and $\rho_{g}$. A phase
index $\phi\equiv\phi(\mathbf{x},\,t)$ is introduced which can vary
between 0 and 1 with $\phi=0$ (respectively $\phi=1$) corresponding
to fluid $l$ (resp. $g$) which is characterized by its density $\rho_{l}$
(resp. $\rho_{g}$) and its kinematic viscosity $\nu_{l}$ (resp.
$\nu_{g}$). All other values of $\phi$ represent the interfacial
zone or a mixture of both fluids $l$ and $g$. When $0<\phi<1$,
 the densities $\rho(\phi)$ and the kinematic viscosities $\nu(\phi)$ are
respectively interpolated by

\begin{subequations}

\begin{align}
\rho(\phi) & =\phi(\mathbf{x},\,t)\rho_{g}+(1-\phi(\mathbf{x},\,t))\rho_{l},\label{eq:Density}\\
\nu(\phi) & =\frac{\nu_{l}\nu_{g}}{\phi(\mathbf{x},\,t)\nu_{l}+(1-\phi(\mathbf{x},\,t))\nu_{g}}.\label{eq:Viscosity}
\end{align}
Local densities depending on position and time are noted $\tilde{\rho}_{\chi}$
(for $\chi=g,\,l$) and write $\tilde{\rho}_{g}(\mathbf{x},\,t)=\rho_{g}\phi(\mathbf{x},\,t)$
and $\tilde{\rho}_{l}(\mathbf{x},\,t)=(1-\phi(\mathbf{x},\,t))\rho_{l}$.
The total density writes $\rho(\mathbf{x},\,t)=\rho_{g}\phi(\mathbf{x},\,t)+(1-\phi(\mathbf{x},\,t))\rho_{l}$.
The method of harmonic mean is used in this work to interpolate the viscosity (Eq. (\ref{eq:Viscosity})) for simulating flows with viscosity contrast (\citep[Eq. (29c)]{Zu-He_PRE2013}). A comparison of both interpolation methods (linear and harmonic mean) is presented on the double-Poiseuille flow in Section \ref{sub:Without-mass-transfer}.
The local velocity $\mathbf{u}_{\chi}$ of each component $\chi$
is related to the volume averaged velocity $\mathbf{u}$, the constant
bulk density value $\rho_{\chi}$, and the volume diffusive flow rate
$\mathbf{j}_{\chi}$ by \cite{Lee-Liu_DropImpact_LBM_JCP2010} $\rho_{\chi}\mathbf{j}_{\chi}=\tilde{\rho}_{\chi}(\mathbf{u}_{\chi}-\mathbf{u})$
i.e. $\tilde{\rho}_{\chi}\mathbf{u}_{\chi}=\tilde{\rho}_{\chi}\mathbf{u}+\rho_{\chi}\mathbf{j}_{\chi}$.
The mass balance equations for each phase $g$ and $l$ writes

\end{subequations}

\begin{subequations}

\begin{align}
\frac{\partial\tilde{\rho}_{g}}{\partial t}+\boldsymbol{\nabla}\cdot\left(\tilde{\rho}_{g}\mathbf{u}+\rho_{g}\mathbf{j}_{g}\right) & =+\dot{m}'''\mathrm{,}\label{eq:Bilan_PhaseA}\\
\frac{\partial\tilde{\rho}_{l}}{\partial t}+\boldsymbol{\nabla}\cdot\left(\tilde{\rho}_{l}\mathbf{u}+\rho_{l}\mathbf{j}_{l}\right) & =-\dot{m}'''\mathrm{,}\label{eq:Bilan_PhaseB}
\end{align}
where $\dot{m}'''$ is the volumic production term ($+$) or sink
term ($-$) due to phase change. Its physical dimension is M.L$^{-3}$.T$^{-1}$
and its computation will be discussed in Section \ref{sub:Alternative-forms}.
In Eqs. (\ref{eq:Bilan_PhaseA}) and (\ref{eq:Bilan_PhaseB}), signs
are chosen such as the phase change produces gas phase $g$ to the
detriment of liquid phase $l$. The mass flux relative to advection
in each phase is $\tilde{\rho}_{\chi}\mathbf{u}$. In interfacial
region, the mass flux $\rho_{\chi}\mathbf{j}_{\chi}$ has a diffusive
origin and results of a regular transition of composition between
two phases. By expressing Eqs. (\ref{eq:Bilan_PhaseA}) and (\ref{eq:Bilan_PhaseB})
with respect to $\phi(\mathbf{x},\,t)$ and assuming that the fluxes
$\mathbf{j}_{g}$ and $\mathbf{j}_{l}$ are identical and opposite,
$\mathbf{j}=\mathbf{j}_{g}=-\mathbf{j}_{l}$, the following equations
are obtained:

\end{subequations}

\begin{subequations}

\begin{align}
\frac{\partial\phi}{\partial t}+\boldsymbol{\nabla}\cdot\left(\mathbf{u}\phi+\mathbf{j}\right) & =+\frac{\dot{m}'''}{\rho_{g}}\mathrm{,}\label{eq:Bilan_PhaseA_Phi}\\
\frac{\partial(1-\phi)}{\partial t}+\boldsymbol{\nabla}\cdot\left(\mathbf{u}(1-\phi)-\mathbf{j}\right) & =-\frac{\dot{m}'''}{\rho_{l}}\mathrm{,}\label{eq:Bilan_PhaseB_Phi}
\end{align}
which after summing yield

\end{subequations}

\begin{equation}
\boldsymbol{\nabla}\cdot\mathbf{u}=\dot{m}'''\left(\frac{1}{\rho_{g}}-\frac{1}{\rho_{l}}\right).\label{eq:ConservMasse_avecChgtPhase}
\end{equation}

To derive the interface tracking equation, in references \cite{Lee-Liu_DropImpact_LBM_JCP2010,Safari_etal_PRE2013}
the flux $\mathbf{j}$ is assumed to be given by the Cahn-Hilliard
flux defined by $\mathbf{j}=-M_{\phi}\boldsymbol{\nabla}\mu_{\phi}$
where $\mu_{\phi}$ is the chemical potential. In that case Eq. (\ref{eq:Bilan_PhaseA_Phi})
becomes the CH equation with a source term of production
in the second member. The Navier-Stokes/Cahn-Hilliard (NS/CH) model
is very popular for simulations of two-phase flow since more than
twenty years (e.g. without LBM \cite{Jasnow-Vinals_PhysFluids1996,Jacqmin_JCP1999}
and \cite{Kendon_etal_SpinodalDecomp_JFM2001,Zheng_etal_LargeDensityRatio_JCP2006,Zu-He_PRE2013,Lee-Liu_DropImpact_LBM_JCP2010}
with LBM). However the chemical potential can be interpreted as the
product of surface tension $\sigma$ and curvature $\kappa$ (see
details in Section \ref{sub:Chemical-potential-and}), and the CH
equation imposes in its formulation a motion due to $\sigma$ and
$\kappa$ even without coupling with a fluid flow. Here, in order
to eliminate the curvature-driven interface motion inside the phase-field
equation, we assume that the flux is defined by \cite{Sun-Beckermann_JCP2007,Chiu-Lin_JCP2011}
$\mathbf{j}=-M_{\phi}(\boldsymbol{\nabla}\phi-4\phi(1-\phi)\mathbf{n}/W)$
and Eq. (\ref{eq:Bilan_PhaseA_Phi}) becomes the conservative Allen-Cahn
(CAC) model with a source term:

\begin{equation}
\frac{\partial\phi}{\partial t}+\boldsymbol{\nabla}\cdot(\mathbf{u}\phi)=\boldsymbol{\nabla}\cdot\left[M_{\phi}\left(\boldsymbol{\nabla}\phi-\frac{4}{W}\phi(1-\phi)\mathbf{n}\right)\right]+\frac{\dot{m}'''}{\rho_{g}}.\label{eq:Bilan_PhaseA_avecFlux}
\end{equation}
In Eq. (\ref{eq:Bilan_PhaseA_avecFlux}), $M_{\phi}$ is the interface
mobility, $W$ is the diffuse interface width and

\begin{equation}
\mathbf{n}=\frac{\boldsymbol{\nabla}\phi}{\bigl|\boldsymbol{\nabla}\phi\bigr|}\label{eq:NormalVector}
\end{equation}
is the unit normal vector at the interface directed from liquid toward
gas. Eq. (\ref{eq:Bilan_PhaseA_avecFlux}) is the Conservative version
of Allen-Cahn (CAC) equation with a source term for modeling interface
tracking with phase change.
The accuracy of the phase-field simulations depends on two parameters: the interface thickness $W$ and the mobility $M_{\phi}$. In reference \citep[Sec. 5]{Jacqmin_JCP1999}, a discussion is given regarding the numerical convergence of the phase-field method and the choice of those parameters in relation to the discretization step $\delta x$. For the Cahn-Hilliard equation, the mobility affects the thickness and perturbation magnitude of the chemical potential boundary layers. Here, for simulations of film boiling, preliminary sensitivity tests are performed on $M_{\phi}$ and some details of its effects will be given in Section \ref{sec:FilmBoiling}.
The choice of $\dot{m}'''$ will be discussed
in Section \ref{sub:Alternative-forms}. In the original paper \cite{Sun-Beckermann_JCP2007},
this equation is derived by assuming that the total advection velocity
is composed of two terms: the external advective velocity $\mathbf{u}$,
plus the normal velocity to the interface $u_{n}\mathbf{n}$. That
velocity $u_{n}$ is also defined as the sum of one term depending on the
curvature $\kappa$, plus one independent of $\kappa$: $u_{n}\mathbf{n}=(\tilde{v}-M_{\phi}\kappa)\mathbf{n}$.
In the right-hand side of Eq. (\ref{eq:Bilan_PhaseA_avecFlux}), the
first term $\boldsymbol{\nabla}\cdot\mathbf{j}$ is an equivalent
expression to the curvature term that is corrected with a ``counter
term'' $-M_{\phi}\kappa\bigl|\boldsymbol{\nabla}\phi\bigr|$ \cite{Folch_etal_CounterTerm_PRE1999_PhysRevE.60.1724},
in order to cancel the curvature-driven interface motion. The derivation
is reminded in \ref{sec:Derivation-Allen-Cahn} by using the usual
definition of curvature $\kappa=\boldsymbol{\nabla}\cdot\mathbf{n}$
with $\mathbf{n}$ defined by Eq. (\ref{eq:NormalVector}), and introducing
the kernel function

\begin{equation}
\phi=\frac{1}{2}\left[1+\mbox{tanh}\left(\frac{2\zeta}{W}\right)\right]\label{eq:Kernel-Function}
\end{equation}
in order to give an expression of $\bigl|\boldsymbol{\nabla}\phi\bigr|$
(see Eq. (\ref{eq:Norm_gradphi}) in \ref{sec:Derivation-Allen-Cahn}):

\begin{equation}
\bigl|\boldsymbol{\nabla}\phi\bigr|=\frac{4}{W}\phi(1-\phi).\label{eq:Norm_GradPhi}
\end{equation}
That choice of kernel function imposes bulk phases for $\phi=0$ and $\phi=1$.
Similar reasoning that cancels the curvature term can be found in
\cite{Jamet-Misbah_PRE2008} in order to eliminate effects of surface
tension (inherent in phase-field models) for membranes embedded in a Newtonian
fluid. 
Let us notice that in this work the standard convention $0\leq\phi\leq1$ is used. Other conventions are possible, particularly when studying two-phase flow with high density ratio e.g. $-\phi^{\star}\leq\phi\leq\phi^{\star}$ where $\phi^{\star}$ is defined by $\rho_g$ and $\rho_l$ (e.g. \cite[Eq. (31)]{Zheng_etal_LargeDensityRatio_JCP2006}). More generally, the inequality $\phi_l\leq\phi\leq\phi_g$ can be chosen. In that case the kernel function (Eq. (\ref{eq:Kernel-Function})) and the expression of $\bigl|\boldsymbol{\nabla}\phi\bigr|$ must change. Moreover the source term in Eq. (\ref{eq:Bilan_PhaseA_avecFlux}) must be modified by (see \cite[Eq. (188)]{Li_etal_ProgECS2016}): $\dot{m}'''(\phi_g/\rho_g-\phi_l/\rho_l$). Here, that expression is simplified to $\dot{m}'''/\rho_g$ with the standard choice $\phi_g=1$ and $\phi_l=0$.

The temperature equation is derived from the conservation law of total
enthalpy $\rho\mathcal{H}$ where $\mathcal{H}$ is the enthalpy (physical
dimension E.M$^{-1}$ where E is used for Energy) as carried out in crystal growth simulations
\cite{Kobayashi_PhysD1993}:

\begin{equation}
\frac{\partial(\rho\mathcal{H})}{\partial t}+\boldsymbol{\nabla}\cdot(\mathbf{u}\rho\mathcal{H})=\boldsymbol{\nabla}\cdot(\mathcal{K}\boldsymbol{\nabla}T)\label{eq:Conservation_Enthalpie}
\end{equation}
where the diffusive flux is given by the Fourier's law
$\mathbf{j}_{T}=-\mathcal{K}\boldsymbol{\nabla}T$ with $T$ being the
temperature and $\mathcal{K}$ the thermal conductivity (physical dimension
E.T$^{-1}$.L$^{-1}$.$\Theta^{-1}$). The enthalpy is defined by
$\mathcal{H}=\mathcal{C}_{p}T+\phi\mathcal{L}$ where $\mathcal{C}_{p}$ is the
specific heat (E.M$^{-1}$.$\Theta^{-1}$) and $\mathcal{L}$ is the latent heat
of phase change (E.M$^{-1}$).  With this relation, enthalpies of liquid and gas
are respectively equal to $\mathcal{H}_{l}=\mathcal{C}_{p}T$ for $\phi=0$ and
$\mathcal{H}_{g}=\mathcal{C}_{p}T+\mathcal{L}$ for $\phi=1$. With those
notations and definitions the heat equation for temperature writes

\begin{equation}
\frac{\partial T}{\partial t}+\boldsymbol{\nabla}\cdot(\mathbf{u}T)=\alpha\boldsymbol{\nabla}^{2}T-\frac{\mathcal{L}}{\mathcal{C}_{p}}\left[\frac{\partial\phi}{\partial t}+\boldsymbol{\nabla}\cdot(\mathbf{u}\phi)\right],\label{eq:Heat-Equation_dimensionless}
\end{equation}
where $\alpha=\mathcal{K}/(\rho\mathcal{C}_{p})$ is the thermal diffusivity,
the second term in the right-hand side of Eq. (\ref{eq:Heat-Equation_dimensionless})
is interpreted as the release (or production) of latent heat during
the displacement of the interface. When $\mathbf{u}=\mathbf{0}$ the
movement of the interface is only due to phase change between liquid
and gas. Solving only Eq. (\ref{eq:Bilan_PhaseA_avecFlux}) and (\ref{eq:Heat-Equation_dimensionless})
must be equivalent to solve the Stefan problem of phase change (see
validation of Section \ref{sec:Validations}).

Finally, the complete model of two-phase flows with phase change writes:

\begin{subequations}

\begin{align}
\boldsymbol{\nabla}\cdot\mathbf{u} & =\dot{m}'''\left(\frac{1}{\rho_{g}}-\frac{1}{\rho_{l}}\right),\label{eq:ConservMasse_ChgtPhase}\\
\left[\frac{\partial(\rho\mathbf{u})}{\partial t}+\boldsymbol{\nabla}\cdot(\rho\mathbf{uu})\right] & =-\boldsymbol{\nabla}p+\boldsymbol{\nabla}\cdot\left[\eta\left(\boldsymbol{\nabla}\mathbf{u}+\boldsymbol{\nabla}\mathbf{u}^{T}\right)\right]+\mathbf{F}_{tot},\label{eq:ConvservQDM_ChgtPhase}\\
\frac{\partial\phi}{\partial t}+\boldsymbol{\nabla}\cdot(\mathbf{u}\phi) & =\boldsymbol{\nabla}\cdot\left[M_{\phi}\left(\boldsymbol{\nabla}\phi-\frac{4}{W}\phi(1-\phi)\mathbf{n}\right)\right]+\frac{\dot{m}'''}{\rho_{g}},\label{eq:CAC_ChgtPhase}\\
\frac{\partial T}{\partial t}+\boldsymbol{\nabla}\cdot(\mathbf{u}T) & =\alpha\boldsymbol{\nabla}^{2}T-\frac{\mathcal{L}}{\mathcal{C}_{p}}\left[\frac{\partial\phi}{\partial t}+\boldsymbol{\nabla}\cdot(\mathbf{u}\phi)\right].\label{eq:Temperature_ChgtPhase}
\end{align}
Eqs. (\ref{eq:ConservMasse_ChgtPhase}) and (\ref{eq:ConvservQDM_ChgtPhase})
are the Navier-Stokes equations for modeling two Newtonian and incompressible
fluids. In those equations $p$ is the pressure, $\rho(\phi)$ is
the density depending on the phase-field $\phi$ and $\eta(\phi)$
is the dynamic viscosity. $\mathbf{F}_{tot}$ is the total force term
defined as:

\end{subequations}

\begin{equation}
\mathbf{F}_{tot}=\mathbf{F}_{s}+\mathbf{F}_{v}\label{eq:Force_Total}
\end{equation}
where $\mathbf{F}_{s}$ is the surface tension force that is defined
in the next subsection. The volumic force $\mathbf{F}_{v}$ is the
buoyancy force. Among different formulations of that force \cite[Sec. 3.7]{Bower-Lee_CF2010},
in this work the buoyancy is defined such as $\mathbf{F}_{v}=(\rho_{l}-\rho(\phi))\mathbf{g}$
where $\mathbf{g}$ is the constant acceleration due to the gravity.
With that formulation, the gravity acts only on the gas phase for
simulations of film boiling in Section \ref{sec:FilmBoiling}.

\subsection{\label{sub:Chemical-potential-and}Chemical potential and Cahn-Hilliard
equation}

The surface tension force $\mathbf{F}_{s}$ is expressed here in its
potential form \citep{Jacqmin_JCP1999}:

\begin{equation}
\mathbf{F}_{s}=\mu_{\phi}\boldsymbol{\nabla}\phi\label{eq:SuperficialTensionForce}
\end{equation}
where $\mu_{\phi}$ is the chemical potential which is defined as
the change of free energy for a small variation of local composition
of mixture: $\mu_{\phi}=\delta\mathscr{F}/\delta\phi$. When the free
energy is defined such as $\mathscr{F}(\phi)=\int_{v}[\mathscr{V}(\phi)+K\left|\boldsymbol{\nabla}\phi\right|^{2}/2]dv$
with $\mathscr{V}(\phi)=H\phi^{2}(1-\phi)^{2}$, the chemical potential
writes

\begin{equation}
\mu_{\phi}=4H\phi(\phi-1)\left(\phi-\frac{1}{2}\right)-K\boldsymbol{\nabla}^{2}\phi.\label{eq:ChemicalPotential}
\end{equation}
The first term of the right-hand side of Eq. (\ref{eq:ChemicalPotential})
is the derivative of $\mathscr{V}(\phi)$ with respect to $\phi$
and the second term comes from the gradient energy term. The double-well
ensures minima at $\phi=0$ and $\phi=1$. Coefficient $H$ is the
height of double-well and $K$ is the gradient energy coefficient.
It is well-known that the one-dimensional solution at equilibrium
(i.e. $\mu_{\phi}=0$) of Eq. (\ref{eq:ChemicalPotential}) is the
hyperbolic tangent function defined by Eq. (\ref{eq:Kernel-Function}).
A dimensional analysis of $\mathscr{F}(\phi)$ indicates that $H$
has the dimension of energy per volume unit, whereas $K$ has the
dimension of energy per length unit. In this formalism, the surface
tension $\sigma$ and the diffuse interface width $W$ are proportional
to the product and the ratio of both coefficients:

\begin{subequations}

\begin{equation}
\sigma=\frac{1}{6}\sqrt{2KH}\quad\mbox{and}\quad W=\sqrt{\frac{8K}{H}}\label{eq:Sigma-Width}
\end{equation}

We also note that $\sqrt{KH}$ is homogeneous to an energy per surface
unit which corresponds to the physical dimension of surface tension.
The term $\sqrt{K/H}$ is homogeneous to a length as expected for
the interface thickness. For the simulations of section \ref{sec:Validations},
values of $\sigma$ and $W$ will be set and $K$ and $H$ will be
derived by inverting those two relationships:

\begin{equation}
K=\frac{3}{2}W\sigma\quad\mbox{and}\quad H=12\frac{\sigma}{W}.\label{eq:Def_K-H}
\end{equation}

\end{subequations}

Let us notice that, if we use Eqs. (\ref{eq:ChemicalPotential}) and
(\ref{eq:Def_K-H}), the surface tension force $\mathbf{F}_{s}=\mu_{\phi}\boldsymbol{\nabla}\phi$
can be written as $\mu_{\phi}\boldsymbol{\nabla}\phi=-(3/2)W\sigma\left[\boldsymbol{\nabla}^2\phi-16\phi(1-\phi)(1-2\phi)/W^{2}\right]\boldsymbol{\nabla}\phi$.
The term inside the brackets is the curvature term $\kappa\bigl|\boldsymbol{\nabla}\phi\bigr|$
provided that the kernel function Eq. (\ref{eq:Kernel-Function})
is used for the second term (see Eq. (\ref{eq:SecondTerm}) in \ref{sec:Derivation-Allen-Cahn}).
In that case, the surface tension $\sigma$ and the curvature $\kappa$
appear explicitly in the definition of the chemical potential $\mu_{\phi}$
and the surface tension force is $\mathbf{F}_{s}=\mu_{\phi}\boldsymbol{\nabla}\phi=-(3/2)W\sigma\kappa\bigl|\boldsymbol{\nabla}\phi\bigr|\boldsymbol{\nabla}\phi$.
Besides, if we set $K=\epsilon^{2}$ and $H=1/4$ in Eq. (\ref{eq:Sigma-Width}),
then we find $(3/2)W=6\sqrt{2}\epsilon$. The surface tension force
is $\mathbf{F}_{s}=-\sigma(6\sqrt{2}\epsilon)(\boldsymbol{\nabla}\cdot\mathbf{n})\bigl|\boldsymbol{\nabla}\phi\bigr|\boldsymbol{\nabla}\phi$
which is the same relation in \cite[Eq. (13)]{Kim_ContinuousSurfaceTension_JCP2005}
provided that the kernel function Eq. (\ref{eq:Kernel-Function})
is applied for $\kappa$. As mentioned earlier, when the diffusive
flux is proportional to the gradient of the chemical potential, then
the evolution of $\phi$ follows the Cahn-Hilliard equation:

\begin{equation}
\frac{\partial\phi}{\partial t}+\boldsymbol{\nabla}\cdot(\mathbf{u}\phi)=\boldsymbol{\nabla}\cdot(M_{\phi}\boldsymbol{\nabla}\mu_{\phi}),\label{eq:Cahn-Hilliard}
\end{equation}
with $\mu_{\phi}$ defined by Eq. (\ref{eq:ChemicalPotential}). Compared
to the standard CH equation, the main advantage of the conservative
Allen-Cahn model lies in the computation of the right-hand side term.
Indeed, the CH equation involves a fourth-order derivative because
the flux is assumed to be proportional to gradient of chemical potential.
A first Laplacian appears in Eq. (\ref{eq:ChemicalPotential}) and
a second one appears in the conservative equation Eq. (\ref{eq:Cahn-Hilliard}). In the
conservative Allen-Cahn equation (Eq. (\ref{eq:CAC_ChgtPhase})),
only the second-order derivative is involved in its definition.

\subsection{\label{sub:Alternative-forms}Production rate $\dot{m}'''$}

\subsubsection{Interface velocity of phase change}

In sharp interface methods, the surface production rate $\dot{m}''$
(physical dimension M.L$^{-2}$.T$^{-1}$) occurs on the separation
area between liquid and gas. It is usually defined by \cite{Delaye_IJMF1974,Juric-Tryggvason_IJMF1998}
$\dot{m}''=\rho_{g}(\mathbf{u}_{g}-\mathbf{V}_{I})\cdot\mathbf{n}=\rho_{l}(\mathbf{u}_{l}-\mathbf{V}_{I})\cdot\mathbf{n}$
where $\mathbf{V}_{I}$ is the velocity of the interface, and $\mathbf{u}_{l}$
and $\mathbf{u}_{g}$ are respectively the velocities on liquid and
gas sides. This relation is derived by integrating the mass conservation
across the interface. Integration of the energy conservation yields
an additional relation on $\dot{m}''$ which can be calculated in
its simplest form by the difference of heat fluxes, $\dot{m}''=(\left.\mathcal{K}\boldsymbol{\nabla}T\right|_{l}-\left.\mathcal{K}\boldsymbol{\nabla}T\right|_{g})\cdot\mathbf{n}/\mathcal{L}$.
The driving force of evaporation is the heat quantity which is transferred
at the interface. In \cite{Safari_etal_PRE2013}, the liquid is assumed
to be at saturation temperature $T_{sat}$ and in that case, only
the heat quantity of the gas is considered and the temperature equation
is solved only in the gas phase. Because of the diffuse interface,
the rate $\dot{m}''$ is transformed to a volumic quantity $\dot{m}'''$
by $\dot{m}'''=\dot{m}''\bigl|\boldsymbol{\nabla}\phi\bigr|=\mathcal{K}\boldsymbol{\nabla}T\cdot\boldsymbol{\nabla}\phi/\mathcal{L}$
where $\phi$ follows the Cahn-Hilliard equation. The model was extended
in \cite{Safari_etal_PRE2014} to include the gradient of the vapor
concentration at the liquid-vapor interface as the driving force for
vaporization. The model \cite{Safari_etal_PRE2013} was also applied
in \cite{Begmohammadi_etal_ICHMT2015} to simulate nucleate pool boiling,
including the bubble growth on and periodic departure from a superheated
wall. Several other popular mass transfer models are reviewed in \cite[Section 4.2]{Review_FilmBoilingIJHMT2017}
for phase change simulations.

Here, we notice that the source term $\dot{m}'''/\rho_{g}$ in Eq.
(\ref{eq:CAC_ChgtPhase}) can be identified as the normal velocity
of the interface $-\tilde{v}\bigl|\boldsymbol{\nabla}\phi\bigr|$
(see Eq. (\ref{eq:CAC}) in \ref{sec:Derivation-Allen-Cahn}) i.e.
$\dot{m}''/\rho_{g}=-\tilde{v}$ (because $\dot{m}'''=\dot{m}''\bigl|\boldsymbol{\nabla}\phi\bigr|$).
In Eq. (\ref{eq:CAC_ChgtPhase}), the total velocity is the sum of an external velocity $\mathbf{u}$ plus the
interface normal velocity. The latter has also been separated into
one velocity depending on the curvature $-M_{\phi}\kappa$ (which
has been canceled) plus one velocity $\tilde{v}$ independent of the
curvature. That velocity is responsible for the displacement of the
interface because of the phase change. Its expression can be approximated
by \cite[Eq. (A.5)]{Sun-Beckermann_JCP2007}:

\begin{equation}
\tilde{v}=\frac{\alpha}{\mathscr{A}}\frac{\theta_{I}-\theta}{W},\label{eq:InterfaceVelocity}
\end{equation}
where $\theta$ is the dimensionless temperature defined as $\theta=(\mathcal{C}_{p}/\mathcal{L})(T-T_{sat})$,
$\theta_{I}$ is the dimensionless interface temperature and $\mathscr{A}$
is a constant of proportionality that will be specified in section
\ref{sub:Value-of-coefficient_A}. Finally, if the kernel function
$\bigl|\boldsymbol{\nabla}\phi\bigr|=(4/W)\phi(1-\phi)$ is used (see
Eq. (\ref{eq:Norm_GradPhi})), the source term $\dot{m}'''/\rho_{g}$
in Eq. (\ref{eq:CAC_ChgtPhase}) takes the form

\begin{equation}
\frac{\dot{m}'''}{\rho_{g}}=-\tilde{v}\bigl|\boldsymbol{\nabla}\phi\bigr|=-\frac{4\alpha}{\mathscr{A}W^{2}}(\theta_{I}-\theta)\phi(1-\phi).\label{eq:ProductionRate}
\end{equation}

\subsubsection{\label{sub:Value-of-coefficient_A}Value of coefficient $\mathscr{A}$}

In order to derive the value of $\mathscr{A}$ in Eq. (\ref{eq:ProductionRate}),
we proceed by analogy with the model of phase change for solidification
and crystallization \cite{Karma-Rappel_PRE1998}. First, Eq. (\ref{eq:CAC_ChgtPhase})
with Eq. (\ref{eq:ProductionRate}) are re-written in order to make
appear the derivatives of the double-well potential $f(\phi)$ and
the interpolation function $p(\phi)$. Those functions are used in
the solidification models derived from variational formulation based
on the minimization of free energy \cite{Karma-Rappel_PRE1998}. The
interface is tracked by Eq. (\ref{eq:CAC_ChgtPhase}) by assuming
that the movement due to curvature is cancelled. That equation can
be re-written (see \ref{sec:Derivation-Allen-Cahn}):

\begin{equation}
\frac{\partial\phi}{\partial t}+\boldsymbol{\nabla}\cdot(\mathbf{u}\phi)=M_{\phi}\left[\boldsymbol{\nabla}^{2}\phi-\frac{\boldsymbol{\nabla}\phi\cdot\boldsymbol{\nabla}\bigl|\boldsymbol{\nabla}\phi\bigr|}{\bigl|\boldsymbol{\nabla}\phi\bigr|}\right]-M_{\phi}\kappa\bigl|\boldsymbol{\nabla}\phi\bigr|-\frac{4\alpha}{\mathscr{A}W^{2}}(\theta_{I}-\theta)\phi(1-\phi).\label{eq:Equiv_Eq_base}
\end{equation}
If the interface temperature is considered at saturation (i.e. $\theta_{I}=0$),
the source term is simplified to $(4\alpha/\mathscr{A}W^{2})\theta\phi(1-\phi)$.
With the kernel function Eq. (\ref{eq:Kernel-Function}), the second
term in the brackets writes (see Eq. (\ref{eq:SecondTerm})) $\boldsymbol{\nabla}\phi\cdot\boldsymbol{\nabla}\bigl|\boldsymbol{\nabla}\phi\bigr|/\bigl|\boldsymbol{\nabla}\phi\bigr|=(16/W^{2})\phi(1-\phi)(1-2\phi)$.
That term is proportional to the derivative (with respect to $\phi$)
of a double-well potential defined by $f(\phi)=H\phi^{2}(1-\phi)^{2}$
with $H=1$, hence $\boldsymbol{\nabla}\phi\cdot\boldsymbol{\nabla}\bigl|\boldsymbol{\nabla}\phi\bigr|/\bigl|\boldsymbol{\nabla}\phi\bigr|=(8/W^{2})\partial f/\partial\phi$.
Besides if we set $K\equiv\varepsilon^{2}$, then the two relationships
Eqs. (\ref{eq:Def_K-H}) with $H=1$ yields $\varepsilon^{2}=W^{2}/8$.
We also set $M_{\phi}=\varepsilon^{2}/\mathcal{T}$ where $\mathcal{T}$
is the kinetic time, then Eq. (\ref{eq:Equiv_Eq_base}) becomes

\begin{equation}
\mathcal{T}\left[\frac{\partial\phi}{\partial t}+\boldsymbol{\nabla}\cdot(\mathbf{u}\phi)\right]=\varepsilon^{2}\boldsymbol{\nabla}^{2}\phi-\frac{\partial f}{\partial\phi}-\varepsilon^{2}\kappa\bigl|\boldsymbol{\nabla}\phi\bigr|-\frac{4\mathcal{T}\alpha}{\mathscr{A}W^{2}}(\theta_{I}-\theta)\frac{\partial p}{\partial\phi}.\label{eq:Equiv_Eq_Interm}
\end{equation}

In the right-hand side of Eq. (\ref{eq:Equiv_Eq_Interm}), the second
term is the derivative of the double-well and the third term is the
counter term. The last term is the coupling with temperature which
involves the derivative (with respect to $\phi$) of an interpolation
function defined as $p(\phi)=\phi^{2}/2-\phi^{3}/3$. The factor 4
comes from the choice $a=1/2$ in the kernel function (Eq. (\ref{eq:Phi_eq}))
and we set $W_{0}=W/2$. If we compare the coupling term of reference
\cite{Karma-Rappel_PRE1998} with the last term of Eq. (\ref{eq:Equiv_Eq_Interm}),
we can identify

\begin{equation}
\lambda^{\star}=\frac{\mathcal{T}\alpha}{\mathscr{A}W_{0}^{2}},\label{eq:Lambda_CouplingTemperature}
\end{equation}
where $W_{0}^{2}=W^{2}/4$ and $\lambda$ is the coupling coefficient
in solidification/crystallization phase-field models. The star of $\lambda^{\star}$
means it is the particular value of $\lambda$ that cancels the
kinetic coefficient in the Gibbs-Thomson condition recovered by the
matched asymptotic analysis of the phase-field model. Hence, that coupling
term (Eq. (\ref{eq:Lambda_CouplingTemperature})) means this is the
particular model of phase change which cancels the kinetic coefficient
in the Gibbs-Thomson equation. Besides, the curvature term is also
removed by the counter term $-\varepsilon^{2}\kappa\bigl|\boldsymbol{\nabla}\phi\bigr|$.
Finally, the coefficient $\mathscr{A}$ is identified to the coefficient
$a_{2}$ in reference \cite{Karma-Rappel_PRE1998}. Its value is $a_{2}=0.6267$
when the phase-field varies between $-1\leq\phi\leq+1$ and when the
derivative of the interpolating function of temperature is $p_{\phi}(\phi)=1-\phi^{2}$
(the index $\phi$ indicates the derivative with respect to $\phi$).
In the present paper, the phase-field $\phi$ varies between $0$
and $1$ and the derivative of the polynomial function is $p_{\phi}=\phi(1-\phi)$.
Because of those differences, the value of $\mathscr{A}$ must be
computed from integrals obtained from the matched asymptotic expansion
of the phase-field model. In \ref{sec:Numerical-value-of_A}, details
are given to obtain $\mathscr{A}=10/48\approx0.21$, value that will
be used for all simulations of this paper.

\section{\label{sec:Lattice-Boltzmann-methods}Lattice Boltzmann schemes}

In this Section, we detail the lattice Boltzmann methods that are
used to simulate the phase change model of Section \ref{sec:Phase-field_Model}
composed of Eqs (\ref{eq:ConservMasse_ChgtPhase})--(\ref{eq:Temperature_ChgtPhase})
with Eq (\ref{eq:SuperficialTensionForce}) for surface tension force
and Eq. (\ref{eq:ProductionRate}) for mass production rate. Simulations
are performed by using three distribution functions $\vartheta_{i}(\mathbf{x},\,t)\equiv\vartheta_{i}$
for $\vartheta=f,\,h,\,s$ where $i=0,\,...,\,N_{pop}$ and $N_{pop}$
is the total number of moving directions $\mathbf{e}_{i}$ on a lattice
(defined below). The first distribution function $f_{i}$ is used
to recover the Navier-Stokes model (subsection \ref{sub:LBM-for-incompressible});
the second one $g_{i}$ is used for the phase-field equation (subsection
\ref{sub:LBM-for-phase-field}) and the last one $s_{i}$ is used
for the temperature equation (subsection \ref{sub:LBM-for-Temperature}).
Each distribution function follows its own discrete lattice Boltzmann
equation in which the collision term is considered with the Bhatnagar-Gross-Krook
(BGK) approximation.
That collision operator uses a unique relaxation parameter that is related to the diffusive parameter of the PDE (kinematic viscosity, mobility or diffusion coefficient). Several improvements exist such as the TRT \citep{Ginzburg_1_AdvWR2005} or MRT \citep{DHumieres_ProgAstr1992, dHumieres_etal_PhilTranRoySoc2002} collision operators. They both use additional relaxation parameters (only one for TRT). With MRT, some of them can be related to physical parameters (e.g. anisotropic diffusion coefficient for transport equation) and the other ones control the stability of the algorithm when increasing the Reynolds number or Péclet number. Hence a wider range of parameters can be reached when simulations are performed with TRT and MRT. Let us mention that other alternatives exist in the literature (entropic, central moments, cumulants, ...) but an in-depth discussion of their benefits and drawbacks is out of the scope of this work. In Eq. (\ref{eq:LBE}), each discrete Boltzmann
equation is expressed in terms of new variables $\overline{f}_{i}$,
$\overline{g}_{i}$ and $\overline{s}_{i}$, each one of them being
defined by an appropriate variable change \cite{He-Chen-Doolen_JCP1998}
(see details in \ref{sec:DLBE}):

\begin{equation}
\overline{\vartheta}_{i}=\vartheta_{i}+\frac{\delta t}{2\tau_{\vartheta}}\left(\vartheta_{i}-\vartheta_{i}^{eq}\right)-\frac{\delta t}{2}\mathcal{S}_{i}^{\vartheta}\quad\mbox{for }\vartheta=f,\,h,\,s,\label{eq:VariableChange_f}
\end{equation}
where $\tau_{\vartheta}$ and $\mathcal{S}_{i}^{\vartheta}$ are respectively
the collision time and the source term relative to the distribution
function $\vartheta$; $\delta t$ is the time step and $\vartheta_{i}^{eq}$
is the equilibrium distribution function. Two other notations are
introduced: $\overline{\tau}_{\vartheta}$ and $\vartheta_{i}^{\star}$.
The first one is the dimensionless collision rate that is defined
by $\overline{\tau}_{\vartheta}=\tau_{\vartheta}/\delta t$ for each
$\vartheta$. The second one is the distribution function that is
obtained after the stages of collision and streaming: $\vartheta_{i}^{\star}\equiv\vartheta_{i}(\mathbf{x}+\mathbf{c}_{i}\delta t,\,t+\delta t)$.
The use of this variable change (Eq. (\ref{eq:VariableChange_f}))
modifies the calculation of the zeroth-order moment $\mathcal{M}_{0}^{\vartheta}$
of the distribution function $\overline{\vartheta}_{i}$ by (see \ref{sec:DLBE})

\begin{equation}
\mathcal{M}_{0}^{\vartheta}=\sum_{i}\overline{\vartheta}_{i}+\frac{\delta t}{2}\mathcal{S}_{i}^{\vartheta}\quad\mbox{for }\vartheta=f,\,h,\,s.\label{eq:Moment0}
\end{equation}

\begin{figure*}
\begin{centering}
\subfloat[\label{fig:D3Q7}D3Q7]{\protect\begin{centering}
\protect\includegraphics[scale=0.25]{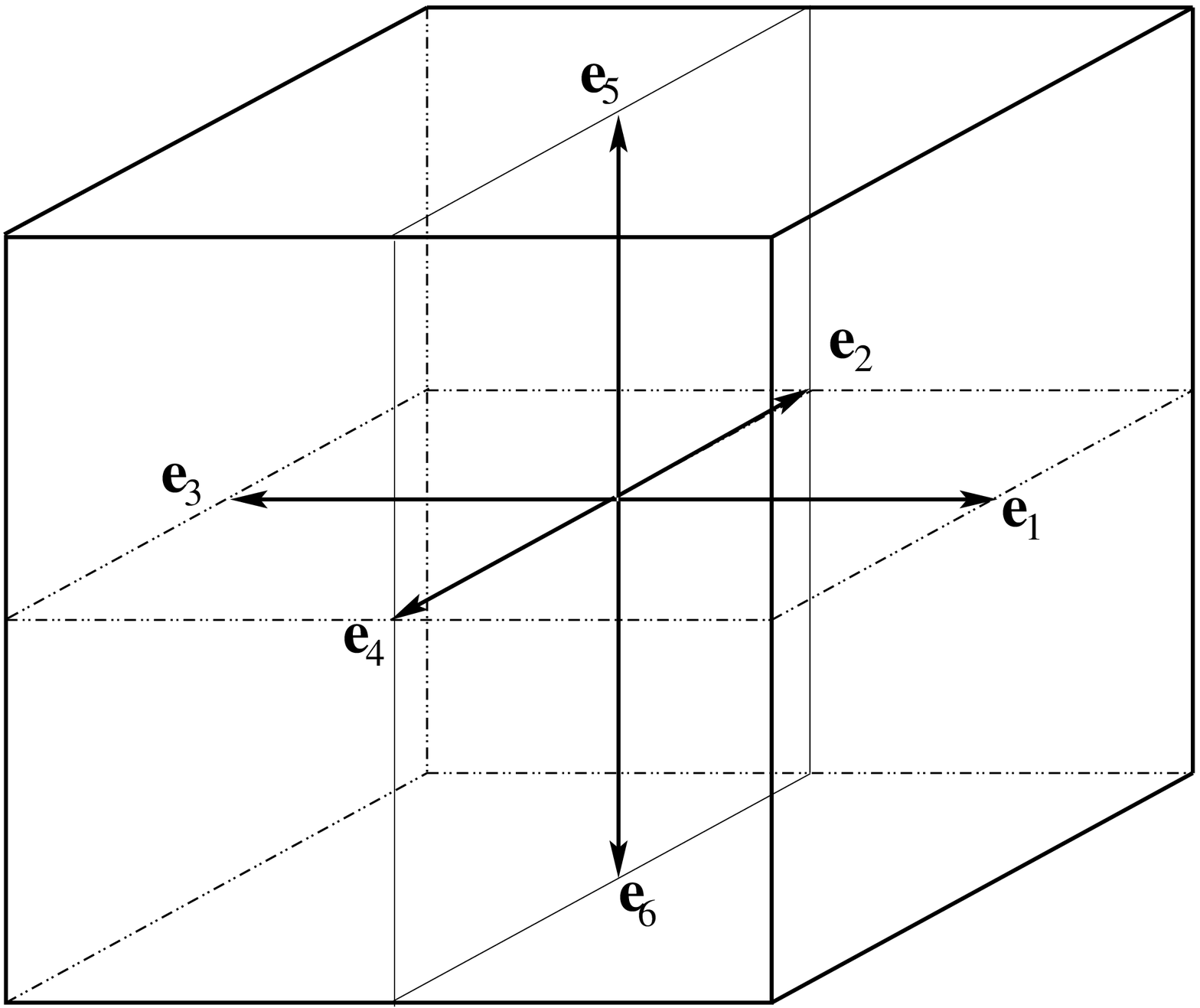}\protect
\par\end{centering}

}$\qquad$\subfloat[\label{fig:D3Q15}D3Q15]{\protect\begin{centering}
\protect\includegraphics[scale=0.25]{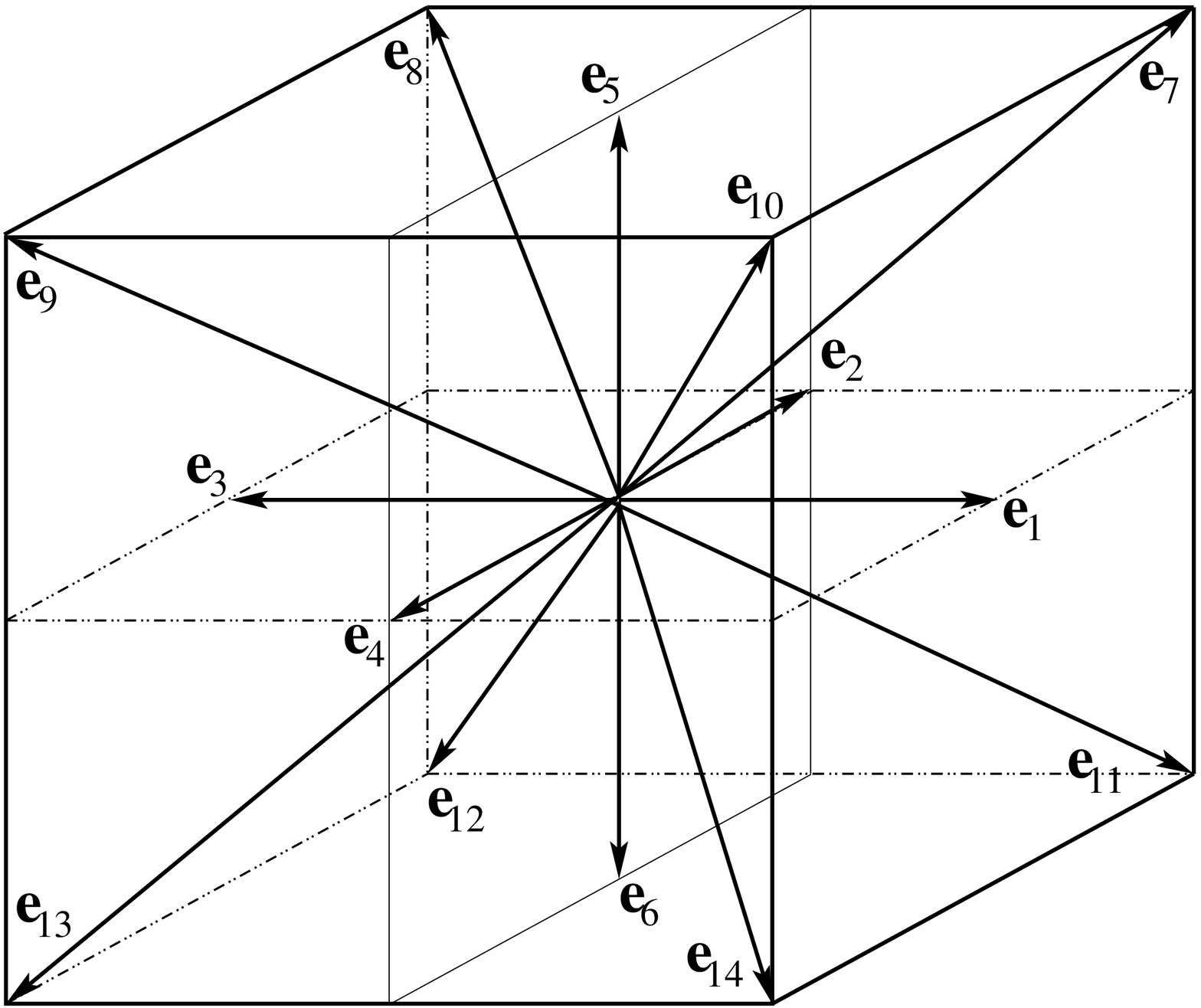}\protect
\par\end{centering}

}$\qquad$\subfloat[\label{fig:D3Q19}D3Q19]{\protect\begin{centering}
\protect\includegraphics[scale=0.25]{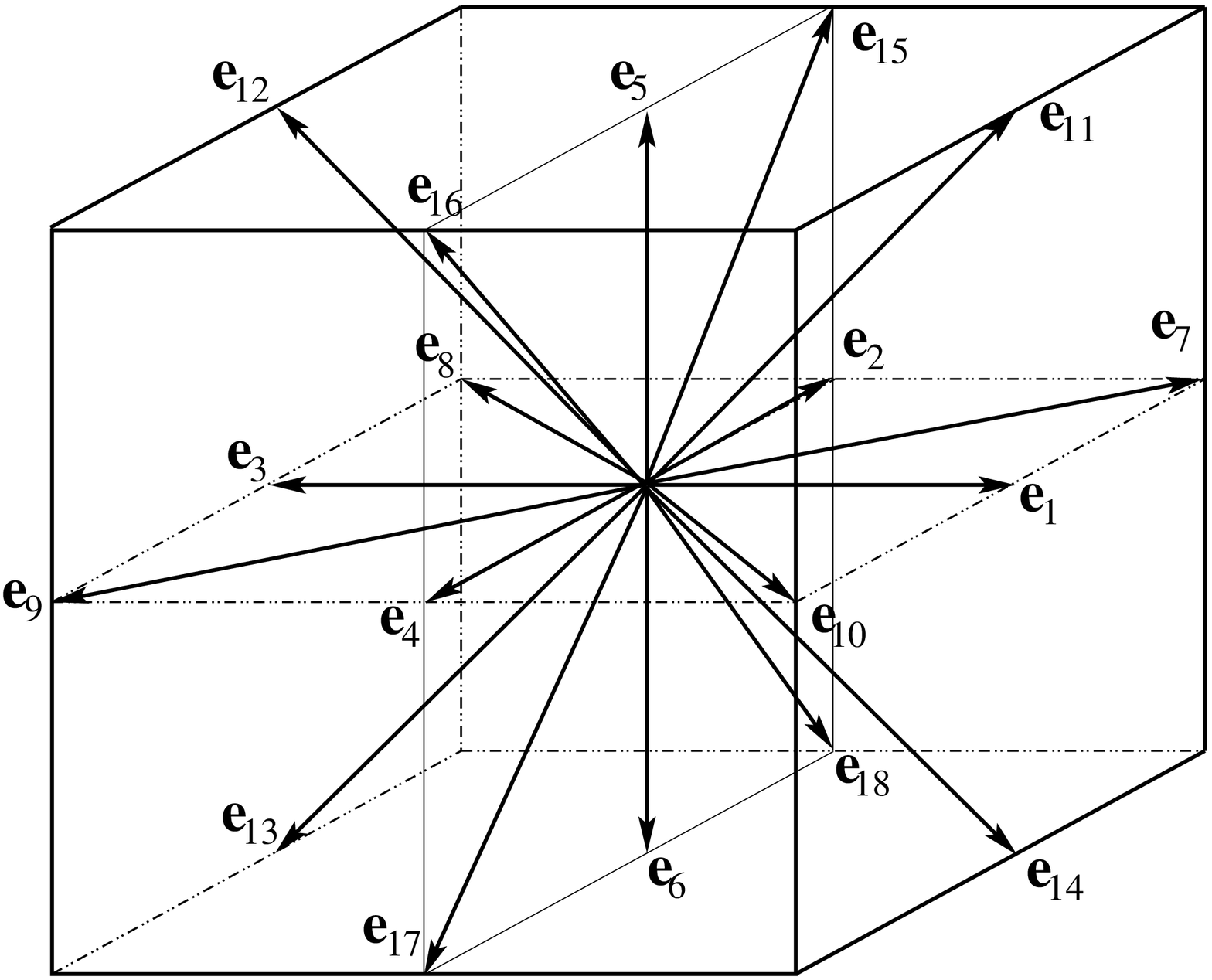}\protect
\par\end{centering}

}
\par\end{centering}

\protect\caption{\label{fig:Lattice-D3Qb}3D lattices of LB scheme.}
\end{figure*}

It is also useful to introduce the variable change for the equilibrium
function (see \ref{sub:Variable-change-for})
\begin{equation}
\overline{\vartheta}_{i}^{eq}=\vartheta_{i}^{eq}-\frac{\delta t}{2}\mathcal{S}_{i}^{\vartheta}\quad\mbox{for }\vartheta=f,\,h,\,s,\label{eq:VariableChange_feq}
\end{equation}
so that, with all those notations, the lattice Boltzmann equation
writes

\begin{equation}
\overline{\vartheta}_{i}^{\star}=\overline{\vartheta}{}_{i}-\frac{1}{\overline{\tau}_{\vartheta}+1/2}\left[\overline{\vartheta}_{i}-\overline{\vartheta}_{i}^{eq}\right]+\mathcal{S}_{i}^{\vartheta}\delta t\label{eq:LBE}
\end{equation}
for each distribution function $\vartheta=f,\,h,\,s$. Before defining
the equilibrium distribution functions and source terms, several lattices
are introduced. In this work, the D2Q9 lattice and three 3D lattices
are used: D3Q7, D3Q15 and D3Q19 (Fig. \ref{fig:Lattice-D3Qb}). For
D2Q9 the moving vectors are defined by $\mathbf{e}_{0}=(0,\,0)$,
$\mathbf{e}_{1,3}=(\pm1,\,0)$, $\mathbf{e}_{2,4}=(0,\,\pm1)$, $\mathbf{e}_{5,6}=(\pm1,\,1)$
and $\mathbf{e}_{7,8}=(\mp1,-1)$. for 3D lattices, the moving vectors
$\mathbf{e}_{i}$ are defined such as $\mathbf{e}_{1}=(1,\,0,\,0)^{T}$,
$\mathbf{e}_{2}=(0,\,1,\,0)^{T}$, $\ldots$, $\mathbf{e}_{6}=(0,\,0,\,-1)^{T}$
for D3Q7 (Fig. \ref{fig:D3Q7}). For D3Q15, additional diagonal vectors
are defined such as (see Fig. \ref{fig:D3Q15}) $\mathbf{e}_{7}=(1,\,1,\,1)^{T}$,
$\mathbf{e}_{8}=(-1,\,1,\,1)^{T}$, $\ldots$, $\mathbf{e}_{14}=(1,\,-1,\,-1)^{T}$.
Finally for D3Q19 (Fig. \ref{fig:D3Q19}): $\mathbf{e}_{7,\,8}=(\pm1,\,1,\,0)^{T}$,
$\mathbf{e}_{9,\,10}=(\pm1,\,-1,\,-0)^{T}$, $\mathbf{e}_{11,\,12}=(\pm1,\,0,\,1)$,
$\mathbf{e}_{13,\,14}=(\pm1,\,0,\,-1)^{T}$, $\mathbf{e}_{15,\,16}=(0,\,\pm1,\,1)^{T}$,
$\mathbf{e}_{17,\,18}=(0,\,\pm1,\,-1)^{T}$. For D3Q7 $N_{pop}=6$,
$e^{2}=1/4$, $w_{\text{0}}=1/4$ and $w_{1,\ldots,6}=1/8$. For D3Q15
$N_{pop}=14$, $e^{2}=1/3$, $w_{0}=2/9$, $w_{1,\ldots,6}=1/9$ and
$w_{7,\ldots,14}=1/72$. For D3Q19 $N_{pop}=18$, $e^{2}=1/3$, $w_{0}=1/3$,
$w_{1,\ldots,6}=1/18$ and $w_{7,...,18}=1/36$. The standard notations
will be used: $\mathbf{c}_{i}=\mathbf{e}_{i}c$ with $c=\delta x/\delta t$
where $\delta x$ and $\delta t$ are the space- and time-steps respectively
and $c_{s}^{2}=c^{2}/3$.

\subsection{\label{sub:LBM-for-incompressible}Incompressible Navier-Stokes}

Several lattice Boltzmann schemes exist for incompressible version
of Navier-Stokes equations. The fully incompressible condition has
already been proposed in literature but necessitates to solve an additional
Poisson equation \citep{Inamuro_etal_JCP2004} or an additional predictor-corrector
step \citep{Zu-He_PRE2013}. Here we prefer to apply the artificial
compressibility method \citep{Chorin_JCP1967} for which the solenoidal
condition $\boldsymbol{\nabla}\cdot\mathbf{u}=0$ is approximated
by $(1/\beta)\partial p/\partial t+\boldsymbol{\nabla}\cdot\mathbf{u}=0$
where $\beta$ is the artificial compressibility coefficient. In LB
framework, the method was derived in \citep{He-Luo_Incompressible_JSP1997}
with $\beta=\rho_{0}c_{s}^{2}$ where $\rho_{0}$ is the constant
density of bulk phase. The LB scheme writes

\begin{subequations}

\begin{align}
\overline{f}_{i}^{\star} & =\overline{f}{}_{i}-\frac{1}{\overline{\tau}_{f}+1/2}\left[\overline{f}_{i}-\overline{f}{}_{i}^{eq}\right]+\mathcal{S}_{i}^{f}\delta t,\label{eq:LBE_NS}\\
f_{i}^{eq} & =w_{i}\left[p+\rho(\phi)c_{s}^{2}\left(\frac{\mathbf{c}_{i}\cdot\mathbf{u}}{c_{s}^{2}}+\frac{(\mathbf{c}_{i}\cdot\mathbf{u})^{2}}{2c_{s}^{4}}-\frac{\mathbf{u}\cdot\mathbf{u}}{2c_{s}^{2}}\right)\right],\label{eq:Feq_NS_Incompr}
\end{align}
with $\overline{f}_{i}^{eq}=f_{i}^{eq}-\mathcal{S}_{i}^{f}\delta t/2$
and $\rho(\phi)$ is given by Eq. (\ref{eq:Density}). In Eq. (\ref{eq:LBE_NS})
$\overline{\tau}_{f}$ is the collision rate which is related to the
kinematic viscosity by $\nu=\overline{\tau}_{f}c_{s}^{2}\delta t$.
Hence, the collision rate is obtained by $\overline{\tau}_{f}(\phi)=3\nu(\phi)(\delta t/\delta x^{2})$
with the kinematic viscosity $\nu(\phi)$ interpolated by Eq. (\ref{eq:Viscosity}).
In Eq. (\ref{eq:LBE_NS}), the source term $\mathcal{S}_{i}^{f}$
contains contributions of external forces (involving $\mathbf{F}_{tot}$)
plus the production term in mass conservation (involving $\dot{m}'''$):

\end{subequations}

\begin{subequations}

\begin{equation}
\mathcal{S}_{i}^{f}=\mathcal{F}_{i}^{f}+\mathcal{P}_{i}^{f}\label{eq:SourceTerm_LBE_F}
\end{equation}
with \cite{MohammadiShad-Lee_LBM-SharpInterface_PRE2017}

\begin{align}
\mathcal{F}_{i}^{f} & =(\mathbf{c}_{i}-\mathbf{u})\cdot\left[(\Gamma_{i}-w_{i})\boldsymbol{\nabla}\rho(\phi)c_{s}^{2}+\Gamma_{i}\mathbf{F}_{tot}\right],\label{eq:Forcing}\\
\mathcal{P}_{i}^{f} & =w_{i}\rho c_{s}^{2}\dot{m}'''\left(\frac{1}{\rho_{g}}-\frac{1}{\rho_{l}}\right).\label{eq:MicroProduction}
\end{align}
In Eq. (\ref{eq:Forcing}), $\mathbf{F}_{tot}$ is the external force
defined by Eq. (\ref{eq:Force_Total}) and the function $\Gamma_{i}\equiv\Gamma_{i}(\mathbf{u})$
is defined by:

\begin{equation}
\Gamma_{i}=w_{i}\left[1+\frac{\mathbf{c}_{i}\cdot\mathbf{u}}{c_{s}^{2}}+\frac{(\mathbf{c}_{i}\cdot\mathbf{u})^{2}}{2c_{s}^{4}}-\frac{\mathbf{u}\cdot\mathbf{u}}{2c_{s}^{2}}\right].\label{eq:FunctionGamma}
\end{equation}

After the stages of collision and streaming, the first-order moment
(momentum) and the zeroth-order moment (pressure) are updated by \cite{MohammadiShad-Lee_LBM-SharpInterface_PRE2017}

\end{subequations}

\begin{subequations}

\begin{align}
\rho\mathbf{u} & =\frac{1}{c_{s}^{2}}\sum_{i}\overline{f}_{i}\mathbf{c}_{i}+\frac{\delta t}{2}\mathbf{F}_{tot},\label{eq:Velocity}\\
p & =\sum_{i}\overline{f}_{i}+\frac{\delta t}{2}\left\{ \mathbf{u}\cdot\boldsymbol{\nabla}\rho c_{s}^{2}+\rho c_{s}^{2}\dot{m}'''\left(\frac{1}{\rho_{g}}-\frac{1}{\rho_{l}}\right)\right\} .\label{eq:Pressure}
\end{align}

\end{subequations}

\subsection{\label{sub:LBM-for-phase-field}Conservative Allen-Cahn model}

The lattice Boltzmann equation for the conservative Allen-Cahn model
acts on the distribution function $\overline{g}_{i}$. The evolution
equation is

\begin{subequations}

\begin{align}
\overline{g}_{i}^{\star} & =\overline{g}_{i}-\frac{1}{\overline{\tau}_{g}+1/2}\left[\overline{g}_{i}-\overline{g}_{i}^{eq}\right]+\mathcal{S}_{i}^{g}\delta t\mathrm{,}\label{eq:LBE_PhaseField_Eq}\\
g^{eq} & =\phi\Gamma_{i}\mathrm{,}\label{eq:Geq_Allen-Cahn_Eq}
\end{align}
with the variable change $\overline{g}_{i}^{eq}=g^{eq}-\delta t\mathcal{S}_{i}^{g}/2$.
The mobility coefficient is related to the collision rate by $M_{\phi}=\overline{\tau}{}_{g}c_{s}^{2}\delta t$.
The source term $\mathcal{S}_{i}^{g}$ contains two contributions:

\end{subequations}

\begin{subequations}

\begin{equation}
\mathcal{S}_{i}^{g}=\mathcal{F}_{i}^{g}+\mathcal{P}_{i}^{g},\label{eq:SourceTerm_LBE_G}
\end{equation}
where the first one $\mathcal{F}_{i}^{g}$ involves the counter term
with the normal vector $\mathbf{n}$ \cite{Mitchell_etal_IJMF2018},
and the second one $\mathcal{P}_{i}^{g}$ involves the mass production
term $\dot{m}'''$:

\begin{equation}
\mathcal{F}_{i}^{g}=\frac{4}{W}\phi(1-\phi)w_{i}\mathbf{c}_{i}\cdot\mathbf{n}\quad\mbox{and}\quad\mathcal{P}_{i}^{g}=w_{i}\frac{\dot{m}'''}{\rho_{g}}.\label{eq:SourceTerm_AllenCahn}
\end{equation}

\end{subequations}

\begin{subequations}

Let us notice that the scheme is equivalent (see \ref{sub:Equiv_AllenCahn})
to the lattice Boltzmann equation

\begin{equation}
\overline{g}_{i}^{\star}=\overline{g}_{i}-\frac{1}{\overline{\tau}_{g}+1/2}\left[\overline{g}_{i}-\overline{g}_{i}^{eq,\,CAC}\right]+\mathcal{P}_{i}^{g}\delta t\label{eq:LBE_PhaseField_Eq_Alternative}
\end{equation}
where only the source term $\mathcal{P}_{i}^{g}$ appears in the source
term and the equilibrium distribution function is redefined as \cite{Fakhari_etal_JCP2017}

\begin{equation}
g_{i}^{eq,\,CAC}=\phi\Gamma_{i}+M_{\phi}\frac{4}{W}\phi(1-\phi)w_{i}\frac{\mathbf{c}_{i}\cdot\mathbf{n}}{c_{s}^{2}}\label{eq:Geq_AC-1}
\end{equation}
with $\overline{g}_{i}^{eq,\,CAC}=g_{i}^{eq,\,CAC}-\delta t\mathcal{P}_{i}^{g}/2$.

\end{subequations}

After the stages of collision and streaming, the new phase-field is
obtained by the zeroth-order moment of $\overline{g}_{i}$ which must
be corrected with the production term:

\begin{equation}
\phi(\mathbf{x},\,t)=\sum_{i}\overline{g}_{i}+\frac{\delta t}{2}\sum_{i}\mathcal{P}_{i}^{g}.\label{eq:Moment_Phi}
\end{equation}
This relation holds for both formulations that use $\overline{g}_{i}^{eq}$
and $\overline{g}_{i}^{eq,\,CAC}$ because $\sum_{i}\mathcal{F}_{i}^{g}\delta t/2=0$.

\subsection{\label{sub:LBM-for-Temperature}Temperature equation}

The lattice Boltzmann scheme for temperature equation writes:

\begin{subequations}

\begin{eqnarray}
\overline{s}_{i}^{\star} & = & \overline{s}_{i}-\frac{1}{\overline{\tau}_{s}+1/2}\left[\overline{s}_{i}-\overline{s}{}_{i}^{eq}\right]+\mathcal{S}_{i}^{s}\delta t\label{eq:LBE_Temp_Eq}\\
s_{i}^{eq} & = & T\Gamma_{i}\label{eq:Seq_Temp}
\end{eqnarray}
where the thermal diffusivity $\alpha$ is related to the collision
rate by $\alpha=\overline{\tau}_{s}c_{s}^{2}\delta t$. The source
term $\mathcal{S}_{i}^{s}$ is defined such as:

\begin{equation}
\mathcal{S}_{i}^{s}=\mathcal{F}_{i}^{s}+\mathcal{P}_{i}^{s}\label{eq:Source_Term_LBE_S}
\end{equation}
where

\begin{equation}
\mathcal{F}_{i}^{s}=w_{i}\frac{\mathcal{L}}{\mathcal{C}_{p}}\boldsymbol{\nabla}\cdot(\mathbf{u}\phi)\quad\mbox{and}\quad\mathcal{P}_{i}^{s}=w_{i}\frac{\mathcal{L}}{\mathcal{C}_{p}}\frac{\partial\phi}{\partial t}\label{eq:Def_SourceTerms_Temp}
\end{equation}

\end{subequations}

Finally, the new temperature is computed by

\begin{equation}
T=\sum_{i}\overline{s}_{i}-\frac{\delta t}{2}\frac{\mathcal{L}}{\mathcal{C}_{p}}\left[\frac{\partial\phi}{\partial t}+\boldsymbol{\nabla}\cdot(\mathbf{u}\phi)\right].\label{eq:Moment_Temp}
\end{equation}

In Sections \ref{sec:Validations} and \ref{sec:FilmBoiling}, simulations
will be carried out with Dirichlet boundary conditions applied on
temperature $T$ and phase-field $\phi$. In order
to impose such a condition, for example on temperature $T_{w}$ on
left boundary of a D2Q9 lattice, the unknown distribution functions
$\left.\overline{s}_{i}\right|_{unknown}$ are updated with the anti
bounce-back method \citep{Ginzburg_AdvWatRes2005}: $\overline{s}_{i}|_{unknown}=-\overline{s}_{i'}+2w_{i}T_{w}$
where $i'$ is the opposite direction of $i$.

\subsection{Computations of gradients and Laplacian}
The unit normal vector $\mathbf{n}$ and force term $\mathbf{F}_{s}$
require computation of gradients. Moreover the chemical potential
$\mu_{\phi}$ necessitates to calculate the Laplacian of $\phi$.
Gradients and Laplacian that are involved in definitions of $\mathbf{n}$ (Eq. (\ref{eq:NormalVector}))
and $\mu_{\phi}$ (Eq. (\ref{eq:ChemicalPotential})) are discretized
by using the directional derivatives methods. The method has already
demonstrated its performance for hydrodynamics problem in order to
reduce parasitic currents for two-phase flow problem \citep{Lee-Fischer_PRE2006,Lee_Parasitic_CAMWA2009,Lee-Liu_DropImpact_LBM_JCP2010}.
The directional derivative is the derivative along each moving direction
on the lattice. Taylor's expansion at second-order of a differentiable
scalar function $\phi(\mathbf{x})$ at $\mathbf{x}+\mathbf{e}_{i}\delta x$
and $\mathbf{x}-\mathbf{e}_{i}\delta x$ yields the following approximation
of directional derivatives:

\begin{subequations}

\begin{equation}
\mathbf{e}_{i}\cdot\boldsymbol{\nabla}\phi\bigr|_{\mathbf{x}}=\frac{1}{2\delta x}\left[\phi(\mathbf{x}+\mathbf{e}_{i}\delta x)-\phi(\mathbf{x}-\mathbf{e}_{i}\delta x)\right]\label{eq:DerivDirec_Grad}
\end{equation}

The number of directional derivatives is equal to the number of moving
direction $\mathbf{e}_{i}$ on the lattice i.e. $N_{pop}$. The gradient
is obtained by

\begin{equation}
\boldsymbol{\nabla}\phi\bigr|_{\mathbf{x}}=3\sum_{i=1}^{N_{pop}}w_{i}\mathbf{e}_{i}\left(\mathbf{e}_{i}\cdot\boldsymbol{\nabla}\phi\bigr|_{\mathbf{x}}\right).\label{eq:Grad}
\end{equation}

\end{subequations}

The three components of the gradient $\partial_{x}\phi$, $\partial_{y}\phi$
and $\partial_{z}\phi$ are obtained by calculating each directional
derivative $\mathbf{e}_{i}\cdot\boldsymbol{\nabla}\phi\bigr|_{\mathbf{x}}$
and next, by calculating the moment of first-order $\boldsymbol{\nabla}\phi\bigr|_{\mathbf{x}}$.
For the calculation of $\boldsymbol{\nabla}^{2}\phi$, all directions
of propagation are taken into account by

\begin{subequations}

\begin{equation}
(\mathbf{e}_{i}\cdot\boldsymbol{\nabla})^{2}\phi\bigr|_{\mathbf{x}}=\frac{1}{\delta x^{2}}\left[\phi(\mathbf{x}+\mathbf{e}_{i}\delta x)-2\phi(\mathbf{x})+\phi(\mathbf{x}-\mathbf{e}_{i}\delta x)\right]\mathrm{.}\label{eq:DerivDirect_Laplacian}
\end{equation}

The Laplacian is obtained by summing and weighting each term with

\begin{equation}
\boldsymbol{\nabla}^{2}\phi\bigr|_{\mathbf{x}}=3\sum_{i = 1}^{N_{pop}}w_{i}(\mathbf{e}_{i}\cdot\boldsymbol{\nabla})^{2}\phi\bigr|_{\mathbf{x}}.\label{eq:Laplacian}
\end{equation}

\end{subequations}

Other approximations exist \citep{Lee-Lin_JCP2005,Lee-Liu_DropImpact_LBM_JCP2010}
such as the first-order and second-order upwind schemes (or biased
differences) respectively defined by $\mathbf{e}_{i}\cdot\boldsymbol{\nabla}^{up_{1}}\phi\bigr|_{\mathbf{x}}=[\phi(\mathbf{x}+\mathbf{e}_{i}\delta x)-\phi(\mathbf{x})]$/$\delta x$
and $\mathbf{e}_{i}\cdot\boldsymbol{\nabla}^{up_{2}}\phi\bigr|_{\mathbf{x}}=\left[-\phi(\mathbf{x}+2\mathbf{e}_{i}\delta x)+4\phi(\mathbf{x}+\mathbf{e}_{i}\delta x)-3\phi(\mathbf{x})\right]/(2\delta x)$.
Here, by simplicity, the central difference approximation is applied for all simulations even though that approximation fails to capture the velocity profiles in low density regions \citep{Fakhari_etal_PRE2017} and biased directional derivatives can fix that issue [2]. Those biased differences could be tested in future works with \texttt{LBM\_saclay}.

\subsection{\label{sub:KernelOptimization}Numerical implementation and kernel optimization}

All LBM schemes of this Section were implemented in a new code called
\texttt{LBM\_saclay} written in C++. The main advantage of this new
code is its portability targeting all major HPC platforms and especially
those based on GPU- and CPU-architectures.
Actually, \texttt{LBM\_saclay} can run without modification on any architecture that \texttt{Kokkos} supports. The current compatibilities are indicated in \cite{KokkosLink} and summarized in Tab. \ref{tab:compatibilities}. For more information, the reader can refer to the \texttt{Kokkos} documentation. Let us mention that the current support for AMD GPU is experimental through the C++ library \texttt{HIP} (Heterogeneous-Compute Interface for Portability) and it is planned to be supported at the end of 2020.

\begin{figure}
\begin{centering}
  \includegraphics[scale=0.25]{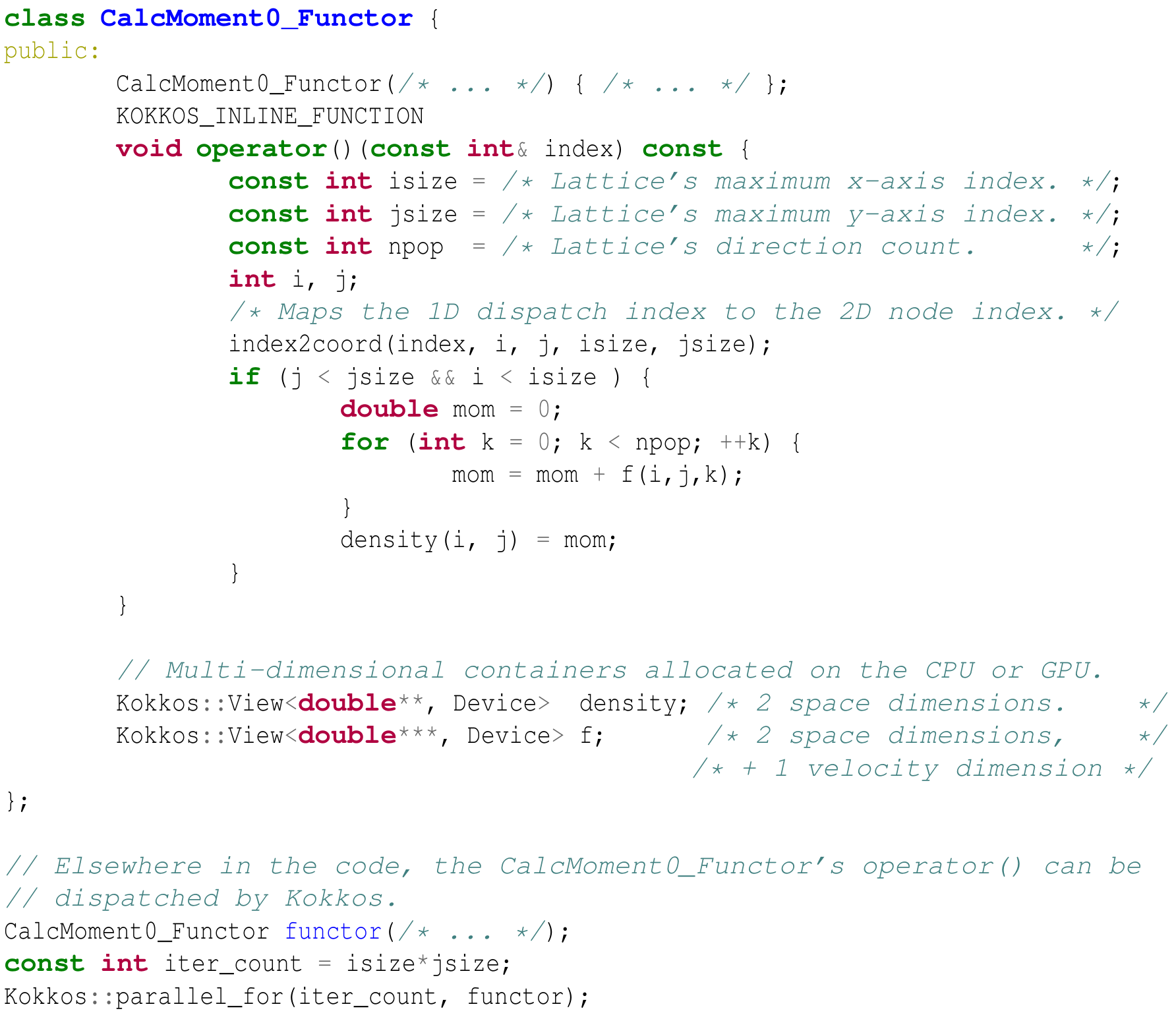}
\par\end{centering}

\protect\caption{\label{fig:Kokkos-Example}Example of using the \texttt{Kokkos} library
to compute the zeroth-order moment of distribution function.}
\end{figure}

\begin{table}
\begin{centering}
\begin{tabular}{lllll}
\hline 
\textbf{Intel CPUs} & \textbf{NVidia GPUs} & \textbf{ARM} & \textbf{IBM} & \textbf{AMD}\tabularnewline
\hline 
Sandy/Ivy Bridge & Kepler & ThunderX & Blue gene Q & AMD CPUs\tabularnewline
Haswell & Maxwell & ARMv8.0 & Power7 & \tabularnewline
Skylake & Pascal & ARMv8.1 & Power8 & \tabularnewline
Westmere CPUs & Volta &  & Power9 & \tabularnewline
Knights Landing/Corner Xeon Phi & Turing &  &  & \tabularnewline
Broadwell Xeon E-class &  &  &  & \tabularnewline
\hline 
\end{tabular}
\par\end{centering}
\caption{\label{tab:compatibilities}List of architectures that are currently compatible with the \texttt{Kokkos} library.}
\end{table}
Two levels of parallelism
are implemented in the code. The first one is the intra-node parallelism
(shared memory) with the \texttt{Kokkos} library, an opensource C++
library with parallel algorithmic patterns and data containers. Specific
commands of the \texttt{Kokkos} library optimize loops with \texttt{OpenMP},
\texttt{Pthreads} or \texttt{CUDA} during compilation. An example
of using \texttt{Kokkos}' functionalities is presented on Fig. \ref{fig:Kokkos-Example}
to compute at each time-step the zeroth-order moment of a distribution
function. The second level of parallelism is a standard domain decomposition
performed with \texttt{MPI}: the full computational domain is cut
into several sub-domains associated with each computational node (distributed
memory).

\begin{figure*}
\begin{centering}
\subfloat[\label{fig:GPU_Diffusion}Comparison of computational times for three
NVIDIA$^{\circledR}$ graphical cards: K80 (oldest), P100 and V100
(newest).]{\protect\begin{centering}
\protect\includegraphics[scale=0.53]{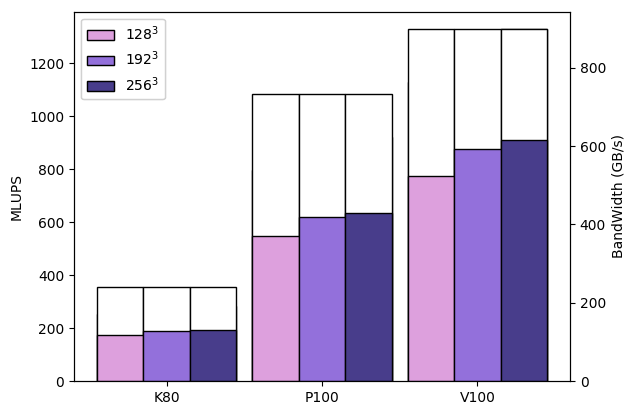}\protect
\par\end{centering}

}$\qquad$\subfloat[\label{fig:CPU_Diffusion}Comparisons of computational times for three
optimizations of LBM kernel for Intel$^{\circledR}$ KNL: fused (left), CSoA (middle)
and CSoA2 (right).]{\protect\begin{centering}
\protect\includegraphics[scale=0.53]{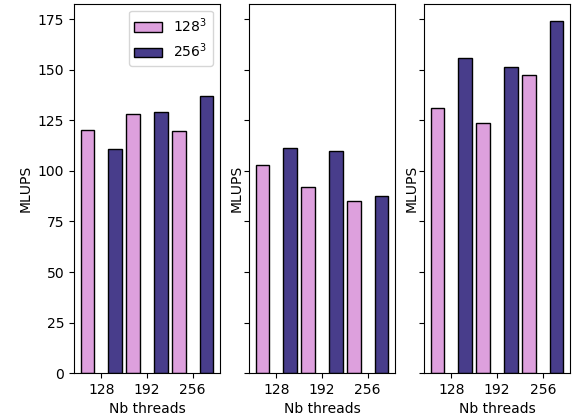}\protect
\par\end{centering}

}
\par\end{centering}

\protect\caption{\label{fig:Computational-times_DiffusiveProblem}Computational times
(in Million Lattice Updates Per Second -- MLUPS) for a diffusive problem
with a D3Q19 lattice. (a) GPU for three mesh sizes and (b) on CPU
(Intel$^{\circledR}$ KNL) for two mesh sizes.}
\end{figure*}

When developing the code, several optimizations were implemented and
compared in particular to enhance its performance on each architecture.
The first way to consider the stages of collision and streaming is
to ``fuse'' those two steps inside a single kernel, i.e. both stages are simply done in one single ``\texttt{for}-loop'' performed on the lattice nodes.
The ``fused'' version does not require an intermediate memory load contrary to standard
implementation for which both stages are well separated.
However, the fused kernel contains more floating point operations per iteration of the ``\texttt{for}-loop''. This is a drawback if the number of floating point operations becomes large enough to exhaust the amount of register memory available on the architecture (this number is significantly lower on GPU than on CPU). So, if the register memory is full, additional variables will be allocated in the external DRAM memory, generating additional traffic on the memory bus and degrading performance.
For GPUs NVIDIA$^{\circledR}$ and CPUs Intel$^{\circledR}$ Skylake, best
performance is obtained with the fused version.

Alternatively, two optimizations were tested which are well suited for Intel$^{\circledR}$
KNL (KNights Landing) processors \cite{Calore_etal_ParalProcAM2018}:
the first one is the ``CSoA'' optimization (Cluster of Structure
of Array) i.e. for each line of the lattice, LBM nodes are stored
in memory modulo $M$ where typically $M=8$ and each line is padded
to be a multiple of $M$. The access of data container is done with
\texttt{data(iMem,j,k,ipop)} where \texttt{iMem} is computed from
the physical node location \texttt{i}. The CSoA optimization improves
vectorization and memory alignment for streaming stage but performance
decreases for large domain on D2Q9 lattice. The second optimization
for KNL is ``CSoA2'', i.e. the population index \texttt{ipop} of
\texttt{data(i,j,k,ipop)} is interverted to \texttt{data(i,ipop,j,k)},
where \texttt{i,j,k} are indices of position. With this permutation,
the memory locality is restored for the collision stage.

Comparisons were performed on a simplified diffusive problem. The CSoA2 optimization
enhances performance on KNL processors, but on Fig. \ref{fig:Computational-times_DiffusiveProblem},
we can see that it remains far below to that obtained on GPUs, even
older generation GPUs (K80). Computational times are expressed in
Million Lattice Updates per Second (MLUPS)
as an effective metric measuring the number of millions of node lattice update per seconds. That performance metric is used by node so that it is independent of the type of lattice (e.g. D3Q7, D3Q15, D3Q19). With that metric, a larger lattice will give a smaller MLUPS.
In the rest of this paper,
most of validations and simulations of Sections \ref{sec:Validations}
and \ref{sec:FilmBoiling} are carried out on GPUs. In Section \ref{sub:Computational-times},
comparisons of computational times on GPU and CPU will be presented
on the test case of film boiling for two mesh sizes.

Finally, let us mention that all kernels (Navier-Stokes, phase-field and temperature equations) have been developed in 2D and in 3D. They all run in 3D separately. However, all coupling terms, i.e. the surface tension force (Eq. (\ref{eq:SuperficialTensionForce})), the chemical potential (Eq. (\ref{eq:ChemicalPotential})) and the advective term in Eq. (\ref{eq:Def_SourceTerms_Temp})) were developed and checked only in 2D. Hence, verification of couplings in Section \ref{sec:Validations} and film boiling simulations of Section \ref{sec:FilmBoiling} will be presented only in 2D. The three-dimensional extension of coupling terms is planned for future works.

\section{\label{sec:Validations}Code verifications}

In this section, the numerical implementation of the LBM schemes of
Section \ref{sec:Lattice-Boltzmann-methods} is checked by comparison
with well-known solutions. Validations are gathered into two parts
in order to check implementations step-by-step. In subsection \ref{sub:Without-mass-transfer},
verifications are done without phase change, i.e. by neglecting the
temperature equation and by assuming that the mass transfer is zero
($\dot{m}'''=0$ in Eq. (\ref{eq:ConservMasse_ChgtPhase}) and (\ref{eq:CAC_ChgtPhase})).
The conservative Allen-Cahn model, and the coupling with fluid flow
are verified successively. In subsection \ref{sub:With-mass-transfer},
the phase change model is checked by considering the phase-field equation
coupled with temperature. The LBM code is compared with an analytical
solution of Stefan's problem with two different diffusivities.

\subsection{\label{sub:Without-mass-transfer}Verifications without phase change}

We first compare implementation of the conservative Allen-Cahn model
on two test cases: Zalesak's slotted disk and interface deformation
inside a vortex. Next the coupling with Navier-Stokes model will be
considered with the layered Poiseuille flow and the Laplace law.

\subsubsection{Verifications of the phase-field model}

Two verifications of phase-field implementation are presented. In the
first one, we check that the contour of a slotted disk is well conserved
inside a rotating fluid \citep{Zalesak_JCP1979}. In the second one,
we check that the simulation retrieves a circle when an initial disk
is deformed inside a vortex that changes its direction of rotation
over time. For both simulations, the mesh is composed of $201\times201\times3$
nodes with periodic boundary conditions applied on all faces, the
time-step is $\delta t=10^{-4}$ and the space-step $\delta x=5\times10^{-3}$.

\paragraph{Zalesak's slotted disk}

\begin{figure*}
\begin{centering}
\subfloat[$t=T_{f}/4$]{\protect\begin{centering}
\protect\includegraphics[scale=0.15]{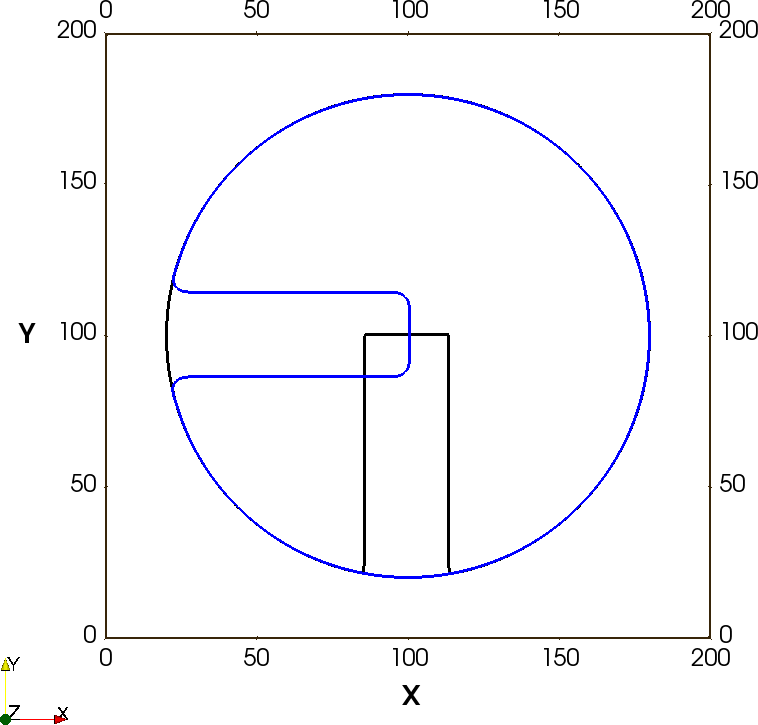}\protect
\par\end{centering}

}\subfloat[$t=T_{f}/2$]{\protect\begin{centering}
\protect\includegraphics[scale=0.15]{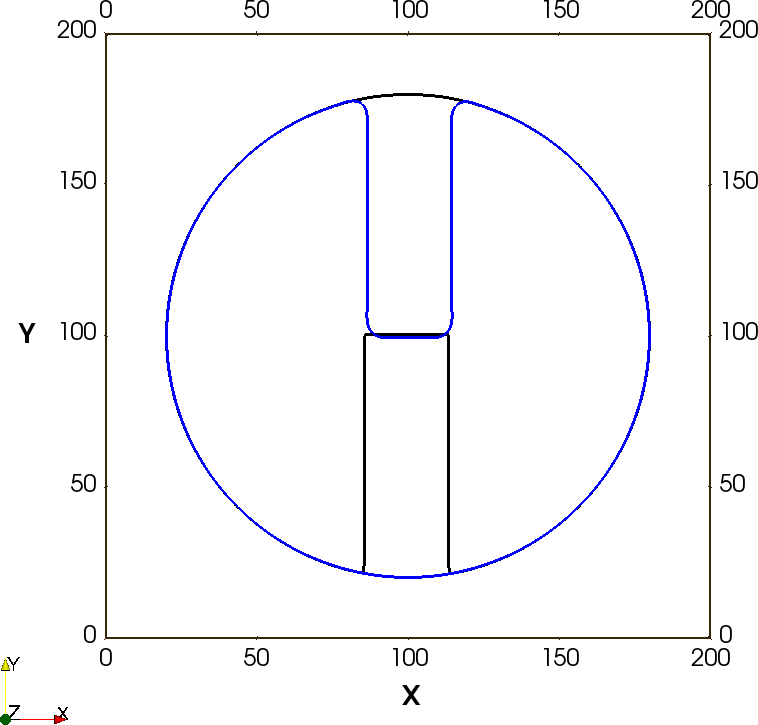}\protect
\par\end{centering}

}\subfloat[$t=3T_{f}/4$]{\protect\begin{centering}
\protect\includegraphics[scale=0.15]{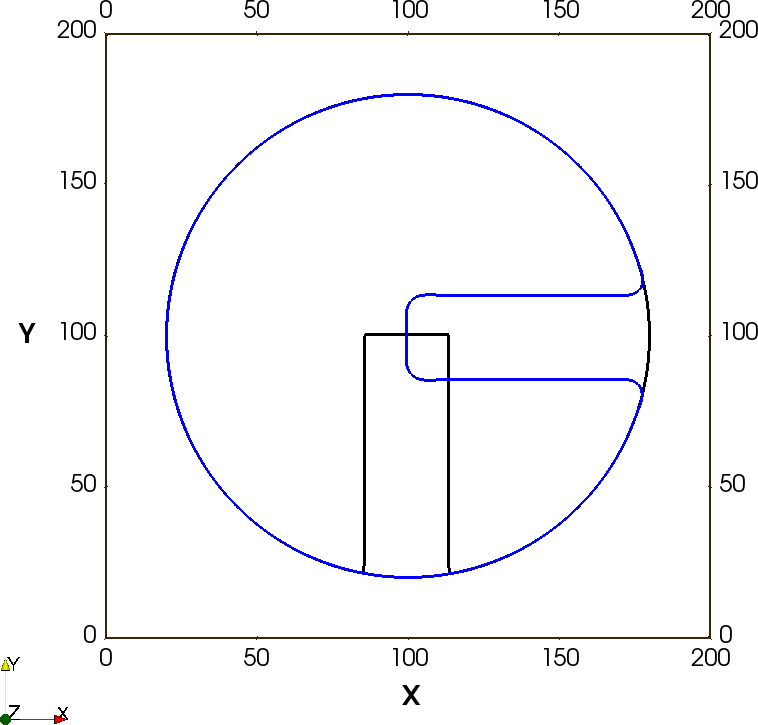}\protect
\par\end{centering}

}\subfloat[{\footnotesize{}\label{fig:Zalesak_d}}$t=T_{f}$]{\protect\begin{centering}
\protect\includegraphics[scale=0.15]{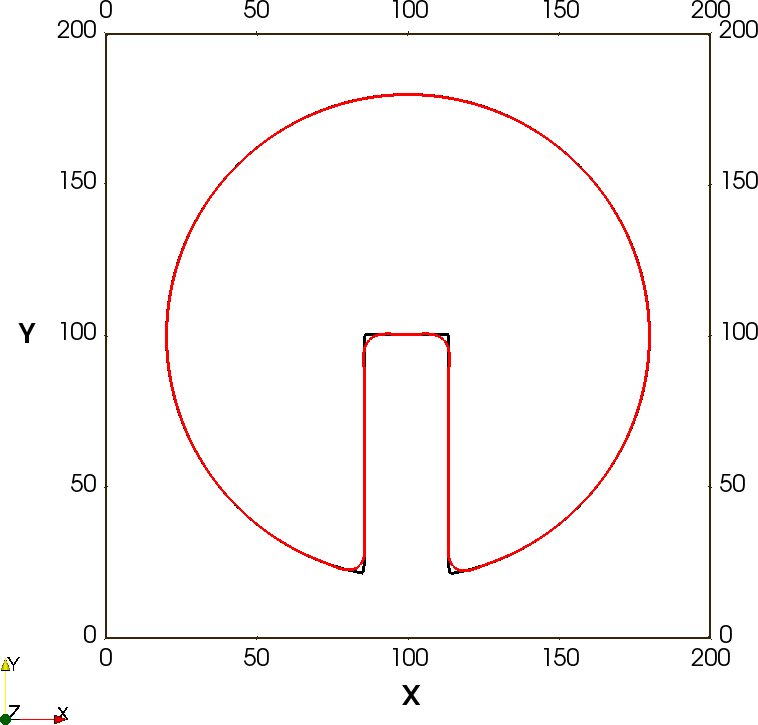}\protect
\par\end{centering}

}
\par\end{centering}

\protect\caption{\label{fig:Zalesak}Zalesak's slotted disk with the conservative Allen-Cahn
model.}
\end{figure*}

Inside a domain of lengths $L_{x}=L_{y}=1$, and $L_{z}=0.01$, a
disk is initialized at the center of the domain $\mathbf{x}_{c}=(100,\,100,\,1)^{T}$
by $\phi(\mathbf{x},\,0)=\left[1+\tanh\left((R-d_{c})/\sqrt{2}W_{0}\right)\right]/2$
with $d_{c}=\sqrt{(x-x_{c})^{2}+(y-y_{c})^{2}+(z-z_{c})^{2}}$, $W_{0}=2$
and $R=80$ l.u. (lattice units). The diffuse disk is slotted by imposing
$\phi(\mathbf{x},\,0)=0$ if $x_{c}-R/6\leq x\leq x_{c}+R/6$ and
$y_{c}-1.1R\leq y\leq y_{c}$. Components of velocity are imposed
by $u_{x}(\mathbf{x})=u_{0}(2y-1)$, $u_{y}(\mathbf{x})=u_{0}(1-2x)$
and $u_{z}(\mathbf{x})=0$. The value of $u_{0}$ is chosen such that
the slotted disk performs one complete rotation at $T_{f}=4$, i.e.
$u_{0}=0.7853975$ and both parameters of CAC model are set as $M_{\phi}=5\times10^{-4}$
and $W=6\delta x$. The rotation of the slotted disk is presented on Fig.
\ref{fig:Zalesak} where the interface position $\phi=1/2$ is superimposed
to the initial condition at four times. At the final time of simulation
$t=T_{f}$ (Fig. \ref{fig:Zalesak_d}), the contour $\phi=0.5$ (red)
is superimposed to the initial one (black) although the slot corners
are slightly rounded. 

\paragraph{Vortex}

\begin{figure*}
\begin{centering}
\subfloat[\label{fig:Vortex-1}Without change of rotation during simulation.
Streamlines of $\mathbf{u}$ (colored lines) and interface $\phi=0.5$
(black line) at three times.]{\protect\begin{centering}
\begin{tabular}{ccc}
{\small{}(i) $t=0$} & {\small{}(ii) $t=T_{f}/2$} & {\small{}(iii) $t=T_{f}$}\tabularnewline[2mm]
\protect\includegraphics[scale=0.18]{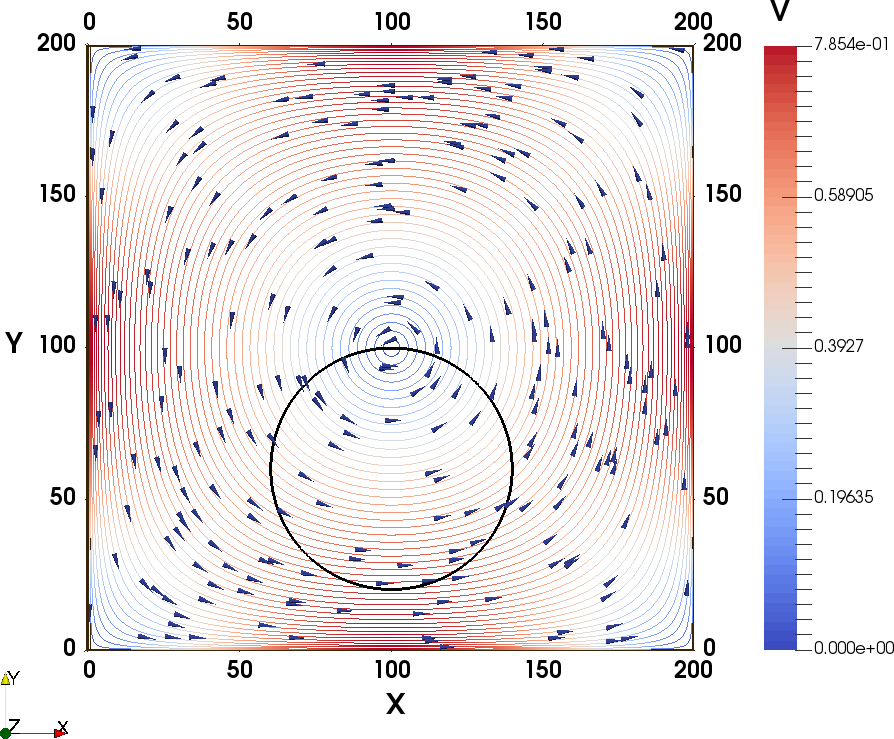} & \protect\includegraphics[scale=0.18]{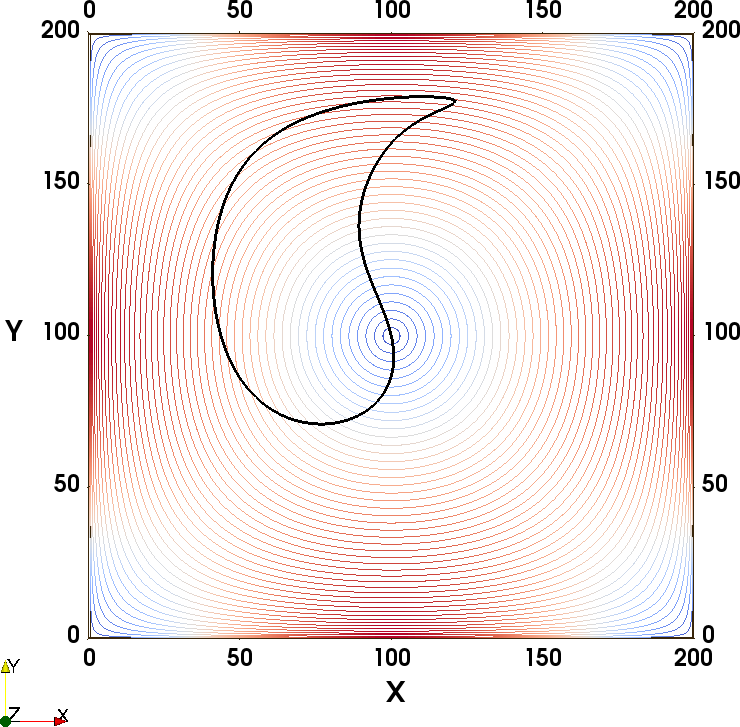} & \protect\includegraphics[scale=0.18]{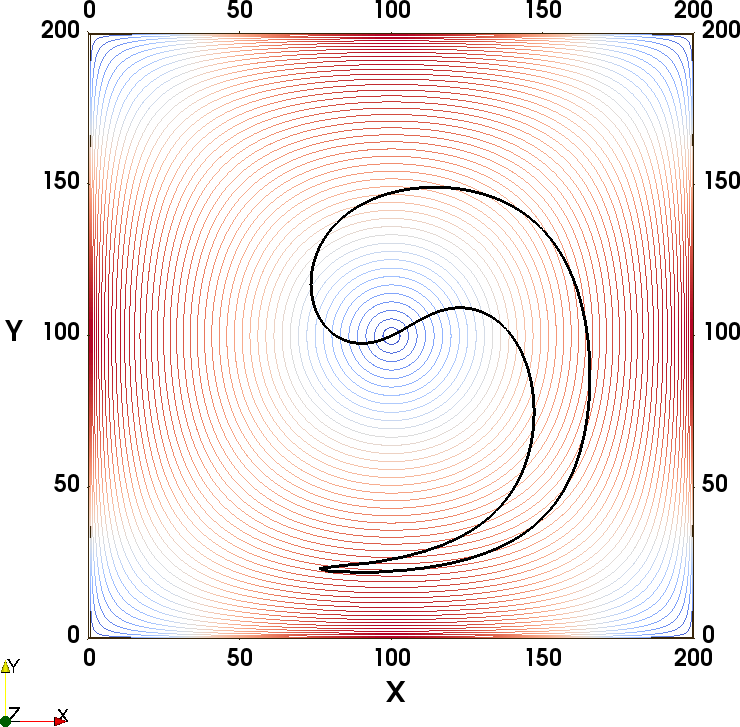}\tabularnewline
\end{tabular}\protect
\par\end{centering}

}
\par\end{centering}

\begin{centering}
\subfloat[\label{fig:Vortex-2}With a change of rotation direction during simulation.
Streamlines of $\mathbf{u}'$ (colored lines) and contours $\phi=0.5$
(black lines) for four times.]{\protect\begin{centering}
\begin{tabular}{cccc}
{\small{}(i) $t=T_{f}/2$} & {\small{}(ii) $t=T_{f}$} & {\small{}(iii) $t=3T_{f}/2$} & {\small{}(iv) $t=2T_{f}$}\tabularnewline[2mm]
\protect\includegraphics[scale=0.15]{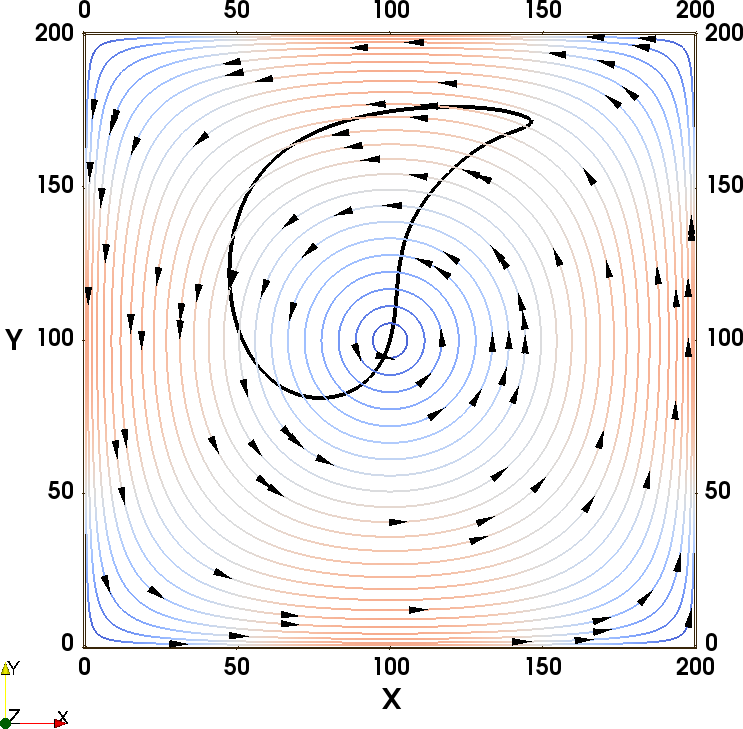} & \protect\includegraphics[scale=0.15]{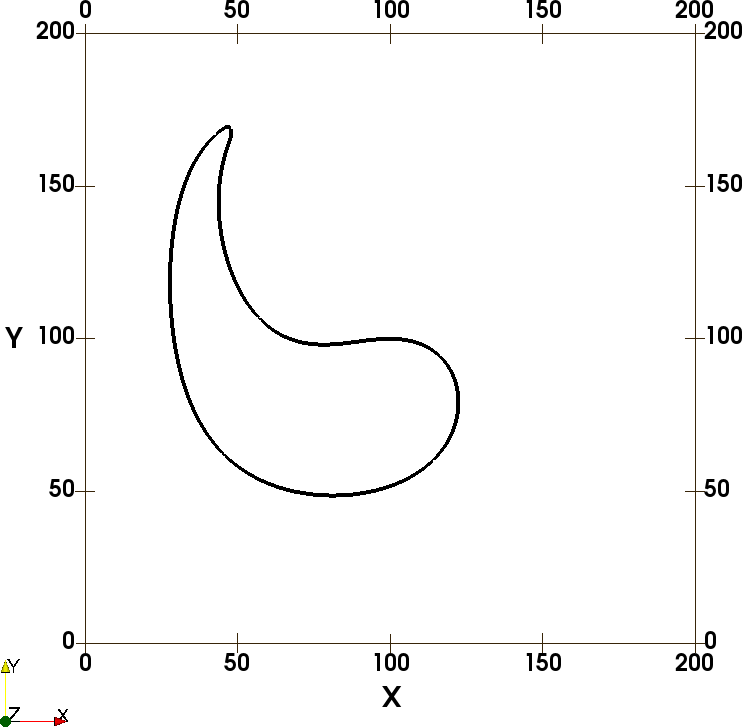} & \protect\includegraphics[scale=0.15]{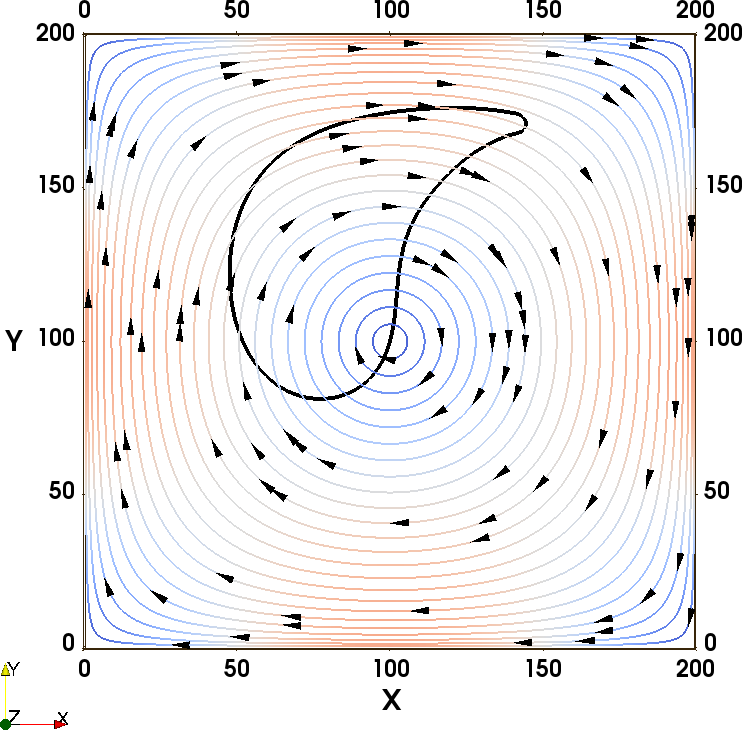} & \protect\includegraphics[scale=0.15]{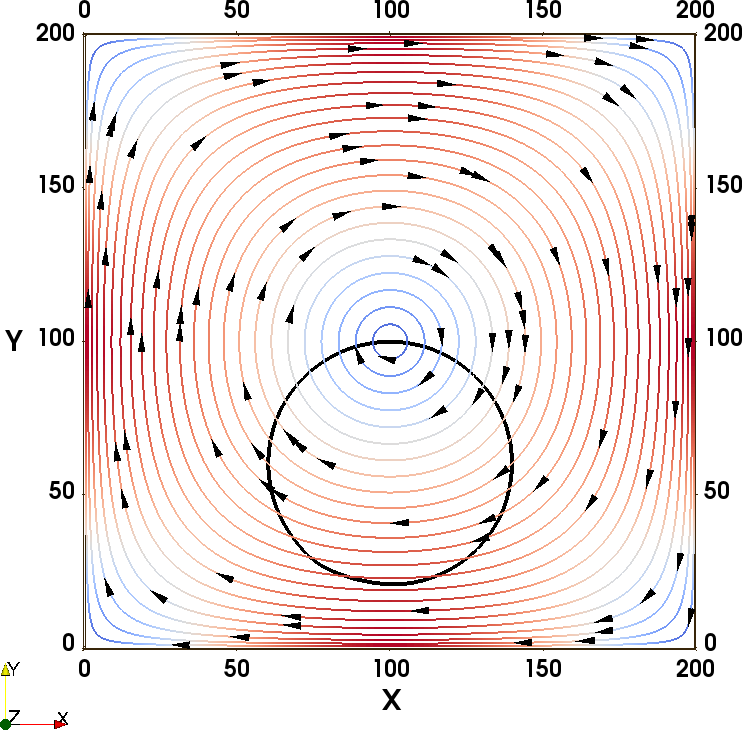}\tabularnewline
\end{tabular}\protect
\par\end{centering}

}
\par\end{centering}

\protect\caption{Deformation of an initial disk standing inside a vortex. (a) Without
change of rotation during simulation. (b) With change of rotation.}
\end{figure*}

We study the deformation of an initial disk standing inside a 2D vortex.
The three components of velocity are defined by $u_{x}(\mathbf{x})=-u_{0}\cos\left[\pi(x-0.5)\right]\sin\left[\pi(y-0.5)\right]$,
$u_{y}(\mathbf{x})=u_{0}\sin\left[\pi(x-0.5)\right]\cos\left[\pi(y-0.5)\right]$
and $u_{z}(\mathbf{x})=0$. LB simulations are performed on a D3Q19
lattice for a 3D domain with a very small thickness in $z$-direction.
The initial condition $\phi(\mathbf{x},\,0)$ is defined by a full
disk centered at $\mathbf{x}_{c}=(100,\,60,\,1)^{T}$, with $W=2$
and $R=40$ l.u. The initial condition ($\phi=0.5$) and streamlines
for $u_{0}=0.7853975$ are presented on Fig. \ref{fig:Vortex-1}-(i).
The rotation is directed counterclockwise. Parameters are $T_{f}=4$,
$W=6\delta x$ and $M_{\phi}=5\times10^{-4}$. For $t=T_{f}/2$ (Fig.
\ref{fig:Vortex-1}-(ii)) and $t=T_{f}$ (Fig. \ref{fig:Vortex-1}-(iii))
black contours $\phi=0.5$ are comparable to those presented in reference
\cite[Fig. 4]{Geier_etal_PRE2015}. Next, the velocity is changed
during the simulation by multiplying $\mathbf{u}(\mathbf{x})$ with
a factor depending on time: $\mathbf{u}'(\mathbf{x},\,t)=\mathbf{u}(\mathbf{x})\times\cos(\pi t/2T_{f})$.
With the cosine function, the velocity $\mathbf{u}'(\mathbf{x},\,t)$
presents three stages during the simulation: when $t<T_{f}$, the
direction of rotation is counterclockwise (Fig. \ref{fig:Vortex-2}-(i));
when $t=T_{f}$ the cosine function cancels the velocity $\mathbf{u}'$
(Fig. \ref{fig:Vortex-2}-(ii)); and when $t>T_{f}$, the sign changes
and the direction of rotation becomes clockwise (Fig. \ref{fig:Vortex-2}-(iii)).
At the end of simulation $t=2T_{f}$, we expect to find the shape
of initial disk. That is what we observe on Fig. \ref{fig:Vortex-2}-(iv)
which confirms that the interface position $\phi=0.5$ is similar
to the initial condition one (Fig. \ref{fig:Vortex-1}-(i)).

\subsubsection{Verifications of phase-field with fluid flow model}

Two classical test cases are presented to check the coupling of phase-field
equation and fluid flow model: the layered Poiseuille flow and the
Laplace law.

\paragraph*{Layered Poiseuille flow}

\begin{figure}
\begin{centering}
\subfloat[\label{fig:Poiseuille_Viscosity}Comparisons between LBM and the double-Poiseuille
analytical solution for three viscosity ratios and $\rho_A/\rho_B=1$.]{\begin{centering}
\includegraphics[angle=-90,scale=0.33]{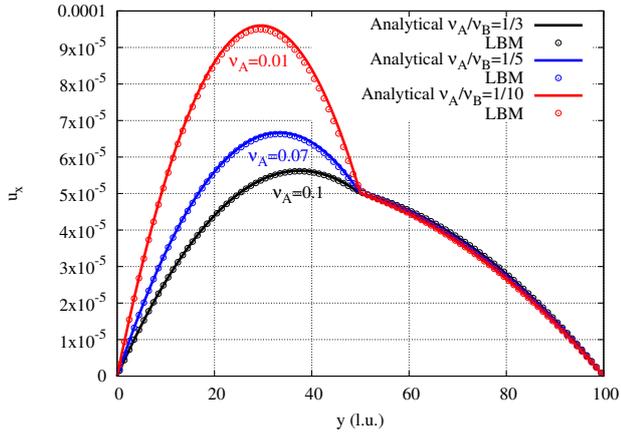}
\par\end{centering}
}$\qquad$\subfloat[\label{fig:Poiseuille_Density}Comparisons between
LBM and the double-Poiseuille analytical solution for three density
ratios and $\nu_A/\nu_B=1$.]{\begin{centering}
\includegraphics[angle=-90,scale=0.33]{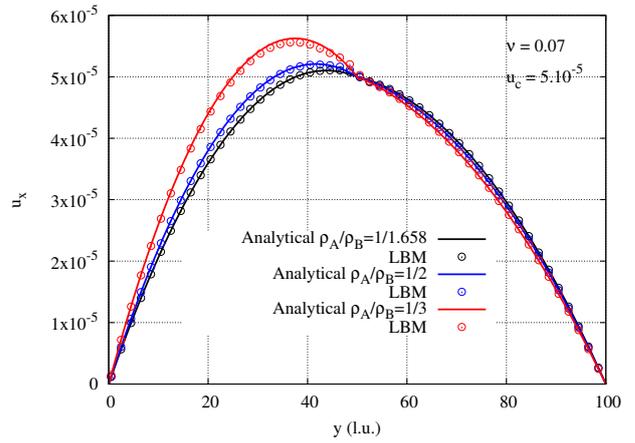}
\par\end{centering}
}
\par\end{centering}
\begin{centering}
\subfloat[\label{fig:Poiseuille_ComparisonInterpMeth}Red line --
comparisons between two interpolation methods of viscosity: linear
(black squares) and harmonic mean (red circles) defined by
Eqs. (\ref{eq:Visco_lin}) and (\ref{eq:Viscosity}) respectively.
Blue line -- verification for $\rho_{A}/\rho_{\text{B}}=1/1.658$.]{\begin{centering}
\includegraphics[angle=-90,scale=0.33]{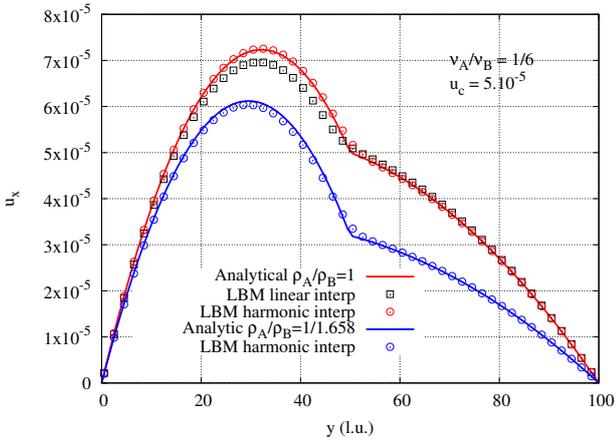}
\par\end{centering}
}$\qquad$\subfloat[\label{fig:Laplace}Laplace's law verification for three values of
surface tension.]{\begin{centering}
\includegraphics[angle=-90,scale=0.33]{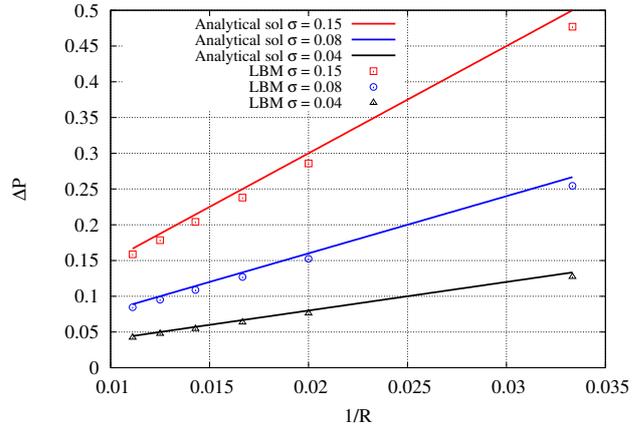}
\par\end{centering}
}
\par\end{centering}
\caption{Verification of coupling between the phase-field equation and fluid
flow model without phase change. (a) Double-Poiseuille flow with three
viscosity ratios. (b) Double-Poiseuille for three density ratios.
(c) Effect of linear interpolation and harmonic mean of viscosity.
(d) Laplace law.}
\end{figure}

The Navier-Stokes implementation is checked with the analytical solution
of a layered Poiseuille flow \citep{Zu-He_PRE2013} for two fluids
named $A$ and $B$:

\begin{equation}
u_{x}(y)=\begin{cases}
\frac{Gh^{2}}{2\eta_{A}}\left[-\left(\frac{y}{h}\right)^{2}-\frac{y}{h}\left(\frac{\eta_{A}-\eta_{B}}{\eta_{A}+\eta_{B}}\right)+\frac{2\eta_{A}}{\eta_{A}+\eta_{B}}\right] & (-h\leq y\leq0)\\
\frac{Gh^{2}}{2\eta_{B}}\left[-\left(\frac{y}{h}\right)^{2}-\frac{y}{h}\left(\frac{\eta_{A}-\eta_{B}}{\eta_{A}+\eta_{B}}\right)+\frac{2\eta_{B}}{\eta_{A}+\eta_{B}}\right] & (0\leq y\leq h)
\end{cases}\label{eq:AnalyticalSol_Poiseuille-1}
\end{equation}
where $\eta_{A}$ and $\eta_{B}$ are the dynamic viscosities and
$2h$ is the channel width. The pressure gradient is defined by $G=u_{c}(\eta_{A}+\eta_{B})/h^{2}$
with $u_{c}=5\times10^{-5}$. For the LB simulation, the mesh is composed
of $101\times101\times3$ nodes and the pressure
gradient is replaced by a force term defined by $\mathbf{F}=(G,\,0,\,0)^{T}$.
Periodic boundary conditions are set for all limits except for planes
of normal vector directed in $y$-direction where no-slip conditions
are imposed with the half bounce-back method. Two layers of different
viscosity are defined as initial condition for $\phi$: $\phi(\mathbf{x},\,0)=0.5\{1+\tanh\left[2(y-y_{0})/W\right]\}$
where $W=6\delta x$ controls the slope of the hyperbolic tangent
function and $y_{0}=(y_{max}+y_{min})/2$. The mobility coefficient
is $M_{\phi}=0.1$. Comparisons between the LBM code
and the analytical solution are presented for two cases. In the first
one, the density is identical for both fluids ($\rho_{A}=\rho_{B}=1$)
and three viscosity ratios are checked on Fig. \ref{fig:Poiseuille_Viscosity}:
$\eta_{B}/\eta_{A}=1/3,\,1/5,\,1/10$. For the first
ratio $\nu_{A}=0.1$ and $\nu_{B}=0.3$; for the second one $\nu_{A}=0.07$
and $\nu_{B}=0.35$ and for the third one $\nu_{A}=0.01$ and $\nu_{B}=0.1$.
For the second test case, the viscosity of each phase is set equal
to $\nu_{A}=\nu_{B}=0.07$ and three density ratios are checked on
Fig. \ref{fig:Poiseuille_Density}: $\rho_{A}/\rho_{B}=1/1.658,\,1/2,\,1/3$. The ratio $1/1.658$ is used in the simulations
of film boiling as well as the viscosity
ratio $\nu_{A}/\nu_{B}=1/6$. In Fig. \ref{fig:Poiseuille_ComparisonInterpMeth}
this viscosity ratio is checked for two cases. In the first simulation (red curve),
the density ratio is equal to one and the viscosity is interpolated by two methods: the linear (black squares)
and the harmonic mean (red circles) defined by
\begin{equation}
\nu(\phi)=[1-\phi(\mathbf{x},\,t)]\nu_{A}+\phi(\mathbf{x},\,t)\nu_{B},\label{eq:Visco_lin}
\end{equation}
and Eq. (\ref{eq:Viscosity}) respectively.
The differences observed with the former method justify the choice of using the latter in
the second simulation (blue curve) which combines both ratios of viscosity and
density.

\paragraph*{Laplace law}

The two-dimensional Laplace law is checked by initializing a drop
at the center of a square domain of length $L_{x}=L_{y}=2.56$ discretized
with $256\times256$ nodes. By varying the radius $R$, the difference
between pressure inside the drop ($p_{in}$) minus the pressure outside
($p_{out}$) must vary proportionally with the surface tension $\sigma$:

\begin{equation}
p_{in}-p_{out}=\frac{\sigma}{R}.\label{eq:LaplaceLaw}
\end{equation}
In order to check that relationship, an initial drop of radius $R$
and surface tension $\sigma$ is initialized at the center of the
domain ($x_{c}=y_{c}=1.28$). The density ratio $\rho_{g}/\rho_{l}$
is set equal to two ($\rho_{g}=2$, $\rho_{l}=1$) and the viscosities
are identical for each phase: $\nu_{l}=\nu_{g}=0.04$. The interface
parameters are $M_{\phi}=0.04$ and $W=0.05=5\delta x$. The LBM code
is run with a time-step equal to $\delta t=10^{-4}$ until the stationary
solution is obtained. At the end of simulation, the difference between
numerical pressures $\Delta p=p_{in}-p_{out}$ is plotted for three
values of surface tension $\sigma=0.04,\,0.08,\,0.15$. For each value
of surface tension, six LBM simulations are run for six values of
radius corresponding to each dot on Fig. \ref{fig:Laplace}. On that
plot, the slopes of LBM vary linearly and fit quite well to the Laplace
law.

\subsection{\label{sub:With-mass-transfer}Verifications with phase change: one-dimensional
Stefan problem}

In this section, we consider the problem of phase change without flow
($\mathbf{u}=\mathbf{0}$). The objective is to validate the coupling
between equations of phase-field and temperature. More precisely,
we check the new approximation (Eq. (\ref{eq:ProductionRate})) of
mass production rate $\dot{m}'''$ in the phase-field equation (Eq. (\ref{eq:CAC_ChgtPhase}))
and the latent heat release in the temperature equation (Eq. (\ref{eq:Temperature_ChgtPhase})),
i.e. the source term $-\partial\phi/\partial t$. Validation is carried
out with the Stefan problem for which several analytical solutions
exist \cite[Chapter 12]{Ozisik_2012}. Here we consider one of the
most general one-dimensional problem where the three unknowns are
the interface position varying with time $x_{I}(t)$, the liquid temperature
$T_{l}(x,\,t)$ and the gas temperature $T_{g}(x,\,t)$. Besides,
the thermal diffusivities of each phase $\alpha_{l}$ and $\alpha_{g}$
can be different. The one-dimensional domain $]0,\,\infty[$, is initially
filled with gas with constant temperature $T_{g}(x,\,t)\bigr|_{x>0,\,t=0}=T_{\infty}$
that is greater than the saturation temperature $T_{sat}$. The left
wall $x=0$ is maintained at $T_{w}$ for $t\geq0$. As a result,
condensation starts at the boundary $x=0$ and the liquid-gas interface
propagates in the positive direction. At $x\rightarrow\infty$, the
temperature is kept at $T_{\infty}$.

\paragraph*{Analytical solutions}

The mathematical formulation of this problem writes \cite[Section 12-3]{Ozisik_2012}

\begin{subequations}

\begin{equation}
\frac{\partial T_{l}}{\partial t}=\alpha_{l}\frac{\partial^{2}T_{l}}{\partial x^{2}}\label{eq:Stefan_EqLiq}
\end{equation}
for $0<x<x_{I}(t)$, with the left boundary condition imposed at $\left.T_{l}(x,\,t)\right|_{x=0}=T_{w}$.
The evolution of the gas phase is formulated as

\begin{equation}
\frac{\partial T_{g}}{\partial t}=\alpha_{g}\frac{\partial^{2}T_{g}}{\partial x^{2}}\label{eq:Stefan_EqGas}
\end{equation}
for $x_{I}(t)<x<\infty$ with $T_{g}(x\rightarrow\infty,\,t)=T_{\infty}$,
with the initial condition $T_{g}(x,\,t=0)=T_{\infty}$ and boundary
condition $T_{g}(x\rightarrow\infty,\,t)=T_{\infty}$. Interfacial
conditions are specified by

\begin{eqnarray}
\left.T_{l}(x,\,t)\right|_{x=x_{I}(t)}=\left.T_{g}(x,\,t)\right|_{x=x_{I}(t)} & = & T_{I},\label{eq:Stefan_CondInterface1}\\
\mathcal{K}_{l}\left.\frac{\partial T_{l}}{\partial x}\right|_{x=x_{I}(t)}-\mathcal{K}_{g}\left.\frac{\partial T_{g}}{\partial x}\right|_{x=x_{I}(t)} & = & \rho\mathcal{L}\frac{dx_{I}(t)}{dt}\mathrm{.}\label{eq:Stefan_CondInterface2}
\end{eqnarray}

\end{subequations}

In Eq. (\ref{eq:Stefan_CondInterface2}), $\mathcal{K}_{l}$ and $\mathcal{K}_{g}$
are the thermal conductivities of each phase. We consider identical
specific heat $\mathcal{C}_{p}^{l}=\mathcal{C}_{p}^{g}=\mathcal{C}_{p}$
and we set $\mathcal{C}_{p}=1$, $\mathcal{L}=1$ and $\rho=1$. Solutions
of interface position and temperature profiles \cite[p. 469]{Ozisik_2012}
are

\begin{subequations}

\begin{eqnarray}
x_{I}(t) & = & 2\xi\sqrt{\alpha_{l}t}\mathrm{,}\label{eq:SolStefan_Position}\\
\theta_{l}(x,\,t) & = & \theta_{w}+(\theta_{I}-\theta_{w})\frac{\mbox{erf}(x/2\sqrt{\alpha_{l}t})}{\mbox{erf}(\xi)}\mathrm{,}\label{eq:SolStefan_Temp_Liq}\\
\theta_{g}(x,\,t) & = & \theta_{\infty}+(\theta_{I}-\theta_{\infty})\frac{\mbox{erfc}(x/2\sqrt{\alpha_{g}t})}{\mbox{erfc}(\xi\sqrt{\alpha_{l}/\alpha_{g}})}\mathrm{,}\label{eq:SolStefan_Temp_Gas}
\end{eqnarray}
where the temperatures are re-written in dimensionless form with $\theta=\mathcal{C}_{p}(T-T_{sat})/\mathcal{L}$.
When $\theta=0$ the temperature of system is at saturation temperature
$T_{sat}$ and when $\theta>0$ (resp. $\theta<0$), the system is
superheated (resp. undercooled). In Eqs. (\ref{eq:SolStefan_Position})\textendash (\ref{eq:Stefan_CondInterface2}),
$\xi$ is solution of the transcendental equation

\begin{equation}
\frac{e^{-\xi^{2}}}{\mbox{erf}(\xi)}+\left(\frac{\alpha_{g}}{\alpha_{l}}\right)^{1/2}\frac{\theta_{I}-\theta_{\infty}}{\theta_{I}-\theta_{w}}\frac{e^{-\xi^{2}(\alpha_{l}/\alpha_{g})}}{\mbox{erfc}(\xi\sqrt{\alpha_{l}/\alpha_{g}})}=-\frac{\xi\sqrt{\pi}}{\theta_{w}}\label{eq:SolStefan_TranscendEq}
\end{equation}
where $\theta_{w}$ in the right-hand side is the Stefan number defined
by $St=\mathcal{C}_{p}(T_{w}-T_{sat})/\mathcal{L}$. Those solutions
are compared with \texttt{LBM\_saclay}, first with identical thermal
diffusivities $\alpha_{l}=\alpha_{g}$ and an interface temperature
$\theta_{I}$ equals to zero. The second validation considers three
ratios of diffusivity $\alpha_{l}^{j}/\alpha_{g}^{j}$ (for $j=1,2,3$)
with an interface temperature which is different of the saturation
one ($\theta_{I}\neq0$).

\end{subequations}

\paragraph*{Data entry of LBM simulations}

For LBM simulations, the two-dimensional D2Q9 lattice is used for
the temperature and phase-field equations. The LBM computational domain
is $[\ell_{x},\,L_{x}]\times[\ell_{y},\,L_{y}]=[0,\,512]\times[0,\,32]$
which is discretized by $N_{x}\times N_{y}=512\times32$ nodes i.e.
$\delta x=1$. The time-step is also set to $\delta t=1$. Boundary
conditions are periodic for $s_{i}$ and $g_{i}$ at $\ell_{y}$ and
$L_{y}$ (bottom and top walls respectively) and Dirichlet boundary
conditions are applied on left ($x=\ell_{x}$) and right ($x=L_{x}$)
walls by anti-bounceback method on $g_{i}$ and $s_{i}$. For phase-field,
the Dirichlet boundary conditions are $\phi(x,\,t)\bigr|_{x=\ell_{x}}=0$
and $\phi(x,\,t)\bigr|_{x=L_{x}}=1$. For the temperature equation, they
are $\theta(x,\,t)\bigr|_{x=\ell_{x}}=\theta_{w}$ and $\theta(x,\,t)\bigr|_{x=L_{x}}=\theta_{\infty}$.
The temperature is initialized with $\theta(x,\,0)=\theta_{\infty}$
for $0<x\leq L_{x}$ and the phase-field with $\phi(x,\,0)=0.5\left[1+\tanh(2x/W)\right]$.
The mobility parameter is $M_{\phi}=0.08$, the interface thickness
is $W=3\delta x$.

\paragraph*{Validations for $\alpha_{l}/\alpha_{g}=1$ and $\theta_{I}=0$}

\begin{figure}[t]
\subfloat[\label{fig:Stefan_Comparison-LBM-Analytical}Comparisons for three
values of thermal diffusivity: $\alpha_{g}^{1}=0.14$ (red), $\alpha_{g}^{2}=0.125$
(blue) and $\alpha_{g}^{3}=0.08$ (black). Left: $x$-profiles of
temperature $\theta$ at the end of simulation $t_{f}=2\times10^{5}$.
Right: evolution of interface position $x_{I}(t)$ tracked by $\phi=1/2$.
The temperature interface is $\theta_{I}=0$.]{\protect\begin{centering}
\begin{tabular}{ccc}
\protect\includegraphics[angle=-90,scale=0.33]{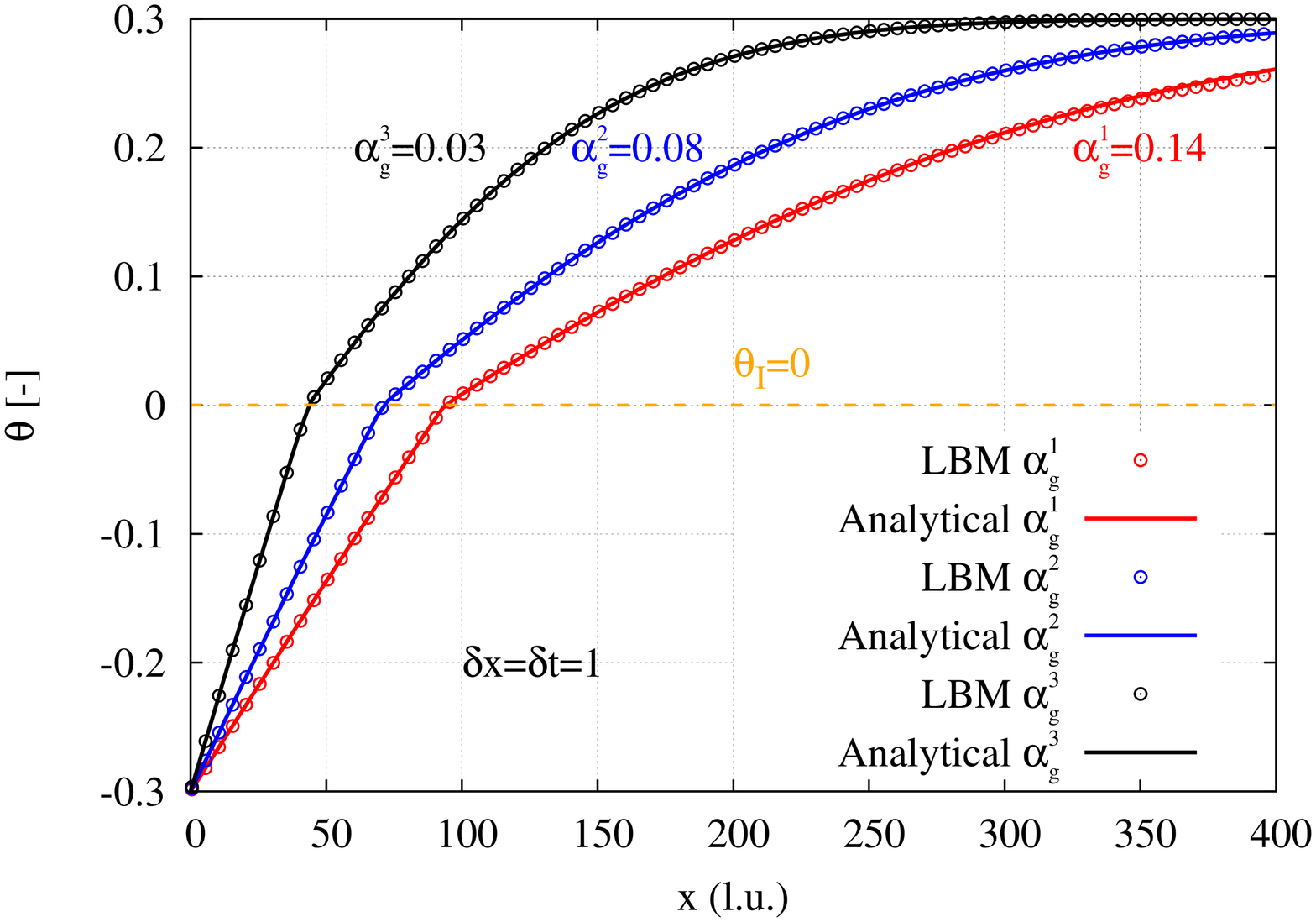} &  & \protect\includegraphics[angle=-90,scale=0.33]{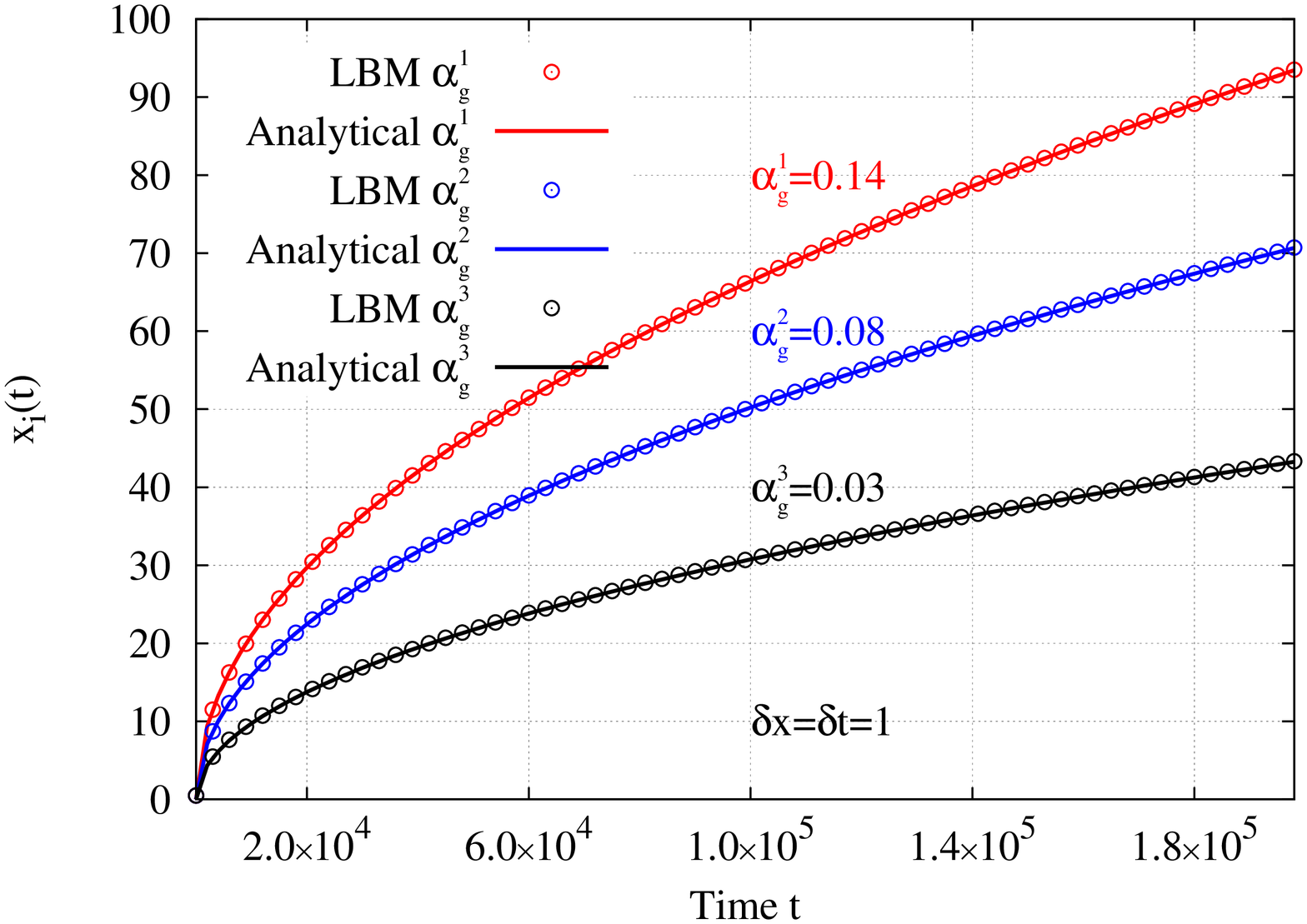}\tabularnewline
\end{tabular}\protect
\par\end{centering}

}

\subfloat[\label{fig:Stefan_Comparison-LBM-Analytical-1}Comparisons for three
ratios $\alpha_{l}^{j}/\alpha_{g}^{j}=10,\,5,\,2$ for $j=1,2,3$
with $\alpha_{l}^{1}=0.14$ (red), $\alpha_{l}^{2}=0.125$ (blue)
and $\alpha_{l}^{3}=0.08$ (black). Left: $x$-profiles of temperature
$\theta$ at the end of simulation $t_{f}=2\times10^{5}$. Right:
evolution of interface position $x_{I}(t)$ tracked by $\phi=1/2$.
The temperature interface is $\theta_{I}=0.05$.]{\protect\begin{centering}
\begin{tabular}{ccc}
\protect\includegraphics[angle=-90,scale=0.33]{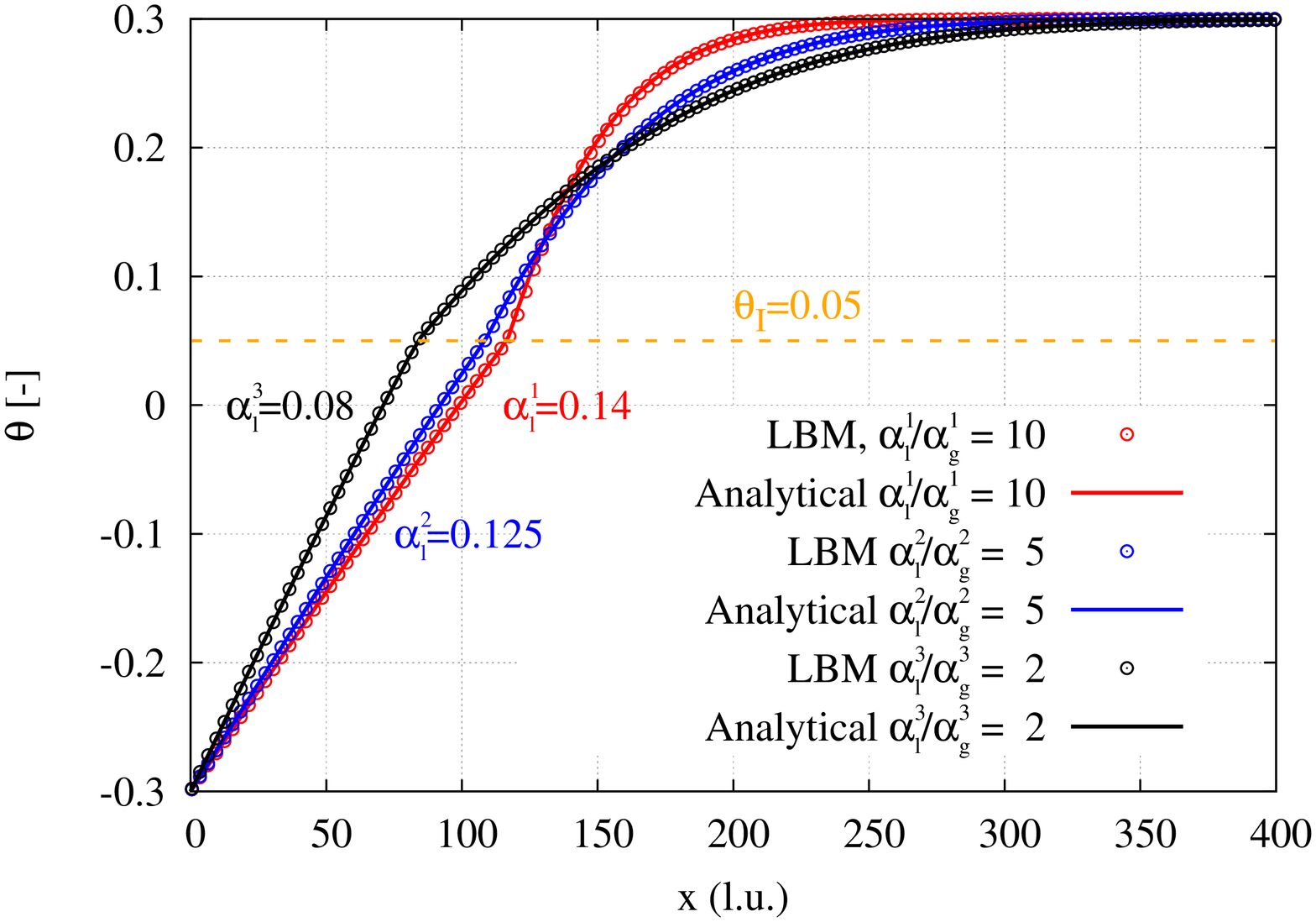} &  & \protect\includegraphics[angle=-90,scale=0.33]{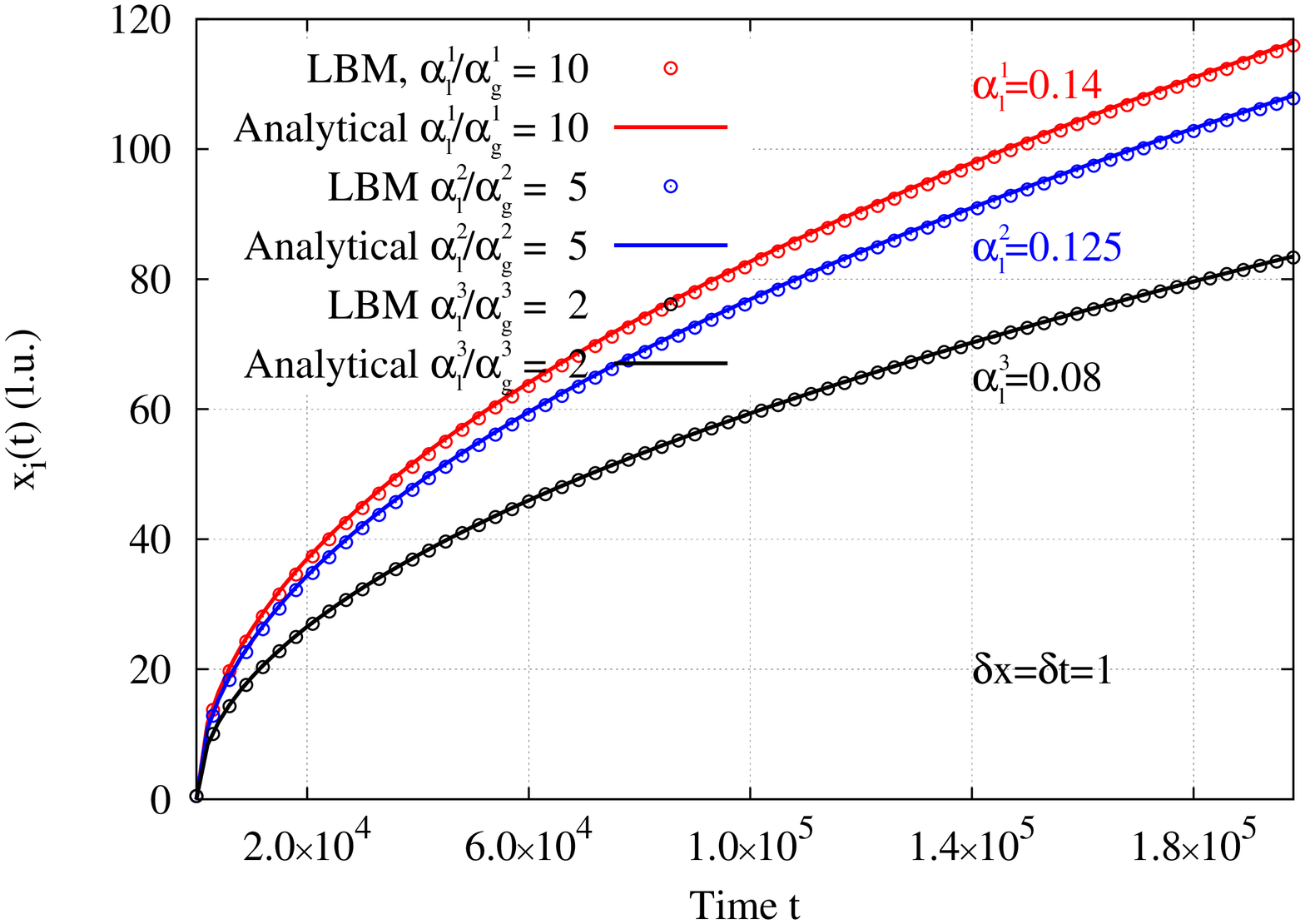}\tabularnewline
\end{tabular}\protect
\par\end{centering}

}

\protect\caption{Comparisons between LBM (dots) and analytical solution of Stefan problem
(solid lines). (a) With $\alpha_{l}/\alpha_{g}=1$ and $\theta_{I}=0$.
(b) With $\alpha_{l}/\alpha_{g}\protect\neq1$ and $\theta_{I}\protect\neq0$.}
\end{figure}

Before considering the more general case $\alpha_{l}/\alpha_{g}\neq1$
and $\theta_{I}\neq0$, we assume that thermal diffusivities are the
same in liquid and gas ($\alpha_{l}=\alpha_{g}=\alpha$) and the interface
temperature is at saturation ($\theta_{I}=0$). In that case, whatever
the diffusivity value $\alpha$, the solution of the transcendental
equation (Eq. (\ref{eq:SolStefan_TranscendEq})) depends only on $\theta_{w}$
and $\theta_{\infty}$. With $\theta_{w}=-0.3$ and $\theta_{\infty}=0.3$,
its solution is $\xi=0.280680$. Comparisons between analytical solutions
and LBM simulations are presented on Fig. \ref{fig:Stefan_Comparison-LBM-Analytical}
for three values of thermal diffusivity $\alpha_{g}^{j}=0.14,\,0.08,\,0.03$
with $j=1,\,2,\,3$. LBM temperature profiles are superimposed with
the analytical solution (Eqs. (\ref{eq:SolStefan_Temp_Liq}) and (\ref{eq:SolStefan_Temp_Gas}))
at the final time of simulation $t_{f}=2\times10^{5}$ (Fig. \ref{fig:Stefan_Comparison-LBM-Analytical},
left). Successive positions of vapor/liquid interface also fit with
the analytical solution (Fig. \ref{fig:Stefan_Comparison-LBM-Analytical},
right) for three values of thermal diffusivity.

\paragraph*{Validations for $\alpha_{l}/\alpha_{g}\protect\neq1$ and $\theta_{I}\protect\neq0$}

Now we consider a more general case for which the diffusivities of
liquid and gas can be different. Three ratios are simulated $\alpha_{l}^{j}/\alpha_{g}^{j}=10,\,5,\,2$
for $j=1,2,3$ with $\alpha_{l}^{1}=0.14$, $\alpha_{l}^{2}=0.125$
and $\alpha_{l}^{3}=0.08$. Same values of $\theta_{w}=-0.3$ and
$\theta_{\infty}=0.3$ are kept, and the interface temperature is
now equal to $\theta_{I}=0.05$. For those values, the corresponding
solutions of the transcendental equation are $\xi^{1}=0.349635$,
$\xi^{2}=0.343882$ and $\xi^{3}=0.331864$. For LBM simulations,
all numerical values are identical except for interface temperature
and diffusivities of each phase. As confirmed by temperature profiles
(Fig. \ref{fig:Stefan_Comparison-LBM-Analytical-1}, left) and the
evolution of interface position (Fig. \ref{fig:Stefan_Comparison-LBM-Analytical-1},
right), the model of phase change is well adapted to simulate the
phase change problem with different diffusivities in each phase and
an interface temperature not equal to zero. Finally this test case
validates the approximation of the mass production rate $\dot{m}'''$
defined by Eq. (\ref{eq:ProductionRate}) and implementation of LBM
for the phase-field and temperature equations.

\section{\label{sec:FilmBoiling}Simulations of film boiling}

Film boiling is a classical problem of two-phase flows with phase
change. It has already been simulated with a lot of different numerical
techniques (see \cite{Review_FilmBoilingIJHMT2017} for a recent review)
for studying the effect of geometries such as an horizontal cylinder
\cite{Esmaeeli-Tryggvason_IJMF2004} or for studying the effect of
an electric field \cite{Vinod_PhysFluids2016}. With the lattice Boltzmann
method, several simulations use the Cahn-Hilliard model or the pseudo-potential
method (respectively in \cite[and references therein]{Begmohammadi_etal_CAMWA2016,Hu-Liu_Boiling-PseudoPot_ApplThermEngi2019}).
Here we present the capability of the conservative Allen-Cahn equation
with a production rate defined by Eq. (\ref{eq:ProductionRate}) to
simulate that problem. In section \ref{sub:FilmBoiling_Physical-problem},
the physical configuration is reminded; in section \ref{sub:FilmBoiling_Nodes-Antinodes}
one simulation of bubbles detachment on nodes and anti-nodes is detailed;
in section \ref{sub:Computational-times}, indications will be given
on computational times for two mesh sizes: $1024^{2}$ for GPU and
CPU and $4096\times3072$ for multi-GPUs.

\subsection{\label{sub:FilmBoiling_Physical-problem}Physical configuration}

Inside a two-dimensional domain $\Omega=\Pi_{\upsilon=x,y}[\ell_{\upsilon},\,L_{\upsilon}]$,
a thin film of gas of height $y_{0}$ is initialized near the bottom
wall $y=\ell_{y}$ which is heated by applying a constant temperature
$\theta|_{y=\ell_{y}}=\theta_{w}$. The liquid is above the thin film
and the gravity acts downward $\mathbf{g}=(0,\,-g_{y})^{T}$. On the
top wall $y=L_{y}$, the temperature is imposed at saturation and
the phase-field is equal to $\phi=+1$ (i.e. gas phase). The left
and right walls are periodic. If the interface is destabilized by
an initial condition defined by 

\begin{equation}
y=y_{0}+y_{1}\sin\left(\frac{2\pi x}{\lambda}\right),\label{eq:FilmBoiling_InitCond}
\end{equation}
where $y_{1}$ and $\lambda$ are respectively the amplitude and the
wavelength of the perturbation, then we can observe bubbles of gas
that grow, detach and rise in the domain, provided that the wavelength
of perturbation $\lambda$ is greater than a critical value $\lambda_{c}$
defined by

\begin{equation}
\lambda_{s}=\sqrt{\frac{\sigma}{(\rho_{l}-\rho_{g})g_{y}}},\qquad\lambda_{c}=2\pi\lambda_{s}.\label{eq:Critical_wavelength}
\end{equation}
The thermal-hydrodynamics of this problem is controlled by several
dimensionless numbers: the Grashof number $Gr=\rho_{g}g_{y}(\rho_{l}-\rho_{g})\lambda_{s}^{3}/\rho_{g}^{2}\nu_{g}^{2}$,
the Prandtl number $Pr=\nu_{g}/\alpha_{g}$ and the Jacob number $Ja=\mathcal{C}_{p}(T_{w}-T_{sat})/\mathcal{L}$.
Moreover the solution is sensitive to parameters that are involved
in Eq. (\ref{eq:FilmBoiling_InitCond}). Several sensitivity simulations
on parameters of the initial condition can be found in \cite{Singh-Premachandran_IJHMT2020}.

Simulations of film boiling with \texttt{LBM\_saclay} are first carried
out inside a two-dimensional domain $\Omega=[0,\,1.28]^{2}$ which
is discretized with $N_{x}\times N_{y}=1024\times1024$ nodes. The
space- and time-steps are respectively equal to $\delta x=1.25\times10^{-3}$
and $\delta t=7.5\times10^{-5}$. The D2Q9 lattice is used for all
distribution functions $f_{i}$, $g_{i}$ and $s_{i}$. For parameters
of Table \ref{tab:FilmBoiling_Parameters}, the value of critical
wavelength is $\lambda_{c}=2\pi\lambda_{s}=0.2738$, with $\lambda_{s}=4.358\times10^{-2}$.
The Jacob number is $Ja=0.025$, the Prandtl $Pr=0.2$ and the Grashof
number is $Gr=871.38$.

\begin{table}
\begin{raggedright}
\hspace{-3mm}{\footnotesize{}}%
\begin{tabular}{lll}
\hline 
\textbf{\footnotesize{}Liquid and gas properties} & \textbf{\footnotesize{}Interface properties} & \textbf{\footnotesize{}Other parameters}\tabularnewline
{\footnotesize{}}%
\begin{tabular}{lll}
\hline 
 & {\footnotesize{}Liquid} & {\footnotesize{}Gas}\tabularnewline
\hline 
{\footnotesize{}Density} & {\footnotesize{}$\rho_{l}=1.658$} & {\footnotesize{}$\rho_{g}=1$}\tabularnewline
{\footnotesize{}Kinematic viscosity} & {\footnotesize{}$\nu_{l}=3\times10^{-3}$} & {\footnotesize{}$\nu_{g}=5\times10^{-4}$}\tabularnewline
{\footnotesize{}Thermal diffusivity} & {\footnotesize{}$\alpha_{l}=2.5\times10^{-4}$} & {\footnotesize{}$\alpha_{g}=2.5\times10^{-3}$}\tabularnewline
 &  & \tabularnewline
\end{tabular} & {\footnotesize{}}%
\begin{tabular}{ll}
\hline 
{\footnotesize{}Parameter} & {\footnotesize{}Value}\tabularnewline
\hline 
{\footnotesize{}Surface tension} & {\footnotesize{}$\sigma=5\times10^{-3}$}\tabularnewline
{\footnotesize{}Interface temp.} & {\footnotesize{}$\theta_{I}=0$}\tabularnewline
{\footnotesize{}Mobility} & {\footnotesize{}$M_{\phi}=1.7\times10^{-3}$}\tabularnewline
{\footnotesize{}Interface width} & {\footnotesize{}$W=5\times10^{-3}$}\tabularnewline
\end{tabular} & {\footnotesize{}}%
\begin{tabular}{ll}
\hline 
{\footnotesize{}Parameter} & {\footnotesize{}Value}\tabularnewline
\hline 
{\footnotesize{}Gravity} & {\footnotesize{}$g_{y}=4$}\tabularnewline
{\footnotesize{}Bottom temp.} & {\footnotesize{}$\theta_{y=\ell_{y}}=0.025$}\tabularnewline
{\footnotesize{}Top temp.} & {\footnotesize{}$\theta_{y=L_{y}}=0$}\tabularnewline
{\footnotesize{}Latent/specific heat} & {\footnotesize{}$\mathcal{L}/\mathcal{C}_{p}=1$}\tabularnewline
\end{tabular}\tabularnewline
\hline 
\end{tabular}
\par\end{raggedright}{\footnotesize \par}

\protect\caption{\label{tab:FilmBoiling_Parameters}Parameters for film boiling simulations.}
\end{table}

\subsection{\label{sub:FilmBoiling_Nodes-Antinodes}Simulation of bubble detachment
on nodes and antinodes}

\begin{figure*}
\subfloat[\label{fig:FilmBoiling_a}Detachment of bubbles occurs on nodes at
the early stage of the simulation.]{\protect\begin{centering}
\begin{tabular}{ccc}
$t_{a}^{\star}\simeq23.94$ & $t_{a}^{\star}+\delta t^{\star}\simeq28.73$ & $t_{a}^{\star}+2\delta t^{\star}\simeq33.52$\tabularnewline[2mm]
\protect\includegraphics[scale=0.15]{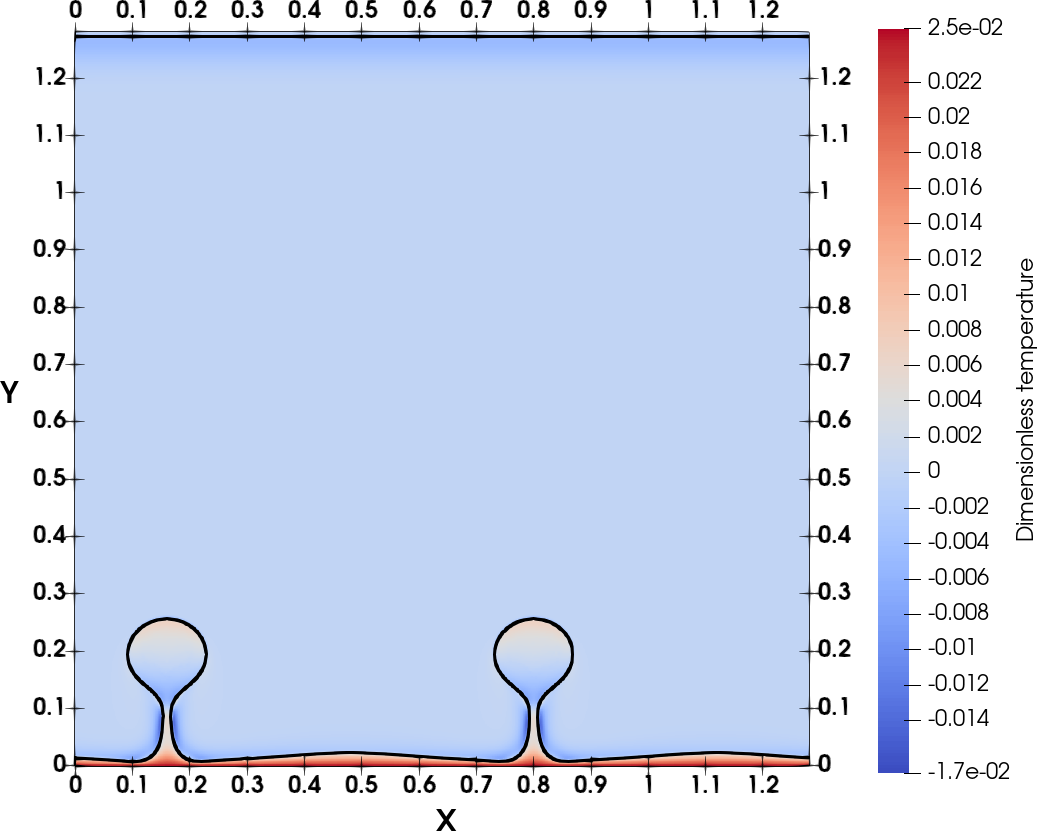} & \protect\includegraphics[scale=0.15]{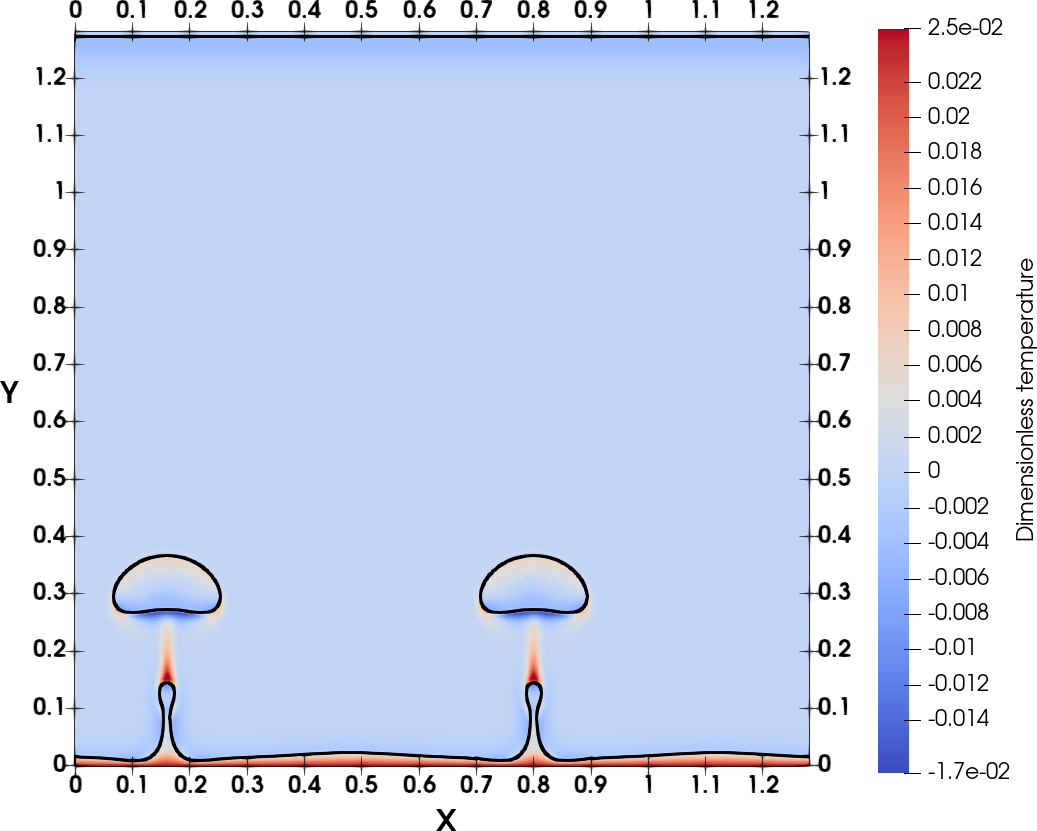} & \protect\includegraphics[scale=0.15]{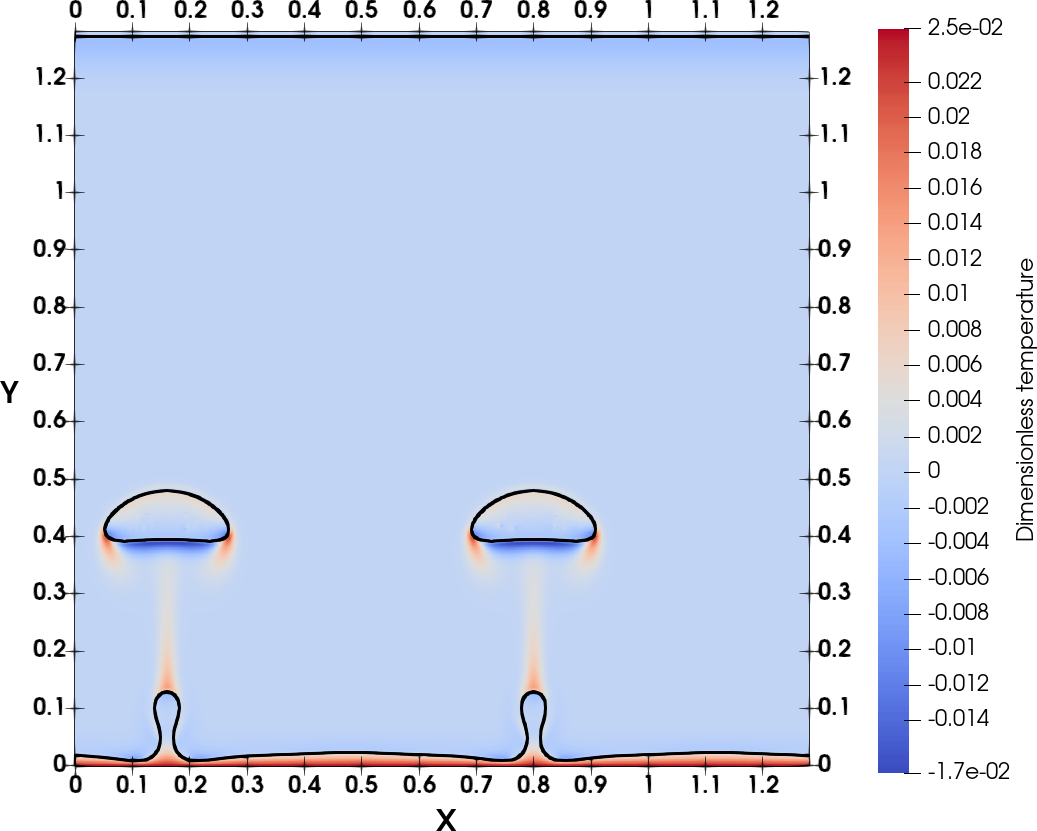}\tabularnewline
\end{tabular}\protect
\par\end{centering}

}

\subfloat[\label{fig:FilmBoiling_b}Coalescence is observed at the top of the
domain for bubbles detached from nodes. It is also observed a detachment
of bubbles at anti-nodes.]{\protect\begin{centering}
\begin{tabular}{ccc}
$t_{b}^{\star}\simeq114.93$ & $t_{b}^{\star}+\delta t^{\star}\simeq119.72$ & $t_{b}^{\star}+2\delta t^{\star}\simeq124.51$\tabularnewline[2mm]
\protect\includegraphics[scale=0.15]{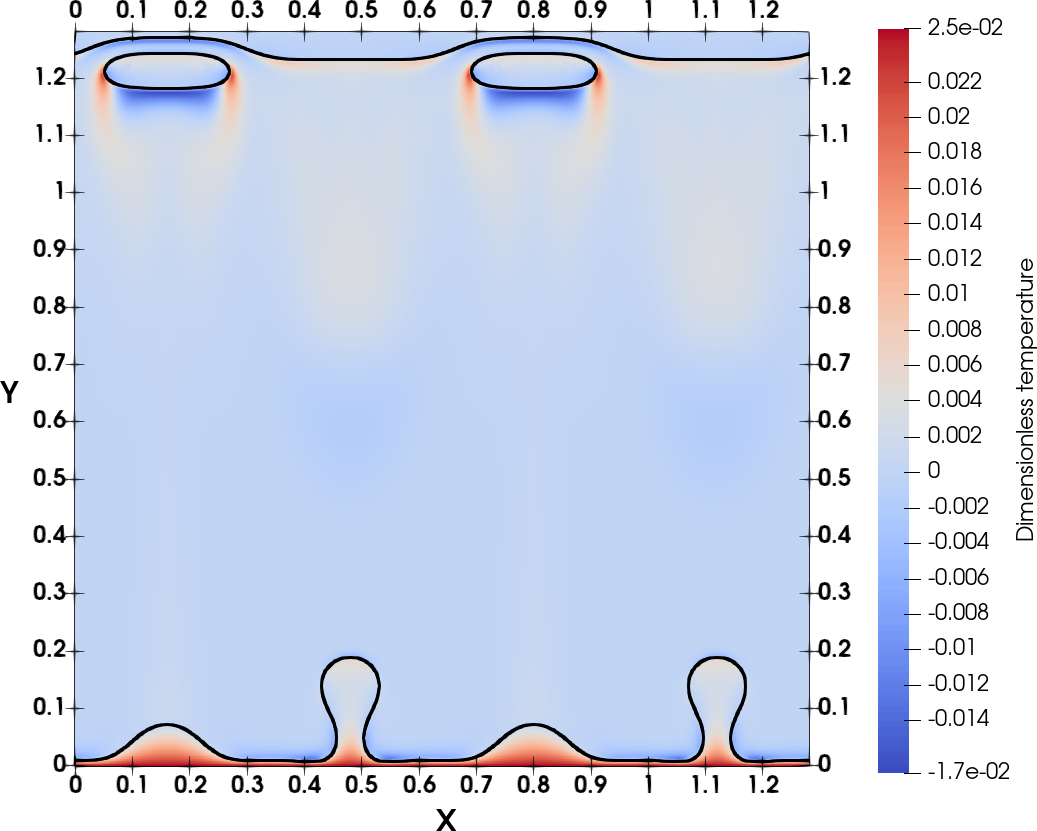} & \protect\includegraphics[scale=0.15]{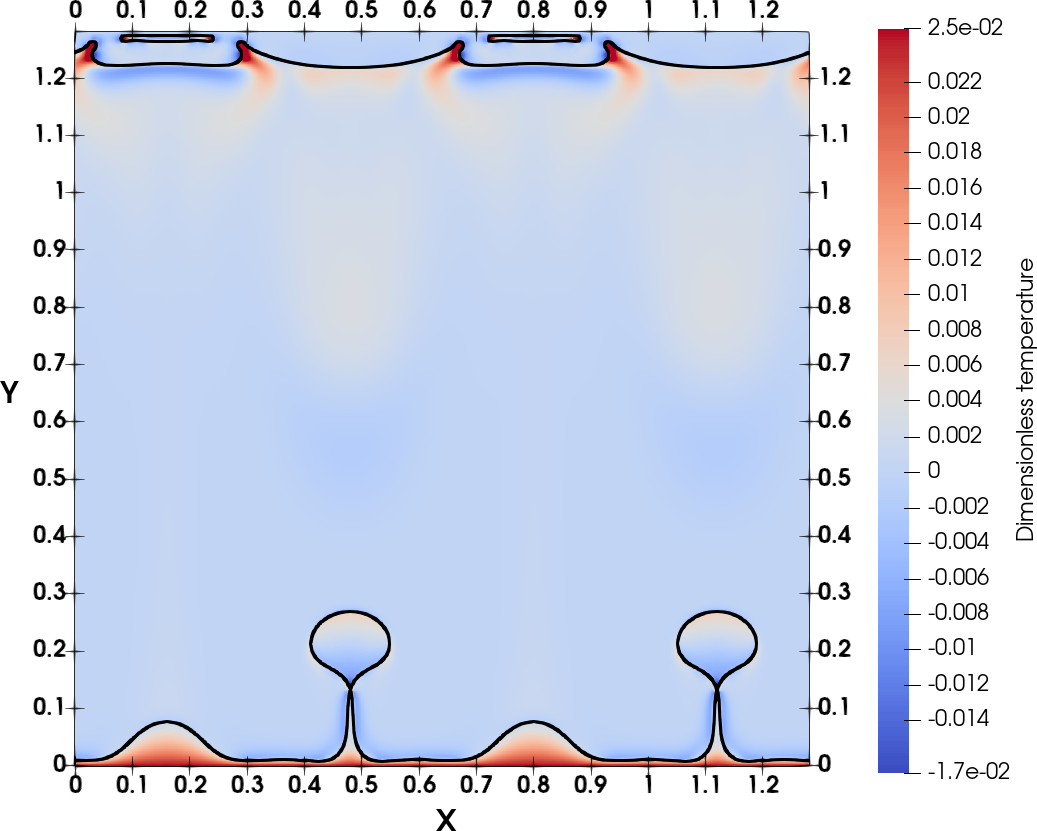} & \protect\includegraphics[scale=0.15]{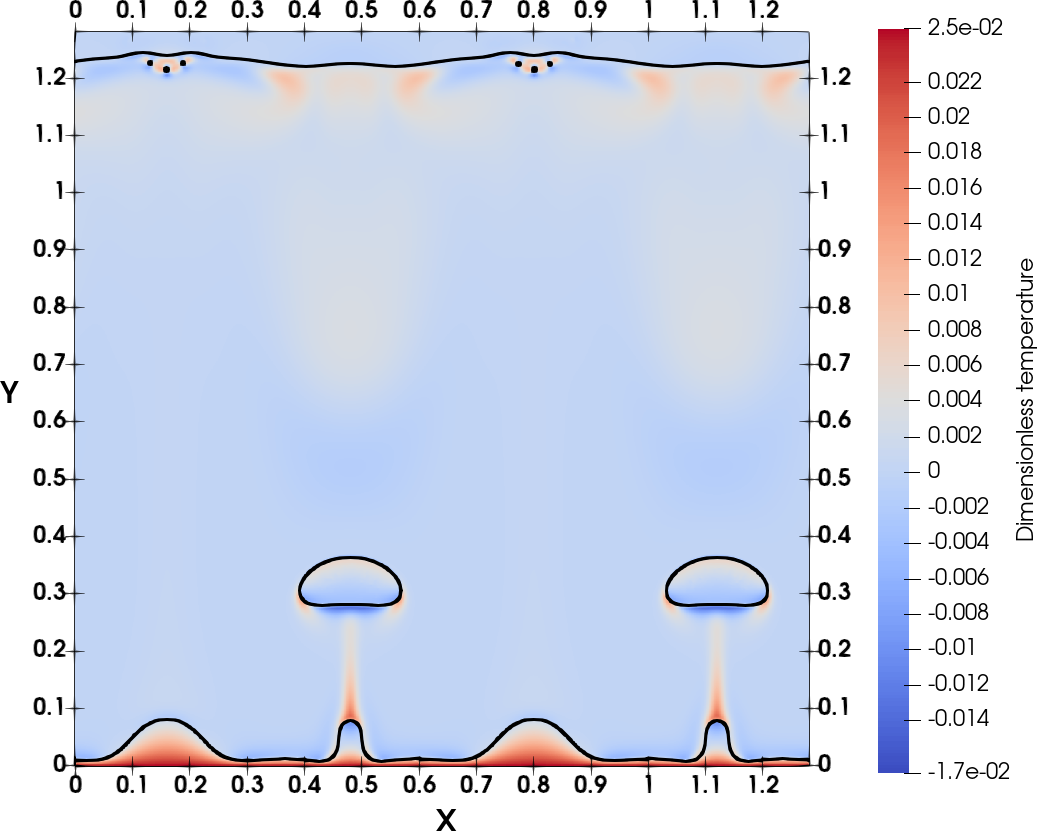}\tabularnewline
\end{tabular}\protect
\par\end{centering}

}

\subfloat[\label{fig:FilmBoiling_c}Later during the simulation, bubbles are
detached on nodes, the cycle is pursued periodically.]{\protect\begin{centering}
\begin{tabular}{ccc}
$t_{c}^{\star}\simeq148.45$ & $t_{c}^{\star}+\delta t^{\star}\simeq153.24$ & $t_{c}^{\star}+2\delta t^{\star}\simeq158.03$\tabularnewline[2mm]
\protect\includegraphics[scale=0.15]{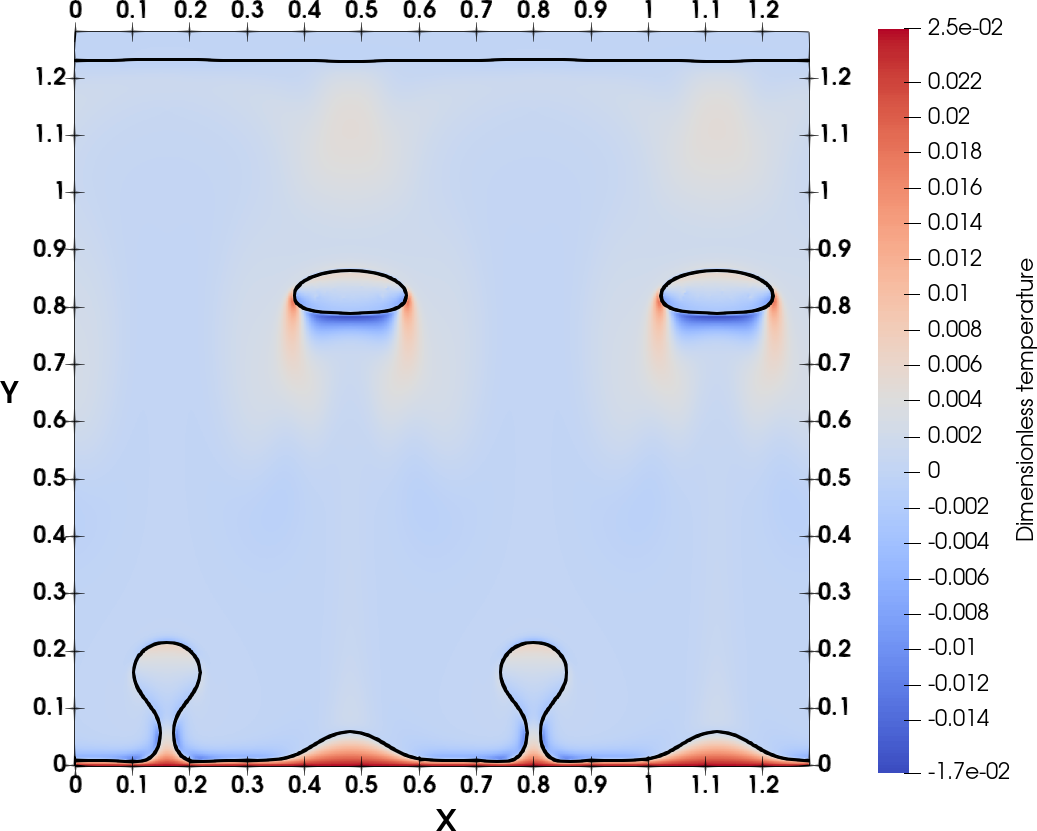} & \protect\includegraphics[scale=0.15]{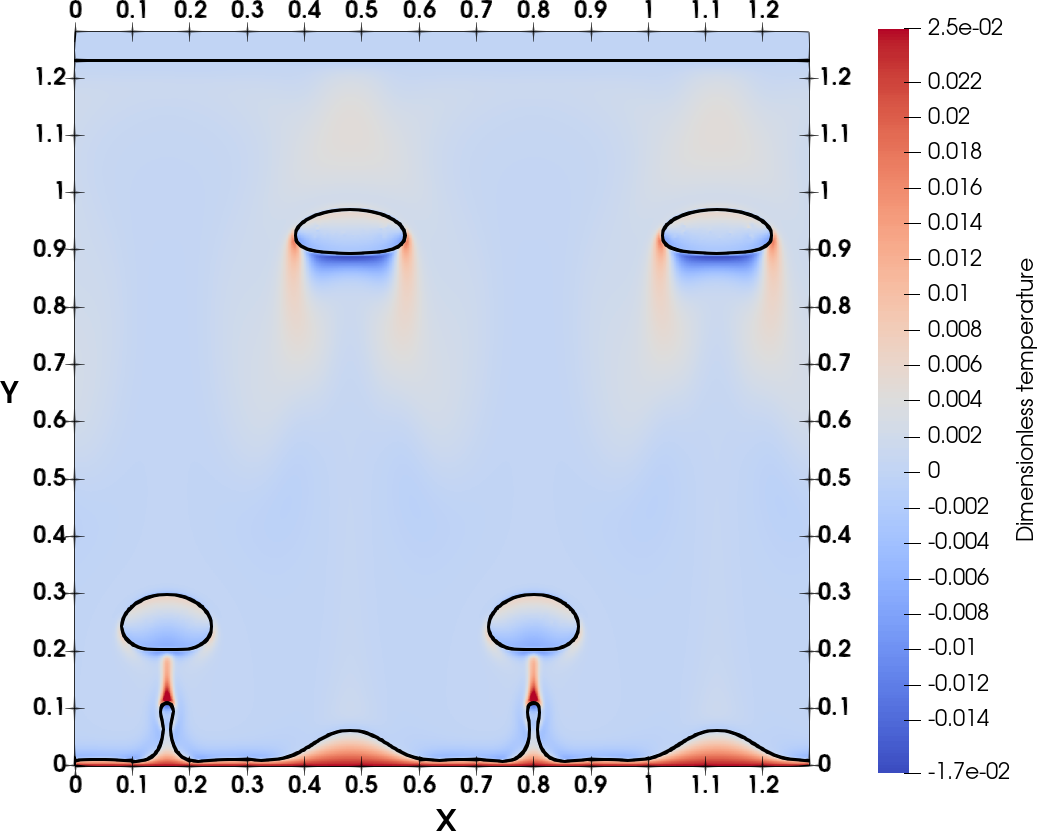} & \protect\includegraphics[scale=0.15]{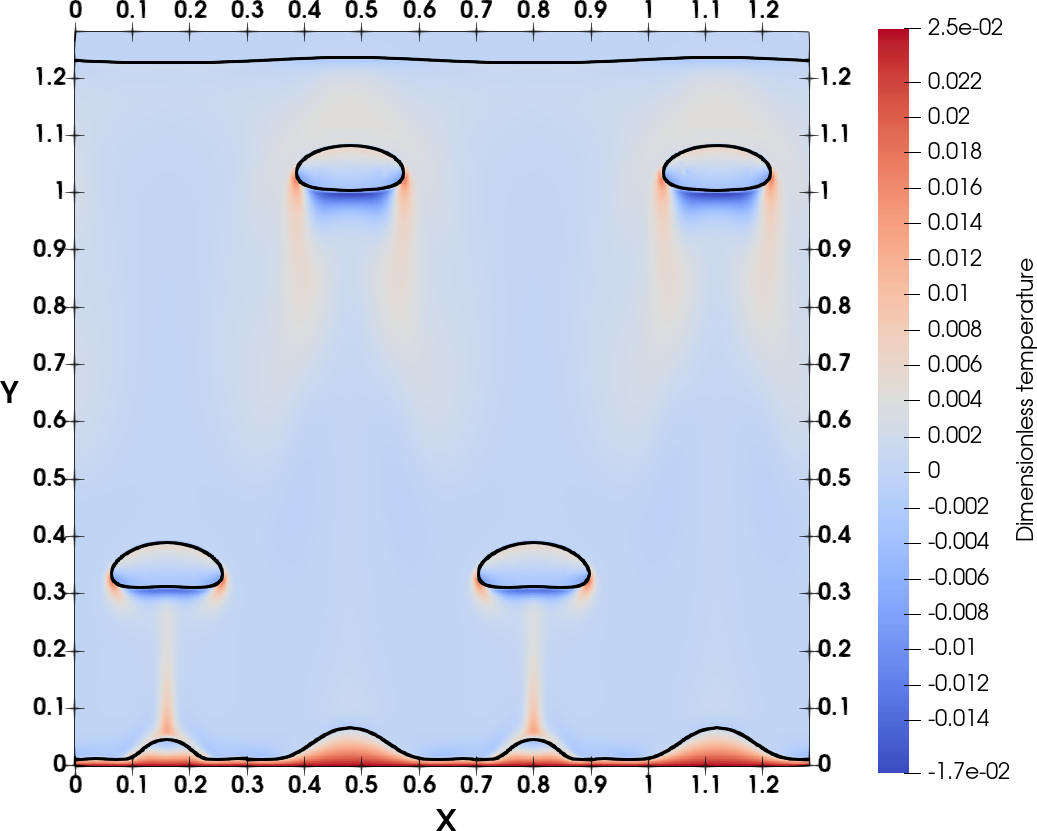}\tabularnewline
\end{tabular}\protect
\par\end{centering}

}

\begin{centering}
\subfloat[\label{fig:Streamlines_FilmBoiling}Streamlines (white lines) and
interface $\phi=1/2$ (black lines) superimposed on the velocity magnitude
(colored field) at $t^{\star}\simeq158.03$.]{\protect\begin{centering}
\protect\includegraphics[scale=0.17]{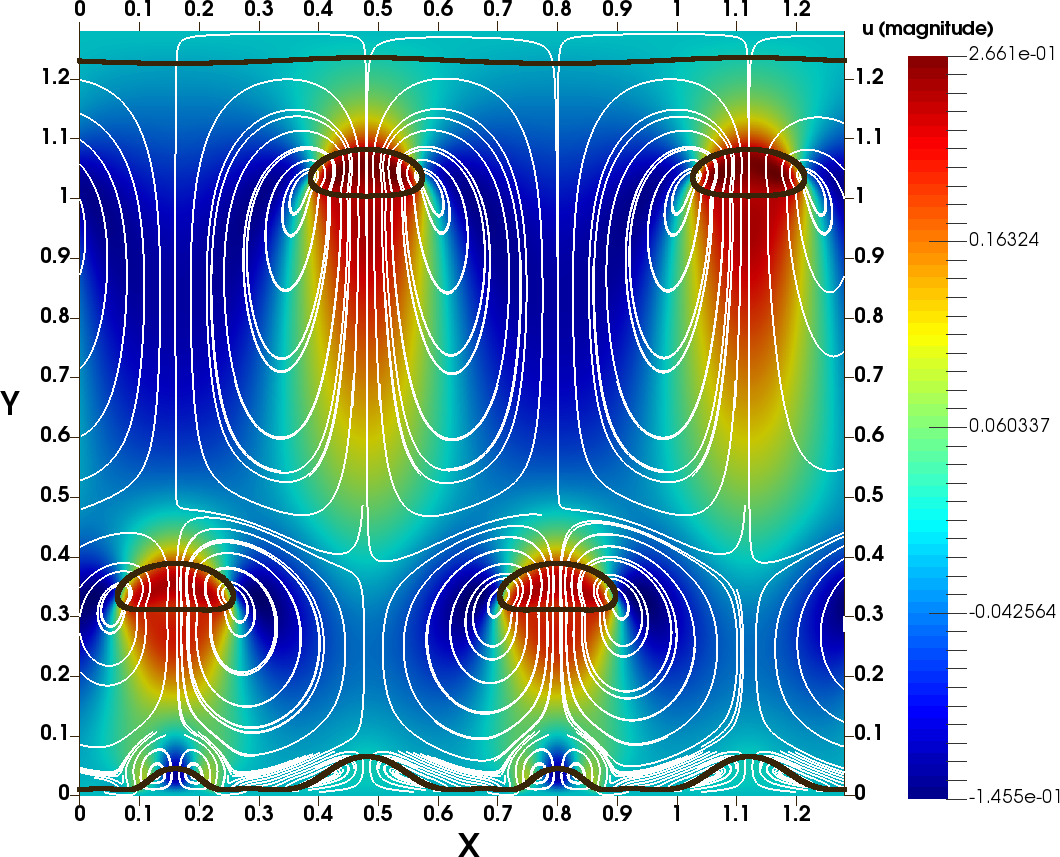}\protect
\par\end{centering}

}
\par\end{centering}

\protect\caption{\label{fig:FilmBoiling_Simulation}Simulation of film boiling for
$Ja=0.025$. Interface position $\phi=1/2$ superimposed on temperature
field and for several dimensionless times of simulation. Three successive
times from (a) $t_{a}^{\star}\simeq23.94$, (b) $t_{b}^{\star}\simeq114.93$
and (c)\textbf{ }$t_{c}^{\star}\simeq148.45$ with $\delta t^{\star}=4.79$.}
\end{figure*}

We present one simulation for which the interface is initialized by
Eq. (\ref{eq:FilmBoiling_InitCond}) with $y_{0}=0.03$, $y_{1}=0.015$
and $\lambda=0.64$. The choice $\lambda=0.64$ was done after one
first preliminary simulation which was performed with $\lambda=0.32$
($>\lambda_{c}=0.2738$) to check detachment of bubbles. For $\lambda=0.64$,
the maximum value of $y$ is $y_{max}=0.045$ for two positions $x_{ymax}^{(1)}=0.16$
and $x_{ymax}^{(2)}=0.8$. Its minimum value is $y_{min}=0.015$ for
two positions $x_{ymin}^{(1)}=0.48$ and $x_{ymin}^{(2)}=1.12$. Positions
$x_{ymax}^{(1),(2)}$ are called ``nodes'' and $x_{ymin}^{(1),(2)}$
are called ``anti-nodes''. Here, we present one simulation to observe
detachment of bubbles alternatively on nodes and anti-nodes. Actually,
it is what we observe on Figs. \ref{fig:FilmBoiling_a}--\ref{fig:FilmBoiling_c}
which present the temperature fields and the iso-values $\phi=1/2$
(black line) at several dimensionless times. The dimensionless time
is defined by $t^{\star}=t/t_{s}$ where $t_{s}=\sqrt{\lambda_{s}/g_{y}}=0.1044$.
At the early stage of simulation (Fig. \ref{fig:FilmBoiling_a}),
we can observe that the detachment of bubbles occurs on nodes. Later
during the simulation (Fig. \ref{fig:FilmBoiling_b}), the bubbles that
are emitted on nodes coalesce on the top on the domain, while two
other bubbles grow and are detached from anti-nodes. Finally (Fig.
\ref{fig:FilmBoiling_c}), the cycle is repeated periodically: bubbles
emitted at anti-nodes coalesce and new bubbles on nodes detach and
rise. Streamlines and velocity magnitude corresponding to the last
time $t^{\star}\simeq158.03$ are presented on Fig. \ref{fig:Streamlines_FilmBoiling}.
In Table \ref{tab:FilmBoiling_Parameters}, the mobility was set to $M_{\phi}=1.7\times10^{-3}$ after a sensitivity analysis. If $M_{\phi}$ is too low, the authors have observed the appearance of parasitic bubbles in the liquid phase. The mobility coefficient is directly related to the relaxation time $\tau_{g}$ and the algorithm can be unstable if its value is too low. It is expected that a wider range of parameter $M_{\phi}$ could be reached with the TRT or MRT collisions operators.

\subsection{\label{sub:Computational-times}Computational times}
A first comparison of computational times between GPUs and CPUs has been indicated in Section \ref{sub:KernelOptimization}, but only for a diffusive problem. For a single- and double-Poiseuille flow of Section \ref{sub:Without-mass-transfer}, the computational times on a $100\times100$ lattice are respectively 56 MLUPS and 38 MLUPS. Those computations have been performed on a computer equipped of one AMD CPU processor (Ryzen 5 2600, 3.4GHz with 12 threads). The MLUPS are higher for the single-phase because the algorithm requires much less floating points computations. There is neither Allen-Cahn equation nor intermediate gradient to update for a single-phase flow. However, let us note that the MLUPS for two distribution functions are higher than half of the value obtained with only one (i.e. 28 MLUPS), which indicates a good code optimization by resolving the phase-field equation.

\begin{figure*}
\begin{centering}
\subfloat[\label{Jacob0.025}$Ja=0.025$.]{\protect\begin{centering}
\protect\includegraphics[scale=0.3]{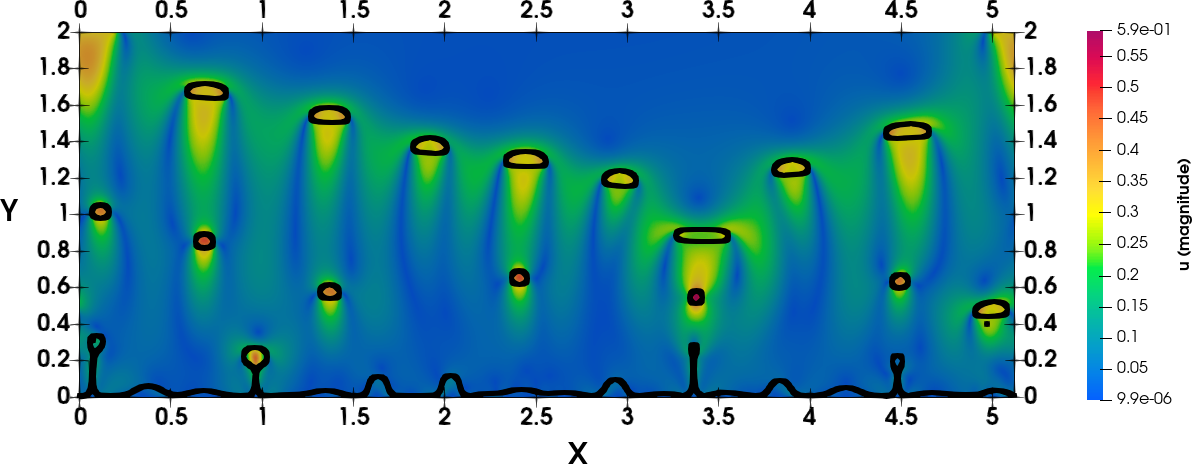}\protect
\par\end{centering}

}
\par\end{centering}

\begin{centering}
\subfloat[\label{Jocob_0.1}$Ja=0.1$.]{\protect\begin{centering}
\protect\includegraphics[scale=0.3]{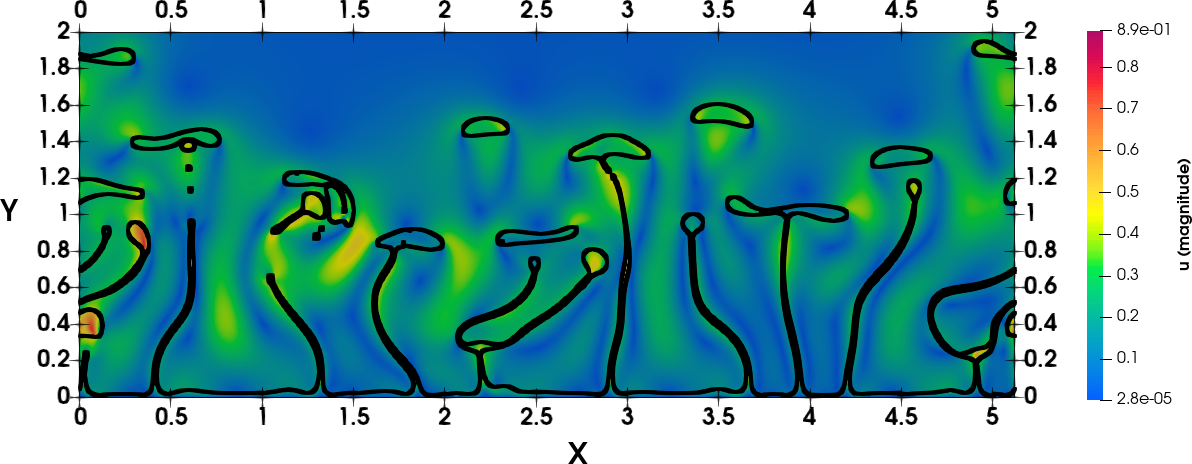}\protect
\par\end{centering}

}
\par\end{centering}

\protect\caption{\label{fig:Jacob-Effect}Velocity magnitude (colored field) and interface
position $\phi=1/2$ (black lines) at $t^{\star}=95.78$ for (a) $Ja=0.025$
and (b) $Ja=0.1$.}
\end{figure*}

Simulation of diffusion or single-phase flow requires only one distribution and the double-Poiseuille flow requires two distribution functions. The film boiling simulation requires three Lattice Boltzmann equations with
three distribution functions and the computation of additional gradients. In that case, to complete $5.33\times10^{5}$ time iterations on a computational domain of
$1024^{2}$ nodes, the simulation took 1h56m (80.96 MLUPS) on a
single GPU NVIDIA$^{\circledR}$ K80. The same simulation took 12h57m
(11.97 MLUPS) on 16-cores Intel$^{\circledR}$
Xeon$^{\circledR}$ CPU E5-2630 v3 2.40GHz. The computation on GPU is quicker
than on CPU as expected after the preliminary diffusion simulation of section
\ref{sub:KernelOptimization}. The ratio is 6.7 times in favor of GPU compared to CPU. Next, the full grid
($1024^{2}$ nodes) is decomposed in four sub-domains composed of
$256\times1024$ nodes, each one of them being taken in charge by one GPU. The
simulation took 38 minutes (249.99 MLUPS) to perform the same
number of time iterations on four parallel GPUs. The computational time is
divided by a factor three compared to a single GPU.
Finally, the computational domain is increased to $\Omega=[0,\,5.12]\times[0,\,3.84]$
and discretized by $N_{x}\times N_{y}=4096\times3072$ nodes, i.e.
the mesh size is twelve times bigger than the previous one. The initial
condition is slightly modified to 

\begin{equation}
y=y_{0}+y_{1}\sum_{i=1}^{16}\sin\left(\frac{2\pi x}{\lambda_{i}}\right)\label{eq:FilmBoiling_InitCond-1}
\end{equation}
where the interface position $y$ is perturbed with several modes
$\lambda_{i}$ which are randomly picked, uniformly distributed between
$0.5\lambda_{c}\leq\lambda_{i}\leq1.5\sqrt{3}\lambda_{c}$. We simulate
two values of wall temperature $\theta_{w}=0.025$ and $\theta_{w}=0.1$
corresponding to Jacob numbers respectively equal to $Ja=0.025$ and
$Ja=0.1$. All other values of physical parameters remain identical
(Table \ref{tab:FilmBoiling_Parameters}). A comparison on shapes
of bubbles is given at $t^{\star}=95.78$ on Fig. \ref{fig:Jacob-Effect}.
When the Jacob number has the value of Section \ref{sub:FilmBoiling_Nodes-Antinodes},
discrete bubbles are released periodically from the initial condition
(Fig. \ref{Jacob0.025}). When the Jacob number is increased to $0.1$,
long vapor jets are observed below bubbles (Fig. \ref{Jocob_0.1}).
That observation is consistent with those simulated with other techniques
and even observed on experiments cited in \cite[Sec 5.1.2 and Fig. 9]{Review_FilmBoilingIJHMT2017}.
The simulation took 80 minutes (713 MLUPS) on 8 parallel GPUs to complete $5.33\times10^{5}$
time iterations.

\section{\label{sec:Conclusion}Conclusion}

In this paper, the LBM implementation of two-phase flows was revisited
by improving two main points. The first one focuses on the model formulation
of phase change and the second one focuses on the portability of the
code on various platforms. The interface is tracked by the conservative
Allen-Cahn model with a source term involving a mass production rate
at the interface. In this work, that source term is simplified compared
to approaches of literature, and the approximation avoids to calculate
the gradients of temperature numerically. The model is able to simulate
two phases of different thermal diffusivities with an interface temperature
which is not necessarily at saturation. The phase-field model is coupled
with the incompressible Navier-Stokes model where a source term was
added in the mass balance equation. The source term is defined as
the product of mass production rate times one term inversely proportional
to densities. An additional equation on temperature completes the
model. The time derivative of phase-field appears in the source term
of that equation. It is interpreted as the release or absorption of
latent heat at the interface.

The Lattice Boltzmann schemes for all equations are implemented in
a new C++ code coupled with the \texttt{Kokkos} library for its performance
portability. The new code, called \texttt{LBM\_saclay}, can be run
with good performance on several architectures such as Graphical Process
Units (GPUs), Central Process Units (CPUs) and even multi-GPUs and
multi-CPUs. Indeed, two levels of parallelism are developed inside
the code. The first one uses \texttt{Kokkos} for intra-node parallelism,
whereas \texttt{MPI} takes in charge the domain decomposition. Preliminary
comparisons between GPUs and CPUs were carried out on a simple diffusive
problem. As expected from literature, those tests show clearly that
best performance is obtained with GPU compared to CPU (Skylake or
KNL) even for best optimization of LBM kernels (CSoA2) which has been
developed for Intel Skylake. Here, comparisons were performed with
the same C++ source code. No low-level language (\texttt{CUDA} or
\texttt{OpenCL}) was used for GPUs.

Numerical implementation was checked with several test cases to validate
step-by-step the full model of fluid flows with phase change. The
conservative Allen-Cahn equation is validated with two test cases:
(i) Zalesak's slotted disk and (ii) interface deformation inside a
vortex. The coupling with Navier-Stokes equations is also checked
with two test cases: the layered Poiseuille flow and Laplace law.
Next, the coupling between equations of phase-field and temperature
were compared to the most general one-dimensional analytical solution
of the Stefan problem. Comparisons were done first by assuming identical
thermal diffusivities, and next by using various ratios of diffusivities
with an interface temperature that is different of the saturation
one. The full model was simulated on the test case of film
boiling on one GPU and one multicore CPU for two mesh sizes. Computational
times are clearly in favor of GPUs. Finally, the film boiling problem
is simulated with 8 parallel GPUs for mesh size that is twelve times
bigger than the previous one.

In this paper, foundations have been laid for improving performance
of lattice Boltzmann simulations in a context of quick evolution of
HPC platforms. In the future, a three-dimensional extension of the coupling terms is planned.
Next \texttt{LBM\_saclay} could be enriched
with other models requiring interface tracking such as crystal growth
and demixing of ternary fluids. Besides, the range of physical parameters
could be increased and the code stability could be enhanced by using
alternative collision operators such as those based on the Two-Relaxation-Times
and Multiple-Relaxation-Times.

\section*{Acknowledgments}
We would like to thank \noun{Mathis Plapp} for the insightful discussions on
theoretical aspects of phase-field models.

\appendix

\section{\label{sec:Derivation-Allen-Cahn}Removal of the driven-curvature
interface motion in Eq. (\ref{eq:Bilan_PhaseA_avecFlux})}

In this Appendix, the derivation of first term in the right-hand side
of Eq. (\ref{eq:Bilan_PhaseA_avecFlux}) is reminded. The advection
of phase index $\phi$ writes 

\begin{equation}
\frac{\partial\phi}{\partial t}+\mathbf{V}\cdot\boldsymbol{\nabla}\phi=0.\label{eq:Adv_Phi}
\end{equation}
If the total velocity \textbf{$\mathbf{V}$ }is defined as the sum of an external
advective velocity $\mathbf{u}$ of an incompressible fluid plus a
normal velocity of the interface $v_{n}\mathbf{n}$, then $\mathbf{V}\cdot\boldsymbol{\nabla}\phi=\mathbf{u}\cdot\boldsymbol{\nabla}\phi+v_{n}\bigl|\boldsymbol{\nabla}\phi\bigr|$.
For the second term, we have used the definition of normal vector
$\mathbf{n}=\boldsymbol{\nabla}\phi/\bigl|\boldsymbol{\nabla}\phi\bigr|$.
If the normal velocity $v_{n}$ is also assumed to be separated into
one term, $-M_{\phi}\kappa$, depending on the curvature $\kappa$
and another one, $\tilde{v}$ independent on $\kappa$ then: $v_{n}\bigl|\boldsymbol{\nabla}\phi\bigr|=-M_{\phi}\kappa\bigl|\boldsymbol{\nabla}\phi\bigr|+\tilde{v}\bigl|\boldsymbol{\nabla}\phi\bigr|$
and Eq. (\ref{eq:Adv_Phi}) writes:

\begin{equation}
\frac{\partial\phi}{\partial t}+\boldsymbol{\nabla}\cdot(\mathbf{u}\phi)=M_{\phi}\kappa\bigl|\boldsymbol{\nabla}\phi\bigr|-\tilde{v}\bigl|\boldsymbol{\nabla}\phi\bigr|.\label{eq:Allen-Cahn_Gen}
\end{equation}

For solidification problems, $\tilde{v}$ is the coupling with temperature
equation and ensures that the Gibbs-Thomson condition is well recovered.
A discussion on $\tilde{v}$ is presented at the end of this appendix.
The next stage of the derivation is to cancel the driven-curvature interface
motion $M_{\phi}\kappa\bigl|\boldsymbol{\nabla}\phi\bigr|$, without
setting $M_{\phi}=0$, but by adding a supplementary counter term:
$M_{\phi}\kappa\bigl|\boldsymbol{\nabla}\phi\bigr|-M_{\phi}\kappa\bigl|\boldsymbol{\nabla}\phi\bigr|=\mathcal{S}(\phi)$.
The purpose is to transform an hyperbolic-type PDE into a parabolic-type
PDE by expanding $\kappa$ in the first term with its definition $\kappa=\boldsymbol{\nabla}\cdot\mathbf{n}=\boldsymbol{\nabla}\cdot(\boldsymbol{\nabla}\phi/\bigl|\boldsymbol{\nabla}\phi\bigr|)$
in order to obtain an expression involving the laplacian of $\phi$:

\begin{equation}
\mathcal{S}(\phi)=M_{\phi}\left[\boldsymbol{\nabla}^{2}\phi-\frac{\boldsymbol{\nabla}\phi\cdot\boldsymbol{\nabla}\bigl|\boldsymbol{\nabla}\phi\bigr|}{\bigl|\boldsymbol{\nabla}\phi\bigr|}\right]-M_{\phi}\kappa\bigl|\boldsymbol{\nabla}\phi\bigr|.\label{eq:Curvature-Canceled}
\end{equation}
The main advantage of this formulation (Eq. (\ref{eq:Curvature-Canceled}))
is that, for a plane interface, i.e. $\kappa=0$, the equilibrium
solution of $\mathcal{S}(\phi)=0$ is an hyperbolic tangent. By using
the definition of $\mathbf{n}$, Eq. (\ref{eq:Curvature-Canceled})
becomes $\mathcal{S}(\phi)=M_{\phi}\left[\boldsymbol{\nabla}^{2}\phi-\mathbf{n}\cdot\boldsymbol{\nabla}\bigl|\boldsymbol{\nabla}\phi\bigr|\right]-M_{\phi}\bigl|\boldsymbol{\nabla}\phi\bigr|\boldsymbol{\nabla}\cdot\mathbf{n}$,
i.e. Eq. (\ref{eq:Allen-Cahn_Gen}) becomes

\begin{equation}
\frac{\partial\phi}{\partial t}+\boldsymbol{\nabla}\cdot(\mathbf{u}\phi)=M_{\phi}\left[\boldsymbol{\nabla}^{2}\phi-\mathbf{n}\cdot\boldsymbol{\nabla}\bigl|\boldsymbol{\nabla}\phi\bigr|-\bigl|\boldsymbol{\nabla}\phi\bigr|\boldsymbol{\nabla}\cdot\mathbf{n}\right]-\tilde{v}\bigl|\boldsymbol{\nabla}\phi\bigr|\label{eq:CAC_withNorm_Form1}
\end{equation}
which, after the straightforward manipulation $-\mathbf{n}\cdot\boldsymbol{\nabla}\bigl|\boldsymbol{\nabla}\phi\bigr|-\bigl|\boldsymbol{\nabla}\phi\bigr|\boldsymbol{\nabla}\cdot\mathbf{n}=-\boldsymbol{\nabla}\cdot(\bigl|\boldsymbol{\nabla}\phi\bigr|\mathbf{n})$
yields

\begin{equation}
\frac{\partial\phi}{\partial t}+\boldsymbol{\nabla}\cdot(\mathbf{u}\phi)=\boldsymbol{\nabla}\cdot\left[M_{\phi}(\boldsymbol{\nabla}\phi-\bigl|\boldsymbol{\nabla}\phi\bigr|\mathbf{n})\right]-\tilde{v}\bigl|\boldsymbol{\nabla}\phi\bigr|.\label{eq:CAC_withNorm_Form2}
\end{equation}
For calculating $\bigl|\boldsymbol{\nabla}\phi\bigr|$, the following
kernel function is used

\begin{equation}
\phi=\frac{1}{2}\left[1+\mbox{tanh}\left(\frac{\zeta}{aW}\right)\right],\label{eq:Phi_eq}
\end{equation}
where $\zeta$ is the normal coordinate of the interface, $a$ controls
the slope of the hyperbolic tangent and $W$ is the interface width.
The above kernel function ensures an hyperbolic tangent profile at
equilibrium. It is consistent with the profile obtained in a thermodynamically
derived phase-field model, such as the one used for computation of
chemical potential (Eq. (\ref{eq:ChemicalPotential})) with bulk phases
$\phi=0$ and $\phi=1$. The normal derivative of Eq. (\ref{eq:Phi_eq})
leads to

\begin{equation}
\bigl|\boldsymbol{\nabla}\phi\bigr|=\frac{\partial\phi}{\partial\zeta}=\frac{2}{aW}\phi(1-\phi).\label{eq:Norm_gradphi}
\end{equation}

Finally by setting $a=1/2$ the conservative Allen-Cahn equation with
a source term is

\begin{equation}
\frac{\partial\phi}{\partial t}+\boldsymbol{\nabla}\cdot(\mathbf{u}\phi)=\boldsymbol{\nabla}\cdot\left[M_{\phi}\left(\boldsymbol{\nabla}\phi-\frac{4}{W}\phi(1-\phi)\mathbf{n}\right)\right]-\tilde{v}\frac{4}{W}\phi(1-\phi).\label{eq:CAC}
\end{equation}

Eq. (\ref{eq:CAC}) is the Allen-Cahn equation for which the curvature-driven
displacement of the interface has been canceled with a counter term.
Let us notice that, if $\tilde{v}$ is chosen such as $\tilde{v}=\alpha(\theta_{I}-\theta)/(\mathscr{A}W)$
then $\tilde{v}\bigl|\boldsymbol{\nabla}\phi\bigr|\approx-(4\alpha/\mathscr{A}W^{2})(\theta_{I}-\theta)\phi(1-\phi)$
can be used in Eq. (\ref{eq:CAC}) for the problem of phase change.
The release or absorption of latent heat at the interface is taken
into account in the temperature equation by the time derivative of
$\phi$. If the physical problem necessitates a curvature-driven interface
motion, the curvature term must be kept in the Allen-Cahn equation
and then only the first term in the right-hand side of Eq. (\ref{eq:Curvature-Canceled})
appears in the derivation. With $a=1/2$, the term $\boldsymbol{\nabla}\phi\cdot\boldsymbol{\nabla}\bigl|\boldsymbol{\nabla}\phi\bigr|/\bigl|\boldsymbol{\nabla}\phi\bigr|$
is equal to

\begin{equation}
\frac{\boldsymbol{\nabla}\phi\cdot\boldsymbol{\nabla}\bigl|\boldsymbol{\nabla}\phi\bigr|}{\bigl|\boldsymbol{\nabla}\phi\bigr|}=\frac{\partial^{2}\phi}{\partial\zeta^{2}}=\frac{16}{W^{2}}\phi(1-\phi)(1-2\phi)\label{eq:SecondTerm}
\end{equation}
The curvatuve-driven term writes

\begin{equation}
M_{\phi}\kappa\bigl|\boldsymbol{\nabla}\phi\bigr|=M_{\phi}\left[\boldsymbol{\nabla}^{2}\phi-\frac{16}{W^{2}}\phi(1-\phi)(1-2\phi)\right].\label{eq:Driven-Curvature_Term}
\end{equation}

\section{\label{sec:Numerical-value-of_A}Numerical value of coefficient $\mathscr{A}$}

When the matched asymptotic expansions are carried out on the one-dimensional
phase-field model, the coefficient $\mathscr{A}$ is defined by four
integrals $\mathscr{I}$, $\mathcal{J}$, $\mathcal{G}$ and $\mathcal{U}$
by (e.g. \cite[Eq. (59)]{Karma-Rappel_PRE1998}):

\begin{equation}
\mathscr{A}=\frac{\mathcal{G}+\mathcal{J}\mathcal{U}}{2\mathscr{I}},\label{eq:Def_A}
\end{equation}
with

\begin{equation}
\mathscr{I}=\int_{-\infty}^{\infty}d\zeta(\partial_{\zeta}\phi_{0})^{2},\quad\mathcal{J}=\int_{-\infty}^{\infty}d\zeta(\partial_{\zeta}\phi_{0})p_{\phi}^{0},\quad\mathcal{G}=\int_{-\infty}^{\infty}d\zeta(\partial_{\zeta}\phi_{0})p_{\phi}^{0}\int_{0}^{\zeta}d\zeta'h^{0},\quad\mbox{and}\quad\mathcal{U}=\int_{-\infty}^{0}d\zeta h^{0}.\label{eq:Def_Integrals}
\end{equation}
In Eq. (\ref{eq:Def_Integrals}), the functions $\phi_{0}$, $p_{\phi}^{0}$
and $h^{0}$ of our model are defined such as

\begin{equation}
\phi_{0}=\frac{1}{2}\left[1+\tanh\left(\frac{2\zeta}{W}\right)\right],\qquad p_{\phi}^{0}=\phi_{0}(1-\phi_{0}),\qquad\mbox{and}\qquad h^{0}=\phi_{0}\label{eq:OurFunctions}
\end{equation}
Those integrals can be computed analytically and yield a numerical value
provided that the interface width $W$ is set. Here, to be consistent
with the rescaling of space and the analysis performed in \cite{Karma-Rappel_PRE1998}, it is enough to set $W=2\sqrt{2}$, and the integrals are:

\begin{equation}
\mathscr{I}=\frac{1}{3\sqrt{2}},\qquad\mathcal{J}=\frac{1}{6},\qquad\mathcal{G}=-\frac{(12\ln2-10)}{72\sqrt{2}},\qquad\mbox{and}\qquad\mathcal{U}=\frac{\ln2}{\sqrt{2}}.\label{eq:Integrals_NumValues}
\end{equation}
Finally Eq. (\ref{eq:Def_A}) yields

\begin{equation}
\mathscr{A}=\frac{10}{48}\cong0.20833.\label{eq:ValueOfA}
\end{equation}

\section{\label{sec:DLBE}Discrete lattice Boltzmann equations}

In this Appendix, the variable change for the discrete lattice Boltzmann
equation is reminded in \ref{sub:Variable-change-for}. In \ref{sub:Equiv_AllenCahn},
we will show that, for CAC model, the formulation with a source term
is equivalent to the formulation with a modification of the equilibrium
distribution function.

\subsection{\label{sub:Variable-change-for}Variable change for discrete lattice
Boltzmann equation}

The discrete lattice Boltzmann equation with an external force or
source term $\mathcal{S}_{i}^{\vartheta}$ can be written with the
BGK collision term:

\begin{equation}
\frac{\partial\vartheta_{i}}{\partial t}+\mathbf{c}_{i}\cdot\boldsymbol{\nabla}\vartheta_{i}=-\frac{\vartheta_{i}-\vartheta_{i}^{eq}}{\tau_{\vartheta}}+S_{i}^{\vartheta}.\label{eq:DLBE}
\end{equation}
In what follows, the calculations will be performed by setting $\vartheta\equiv f$,
$\mathcal{S}_{i}^{\vartheta}=\mathcal{S}_{i}^{f}=\mathcal{S}_{i}$
and $\tau_{\vartheta}\equiv\tau$ but the variable change derivation
holds also for $\vartheta\equiv h$ and $\vartheta\equiv s$. Terms
that are evaluated at position $\mathbf{x}$ and time $t$ are noted
$f_{i}\equiv f_{i}(\mathbf{x},\,t)$, $f_{i}^{eq}\equiv f_{i}^{eq}(\mathbf{x},\,t)$
and $\mathcal{S}_{i}\equiv\mathcal{S}_{i}(\mathbf{x},\,t)$, whereas
terms evaluated at position $\mathbf{x}+\mathbf{c}_{i}\delta t$ and
time $t+\delta t$ are noted with a star: $f_{i}^{\star}\equiv f_{i}(\mathbf{x}+\mathbf{c}_{i}\delta t,\,t+\delta t)$,
$f_{i}^{\star eq}\equiv f_{i}^{eq}(\mathbf{x}+\mathbf{c}_{i}\delta t,\,t+\delta t)$
and $\mathcal{S}_{i}^{\star}\equiv\mathcal{S}_{i}(\mathbf{x}+\mathbf{c}_{i}\delta t,\,t+\delta t)$.
With those notations, integration of Eq. (\ref{eq:DLBE}) over $t$
and $t+\delta t$ yields:

\begin{equation}
f_{i}^{\star}=f_{i}-\frac{\delta t}{2\tau}\left(f_{i}^{\star}-f_{i}^{\star eq}\right)-\frac{\delta t}{2\tau}\left(f_{i}-f_{i}^{eq}\right)+\frac{\delta t}{2}\mathcal{S}_{i}^{\star}+\frac{\delta t}{2}\mathcal{S}_{i}\label{eq:LBE_Trapezoidal}
\end{equation}
where the trapezoidal rule was applied for the right-hand side of
Eq. (\ref{eq:DLBE}). In this expression, the natural variable change
for implicit terms is

\begin{equation}
\overline{f}_{i}^{\star}=f_{i}^{\star}+\frac{\delta t}{2\tau}\left(f_{i}^{\star}-f_{i}^{\star eq}\right)-\frac{\delta t}{2}\mathcal{S}_{i}^{\star}.\label{eq:VariableChange}
\end{equation}
The same variable change is used for $\overline{f}_{i}$:

\begin{equation}
\overline{f}_{i}=f_{i}+\frac{\delta t}{2\tau}\left(f_{i}-f_{i}^{eq}\right)-\frac{\delta t}{2}\mathcal{S}_{i}.\label{eq:VariableChange_Expl}
\end{equation}

By inverting the latter relation in order to express $f_{i}$ with
respect to $\overline{f}_{i}$ , we obtain:

\begin{equation}
f_{i}=\frac{2\tau}{2\tau+\delta t}\left(\overline{f}_{i}+\frac{\delta t}{2\tau}f_{i}^{eq}+\frac{\delta t}{2}\mathcal{S}_{i}\right).\label{eq:f_versus_fbarre}
\end{equation}

With Eqs. (\ref{eq:VariableChange}) and (\ref{eq:f_versus_fbarre}),
Eq. (\ref{eq:LBE_Trapezoidal}) becomes

\begin{equation}
\overline{f}_{i}^{\star}=\overline{f}_{i}-\frac{\delta t}{\tau+\delta t/2}\left(\overline{f}_{i}-f_{i}^{eq}+\frac{\delta t}{2}\mathcal{S}_{i}\right)+\delta t\mathcal{S}_{i}\label{eq:LBE_Intermediate}
\end{equation}

At this stage, if we define a new variable change

\begin{equation}
\overline{f}_{i}^{eq}=f_{i}^{eq}-\frac{\delta t}{2}\mathcal{S}_{i},\label{eq:Feq_barre}
\end{equation}
then Eq. (\ref{eq:LBE_Intermediate}) is equivalent to

\begin{equation}
\overline{f}_{i}^{\star}=\overline{f}_{i}-\frac{\delta t}{\tau+\delta t/2}\left(\overline{f}_{i}-\overline{f}_{i}^{eq}\right)+\delta t\mathcal{S}_{i}.\label{eq:LBE_Form1}
\end{equation}

Without using the previous variable change for $f_{i}^{eq}$, Eq.
(\ref{eq:LBE_Intermediate}) is equivalent to

\begin{equation}
\overline{f}_{i}^{\star}=\overline{f}_{i}-\frac{\delta t}{\tau+\delta t/2}\left(\overline{f}_{i}-f_{i}^{eq}\right)+\frac{\tau\delta t}{\tau+\delta t/2}\mathcal{S}_{i},\label{eq:LBE_Form2}
\end{equation}
where only the factor in front of the source term is modified.

By introducing the dimensionless collision rate which is defined by
$\overline{\tau}=\tau/\delta t$, Eq. (\ref{eq:LBE_Form1}) finally
writes

\begin{equation}
\overline{f}_{i}^{\star}=\overline{f}_{i}-\frac{1}{\overline{\tau}+1/2}\left(\overline{f}_{i}-\overline{f}_{i}^{eq}\right)+\delta t\mathcal{S}_{i},\label{eq:LBE_2ndOrderTime_Form1}
\end{equation}
or alternatively,

\begin{equation}
\overline{f}_{i}^{\star}=\overline{f}_{i}-\frac{1}{\overline{\tau}+1/2}\left(\overline{f}_{i}-f_{i}^{eq}\right)+\frac{\overline{\tau}\delta t}{\overline{\tau}+1/2}\mathcal{S}_{i}.\label{eq:LBE_2ndOrderTime_Form2}
\end{equation}

In Section \ref{sec:Lattice-Boltzmann-methods}, Eq. (\ref{eq:LBE_2ndOrderTime_Form1})
is the starting point for each lattice Boltzmann equation. The variable
change Eq. (\ref{eq:VariableChange_Expl}) leads to the calculation
of the zeroth-order moment:

\begin{equation}
\mathcal{M}_{0}=\sum_{i}\overline{f}_{i}+\frac{\delta t}{2}\sum_{i}\mathcal{S}_{i}\label{eq:Moment_Order0}
\end{equation}

\subsection{\label{sub:Equiv_AllenCahn}Equivalence of lattice Boltzmann formulations
for the Allen-Cahn equation}

The purpose of this Appendix is to prove the equivalence between the
source term and the modification of the equilibrium distribution function.
The lattice Boltzmann scheme for the conservative Allen-Cahn equation
is (Eq. (\ref{eq:LBE_PhaseField_Eq}) with $\mathcal{S}_{i}^{g}$
defined by Eq. (\ref{eq:SourceTerm_LBE_G})):

\begin{eqnarray}
\overline{g}_{i}^{\star} & = & \overline{g}_{i}-\frac{1}{\overline{\tau}_{g}+1/2}\left[\overline{g}_{i}-\overline{g}_{i}^{eq}\right]+\mathcal{P}_{i}^{g}\delta t+\mathcal{F}_{i}^{g}\delta t\label{eq:LBE_PhaseField_Eq-1}
\end{eqnarray}
with the mobility coefficient defined by $M_{\phi}=\overline{\tau}_{g}c_{s}^{2}\delta t$.
By using the definition of $g_{i}^{eq}$ for $\overline{g}_{i}^{eq}=\phi\Gamma_{i}-\delta t\mathcal{P}_{i}^{g}/2-\mathcal{F}_{i}^{g}\delta t/2$
and gathering the term $\mathcal{F}_{i}^{g}\delta t$ inside the bracket,
we obtain

\begin{equation}
\overline{g}_{i}^{\star}=\overline{g}_{i}-\frac{1}{\overline{\tau}_{g}+1/2}\left[\overline{g}_{i}-\phi\Gamma_{i}-\mathcal{F}_{i}^{g}\overline{\tau}_{g}\delta t+\frac{\delta t}{2}\mathcal{P}_{i}^{g}\right]+\mathcal{P}_{i}^{g}\delta t\label{eq:Equiv1}
\end{equation}

Next, the collision rate is replaced by its mobility $\overline{\tau}{}_{g}=M_{\phi}/(c_{s}^{2}\delta t)$:

\begin{equation}
\overline{g}_{i}^{\star}=\overline{g}_{i}-\frac{1}{\overline{\tau}_{g}+1/2}\left[\overline{g}_{i}-\phi\Gamma_{i}-\mathcal{F}_{i}^{g}\frac{M_{\phi}}{c_{s}^{2}}+\frac{\delta t}{2}\mathcal{P}_{i}^{g}\right]+\mathcal{P}_{i}^{g}\delta t\label{eq:Equiv2}
\end{equation}

Finally, if we use the definition of $\mathcal{F}_{i}^{g}$ given
by Eq. (\ref{eq:SourceTerm_AllenCahn}), the Allen-Cahn equilibrium
distribution function $g_{i}^{eq,\,CAC}$ can be defined by \cite{Fakhari_etal_JCP2017}

\begin{equation}
g_{i}^{eq,\,CAC}=\phi\Gamma_{i}+M_{\phi}\frac{4}{W}\phi(1-\phi)w_{i}\frac{\mathbf{c}_{i}\cdot\mathbf{n}}{c_{s}^{2}}\label{eq:Geq_AC}
\end{equation}
and the alternative lattice Boltzmann equation is

\begin{equation}
\overline{g}_{i}^{\star}=\overline{g}_{i}-\frac{1}{\overline{\tau}_{g}+1/2}\left[\overline{g}_{i}-\overline{g}{}_{i}^{eq,\,CAC}\right]+\mathcal{P}_{i}^{g}\delta t\label{eq:Alt_LBE_PhaseField}
\end{equation}
with $\overline{g}_{i}^{eq,\,CAC}=g_{i}^{eq,\,CAC}-\mathcal{P}_{i}^{g}\delta t/2$
with $\mathcal{P}_{i}^{g}$ defined by Eq. (\ref{eq:SourceTerm_AllenCahn}).


\bibliographystyle{elsarticle-num}
\bibliography{CMAME-D-20-00414_BibTeX}

\end{document}